\documentclass[12pt]{article}
\newcommand{\be}{\begin{equation}}
\newcommand{\ee}{\end{equation}}
\newcommand{\bea}{\begin{eqnarray}}
\newcommand{\eea}{\end{eqnarray}}

\usepackage{amsmath,amssymb,color,graphics,amscd,amsfonts,mathrsfs,epsf,indentfirst}
\usepackage{bm}
\usepackage{subfigure}
\usepackage{multirow,graphicx}

\title{What happens at the horizon(s) of an extreme black hole?}

\author{Keiju Murata$^a$, Harvey S. Reall$^b$ and Norihiro Tanahashi$^c$\\ 
\small \sl $^a$ Department of Physics, and Research and Education Center for Natural Sciences,\\
\small \sl Keio University, 4-1-1 Hiyoshi, Yokohama 223-8521, Japan\\
\small \sl $^b$ DAMTP, University of Cambridge, Wilberforce Road, Cambridge CB3 0WA, UK\\
\small \sl $^c$ Kavli Institute for the Physics and Mathematics of the Universe (WPI),\\
\small \sl The University of Tokyo, Kashiwa, Chiba 277-8583, Japan
}

\begin{document}

\maketitle

\begin{abstract}
A massless scalar field exhibits an instability at the event horizon of an extreme black hole. We study numerically the nonlinear evolution of this instability for spherically symmetric perturbations of an extreme Reissner-Nordstrom (RN) black hole. We find that generically the endpoint of the instability is a non-extreme RN solution. However, there exist fine-tuned initial perturbations for which the instability never decays. In this case, the perturbed spacetime describes a time-dependent extreme black hole. Such solutions settle down to extreme RN outside, but not on, the event horizon. The event horizon remains smooth but certain observers who cross it at late time experience large gradients there. Our results indicate that these dynamical extreme black holes admit a $C^1$ extension across an inner (Cauchy) horizon.
\end{abstract}

\section{Introduction}

Recently, Aretakis has demonstrated the existence of an instability of a massless scalar field at the horizon of an extreme Reissner-Nordstrom (RN) black hole \cite{Aretakis:2011ha,Aretakis:2011hc}. For general initial data specified on a surface intersecting the event horizon, the field and its derivatives decay {\it outside} the horizon. However, a conservation law ensures that the first transverse derivative of the field {\it on} the event horizon generically does not decay. Instead it approaches a constant value at late time. This implies that a second derivative of the field generically grows with time on the horizon: an instability. Similar results apply to an axisymmetric scalar field in the extreme Kerr geometry \cite{Aretakis:2011gz,Aretakis:2012ei}. 

This instability can be understood physically as follows. Consider a photon travelling along a null generator of the event horizon of a stationary black hole. The conserved energy of such a photon is zero. However, the energy as measured by a family of identical infalling observers is non-zero and redshifts as $e^{-\kappa v}$ where $\kappa$ is the surface gravity and $v$ a Killing time coordinate. So, for a non-extreme black hole, outgoing radiation at the horizon decays. However, for an extreme black hole, $\kappa=0$ so the horizon redshift effect is absent: outgoing radiation at the horizon does not decay.

Aretakis' results have been extended in several directions. It has been
argued that the massless scalar field instability occurs for any extreme
black hole \cite{Lucietti:2012sf}. A similar instability occurs for
linearized gravitational perturbations of extreme Kerr
\cite{Lucietti:2012sf} and certain higher dimensional extreme vacuum black holes \cite{Murata:2012ct}; and also for massive scalar field, or coupled linearized
gravitational and electromagnetic, perturbations of extreme RN
\cite{Lucietti:2012xr}. It has been observed that if no outgoing radiation is present initially at the horizon then an instability, if one exists, should be milder in the sense that it will afflict quantities involving more transverse derivatives \cite{Dain:2012qw}. Nevertheless, it has been argued in Refs.~\cite{Lucietti:2012xr,Bizon:2012we}, and proved in Ref.~\cite{Aretakis:2012bm} that such an instability does indeed exist: for a massless scalar in extreme RN, an ingoing wavepacket generically results in blow-up of a third transverse derivative of the scalar field at the horizon. Very recently, Aretakis studied a test scalar field with a nonlinear self-interaction in the extreme Kerr geometry~\cite{Aretakis:2013dpa}. He found that nonlinearity makes the instability stronger: generically a second derivative at the horizon blows up in {\it finite} time.

So far, all discussions of this instability have considered the case of test fields in a fixed extreme black hole spacetime. But what happens when one takes into account gravitational backreaction? What is the ``final state'' of the instability? These are the questions we will address in this paper. It has been conjectured by Dafermos \cite{mihalis} that:  the instability will persist when backreaction is included; generically the endpoint will be a stationary non-extreme black hole; there exist non-generic initial perturbations for which the instability never ends. Our results support this conjecture.

\begin{figure}
\begin{center}
\includegraphics[scale=0.4]{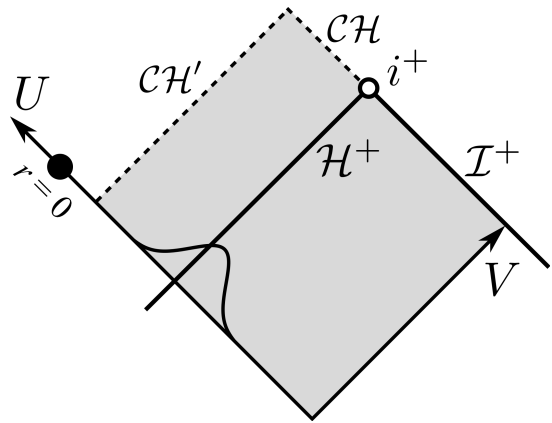}
\end{center}
\caption{
Penrose diagram showing initial data prescribed on a pair of null hypersurfaces, the $U$ and $V$ axes, whose intersection is a sphere. The location of the future event horizon ${\cal H}^+$ and future null infinity ${\cal I}^+$ are also shown. The initial data for the scalar field describes an outgoing wavepacket near ${\cal H}^+$: this data vanishes on the outgoing null hypersurface and is compactly supported near ${\cal H}^+$ on the ingoing null hypersurface. The initial data has a singularity denoted by the heavy dot. We construct a numerical solution in the shaded region. This is bounded by Cauchy horizons ${\cal CH}$ and ${\cal CH}'$.
}
 \label{initial}
\end{figure}

We will consider Einstein-Maxwell theory coupled to a massless scalar field, assuming spherical symmetry. 
We will construct initial data describing an extreme RN black hole perturbed by an outgoing scalar field wavepacket at the event horizon. We will use a characteristic initial value formulation of the problem. Initial data is prescribed on ingoing and and outgoing null hypersurfaces which intersect in a 2-sphere as shown in Fig.~\ref{initial}. After fixing a gauge, initial data is uniquely determined by specifying the (conserved) electric charge $Q>0$ (which just sets a scale), the initial Bondi mass $M_i$, and the scalar field profile on the two hypersurfaces. We take the scalar field to be an outgoing wavepacket as shown. The shape of the wavepacket will be fixed but we will consider different values $\epsilon$ for its amplitude. 

Our initial data coincides with that of a RN solution except where the scalar field is non-trivial. This implies that the ingoing null hypersurface ends at $r=0$ where there is a curvature singularity. For large enough $M_i$, this singularity will be hidden behind an event horizon of the full spacetime but if $M_i$ is taken too low then the singularity will be naked, as for super-extreme RN. We will consider only spacetimes for which there is an event horizon. 

Our initial data reduces to that of extreme RN when $\epsilon = 0$ and $M_i = Q$. Hence the question of stability of extreme RN involves investigating the behaviour of the solution as $\epsilon \rightarrow 0$ and $M_i \rightarrow Q+$. It is convenient to reduce to a 1-parameter family of initial data by taking $M_i$ to be some function of $\epsilon$ such that $M_i(\epsilon) \rightarrow Q+$ as $\epsilon \rightarrow 0$. We have studied two different choices for this function. 

Our first choice has $M_i(\epsilon) = Q + {\cal O}(\epsilon^2)$. The linearization of such a solution has vanishing metric and Maxwell field perturbations: it corresponds to a test scalar field in a fixed extreme RN background, precisely the model considered by Aretakis. Since the metric perturbation is ${\cal O}(\epsilon^2)$, 
one might guess that backreaction is negligible. This is incorrect because, at extremality, a second order metric perturbation causes a first order change in the position of the horizon. Hence it is conceivable that backreaction shifts the instability behind the horizon. 

We find that the solution eventually settles down to a non-extreme RN black hole with surface gravity $\kappa = {\cal O}(\epsilon)$. However, for an (Eddington-Finkelstein-like) time of order $1/\epsilon$, the scalar field exhibits features which are similar to the Aretakis instability. At the horizon, the transverse derivative of the field decays exponentially in time but the coefficient in the exponent is $\kappa$ so this decay is very slow. The second transverse derivative of the field at the event horizon grows for a time of order $1/\epsilon$ before undergoing slow exponential decay. The maximum value of this second derivative at the horizon does not vanish as $\epsilon \rightarrow 0$. This demonstrates that an instability exists in the nonlinear theory. 

Our second choice of initial data has $M_i = Q + {\cal O}(\epsilon)$, i.e., it allows for a first order metric/Maxwell perturbation. Again, the solution eventually settles down to a non-extreme RN black hole, this time with $\kappa = {\cal O}(\epsilon^{1/2})$. In this case, we find that the scalar field and its first two derivatives decay on, and outside, the event horizon. However, the {\it third} transverse derivative of the scalar field at the horizon has a maximum value that does not vanish as $\epsilon \rightarrow 0$. Hence an instability is still present but it is milder than for the case just discussed.

In our numerical solutions, the metric is always close to a non-extreme RN black hole. This suggests that our results should be similar to the case of a test scalar field (of amplitude $\epsilon$) in a non-extreme RN geometry with $\kappa = {\cal O}(\epsilon)$ or ${\cal O}(\epsilon^{1/2})$. We find that this is indeed the case, and this toy model can be used to reproduce the essential features of our numerical solution.

We have considered also the case of an extreme RN black hole perturbed by an {\it ingoing} wavepacket. In this case, we find that generically there is an instability in the third derivative of the scalar field on the event horizon, just as occurs for the test field case \cite{Lucietti:2012xr,Bizon:2012we,Aretakis:2012bm}.

Returning to outgoing perturbations: generically we obtain a spacetime which settles down to a non-extreme RN solution. It is interesting to ask whether it is possible to fine tune the initial data so that the spacetime settles down to {\it extreme} RN. If we reduce $M_i$ slightly then the final RN black hole is closer to extremality. If we take $M_i$ too small then we obtain initial data for a spacetime with no event horizon: a naked singularity. Let $M_*(\epsilon)$ be the critical value of $M_i$ such that an event horizon does not form when $M_i<M_*(\epsilon)$ but one does form when $M_i>M_*(\epsilon)$. It seems plausible that initial data with $M_i = M_*(\epsilon)$ will approach an extreme RN black hole at late time.

One might wonder whether a solution evolving to an extreme black hole at late time contradicts the third law of black hole mechanics: ``a non-extreme black hole cannot become extreme in any physical process''. This is not very precise e.g.\ one needs a definition of ``extreme'' applicable to a time-dependent spacetime. One such definition has been given by Israel \cite{israel} who defined extremality as the absence of trapped surfaces. His version of the third law states that if trapped surfaces are present on some initial Cauchy surface $\Sigma$ then they will also be present on any Cauchy surface lying to the future of $\Sigma$. This does not exclude time-dependent black holes which are extreme forever, i.e., time-dependent black hole spacetimes without trapped surfaces. We will refer to such a spacetime as a {\it dynamical extreme black hole}.

We will argue below that a solution with $M_i = M_*(\epsilon)$ is a dynamical extreme black hole. We find that such a solution settles down to extreme RN {\it outside} the event horizon. However, {\it on} the event horizon, the scalar field behaves exactly as in the test field case studied by Aretakis. At late time, the scalar field decays on and outside the event horizon but its transverse derivative at the horizon approaches a constant dependent on the initial perturbation. This could be regarded as ``hair'' on the horizon of the black hole. The solution remains smooth at the event horizon however the second transverse derivative of the scalar field there grows without bound, i.e., it blows up. 

We have also studied the black hole interior. It is well-known that the Cauchy horizon (${\cal CH}$ in Fig.~\ref{initial}) of a non-extreme RN black hole is unstable against linearized perturbations \cite{chandra}. For nonlinear, spherically symmetric perturbations, analytic \cite{Poisson:1989zz,Poisson:1990eh,Ori:1991zz} and numerical \cite{Brady:1995ni,Burko:1997zy} studies suggest the following picture. Near $i^+$, there exist $C^0$ extensions of the metric, Maxwell and scalar fields across ${\cal CH}$. However, at ${\cal CH}$, the invariant Hawking mass will diverge (``mass inflation''). This implies that ${\cal CH}$ is actually a null curvature singularity (at least near $i^+$). For the theory considered here, Ref.~\cite{Dafermos:2003wr} proved rigorously the existence of a $C^0$ extension across ${\cal CH}$ and, subject to an assumption concerning the decay of the scalar field along the event horizon, that the Hawking mass diverges at ${\cal CH}$.

The significance of these results is that they show that the metric (and other fields) cannot be extended in a way in which the equations of motion are satisfied, even in a weak sense, at the Cauchy horizon. This supports the strong cosmic censorship conjecture.\footnote{
Christodoulou has suggested that strong cosmic censorship should be formulated as non-existence of a $C^0$ extension of the metric for which the Christoffel symbols are locally square integrable  \cite{chr}. This implies non-existence of an extension which is a weak solution of the equations of motion. See the introduction of Ref.~\cite{Dafermos:2012np} for a more detailed discussion and a nice review of results on black hole interiors.}

We find a qualitative difference for dynamical extreme black holes. It is easy to show that the Hawking mass is bounded inside the black hole so there is no mass inflation. Our numerical results indicate that the fields and their derivatives extend continuously to ${\cal CH}$. This implies that there exist (non-unique) $C^1$ extensions across ${\cal CH}$. In particular, it is possible to extend the fields as a (weak) solution of the equations of motion, in contrast with the non-extreme case. Of course, extreme black holes are non-generic so there is no disagreement with strong cosmic censorship.

Marolf and Ori \cite{Marolf:2010nd,Marolf:2011dj} have discussed the experience of observers who fall freely into an extreme black hole. They considered a freely falling observer, with energy $E$, who crosses the event horizon at advanced time $v$. Such an observer can calculate the gradient of, say, our scalar field $\phi$ with respect to proper time $\tau$. They suggested that, in the limit $v \rightarrow \infty$ (at fixed $E$), $d\phi/d\tau$ will diverge within a vanishingly short proper time after crossing the horizon. So a very late time freely falling observer of given energy would regard the horizon as effectively singular. Our results for the interior of a dynamical extreme black hole support this conjecture. 

This paper is organized as follows. In section \ref{setup} we describe the Einstein-Maxwell-scalar field model that we will study, review some results concerning spherically symmetric solutions, and describe our initial data. Section \ref{NIERN} presents our results for the generic case in which the spacetime eventually settles down to non-extreme RN. Results in this section are for initial data describing an outgoing wavepacket. The case of an ingoing wavepacket is discussed in Appendix \ref{app:ingoing}. Our results for dynamical extreme black holes are presented in section \ref{extreme}. We discuss our results further in section \ref{discussion}.

\section{Set up for numerical simulations}

\label{setup}

\subsection{Basic equations}

We consider Einstein-Maxwell theory coupled to a massless scalar field. The action is 
\begin{equation}
 S=\int d^4x \sqrt{-g}\left[
R-\frac{1}{4}F_{\mu\nu}F^{\mu\nu}
-\frac{1}{2}\partial_\mu\phi \partial^\mu\phi
\right]\ .
\end{equation}
We impose spherical symmetry and use double null coordinates:
\begin{equation}
\label{anzatz}
 ds^2=-f(U,V)dUdV+r^2(U,V)d\Omega^2\ ,\quad
 \phi=\phi(U,V)\ ,\quad
 F=\alpha(U,V)dU\wedge dV\ .
\end{equation}
Lines of constant $U$ ($V$) are outgoing (ingoing) radial null geodesics. The equation of motion for the Maxwell field gives $\partial_V(r^2\alpha/f)=\partial_U(r^2\alpha/f)=0$. Thus we have 
\begin{equation}
\alpha=\frac{Qf}{r^2}\ ,
\end{equation}
where the constant $Q$ is the electric charge of the spacetime.
Using this equation, we can eliminate $\alpha$ from the other equations of motion, which are the evolution equations
\begin{align}
& (\log f)_{,UV}=
\frac{f}{2r^2}+\frac{2}{ r^2} r_{,U}r_{,V}-\frac{Q^2f}{r^4}-\frac{1}{2} \phi_{,V}\phi_{,U}\ ,
\label{fUV}\\
&rr_{,UV} + r_{,U} r_{,V} =-\frac{f}{4}\left( 1-\frac{Q^2}{r^2}\right)\
 ,\label{rUV}\\
&\phi_{,UV}=-\frac{1}{r}(r_{,U}\phi_{,V}+r_{,V}\phi_{,U})\ ,
\label{phiUV}
\end{align}
and the constraints (the Raychaudhuri equations for outgoing and ingoing radial null geodesics)
\begin{align}
&C_1\equiv-rf \left [ \left( f^{-1} r_{,V} \right)_{,V} +\frac{1}{4}r f^{-1}  \phi_{,V}^2 \right]=0\ ,
\label{C1}\\
&C_2\equiv-rf \left [ \left( f^{-1} r_{,U} \right)_{,U} +\frac{1}{4}r f^{-1}  \phi_{,U}^2 \right]=0\ .
\label{C2}
\end{align}
The evolution equations imply that the constraints are preserved:
\begin{equation}
 \partial_U C_1=0\ ,\qquad
 \partial_V C_2=0\ .
\label{eqconst}
\end{equation}
We will prescribe initial data on a pair of intersecting null hypersurfaces (Fig.~\ref{initial})
\begin{equation}
 \Sigma=\Sigma_1 \cup \Sigma_2\ ,
\end{equation}
where
\begin{equation}
\Sigma_1=\{(U,V)| (U\geq U_0, V=V_0)\}\ ,\quad
\Sigma_2=\{(U,V)| (U=U_0, V\geq V_0)\}\ .
\end{equation}
$\Sigma_1$ and $\Sigma_2$ intersect in a 2-sphere with coordinates $(U_0,V_0)$. We will integrate numerically the equations of motion to determine the solution in the future domain of dependence of $\Sigma$, which is a subset of the region $\{U \ge U_0, V \ge V_0 \}$. From Eq.~(\ref{eqconst}), we can see that, if the constraint equations are satisfied on $\Sigma$, then they are also satisfied in the whole spacetime.

In Appendix \ref{RNDN}, we explain how to write the RN solution in the above form. 

\subsection{Trapped surfaces, apparent horizon}

\label{trappedapparent}

In this section we will review definitions and some well-known results concerning trapped surfaces and apparent horizons in spherical symmetry \cite{christodoulou}. 

Far from the black hole $r$ will decrease (increase) along ingoing (outgoing) 
null geodesics so $r_{,U}<0$ and $r_{,V}>0$. We will consider only initial data 
for which $r_{,U}<0$ everywhere along $\Sigma_1$ (in particular, $\Sigma_1$ does 
not contain ``anti-trapped'' surfaces such as can occur in white holes). Our initial data will also have $r_{,U}<0$ on $\Sigma_2$. Eq.~(\ref{C2}) then implies that $r_{,U}<0$ in the future domain of dependence of $\Sigma$. In contrast, the sign of $r_{,V}$ can change. The 2-sphere $(U,V)$ is {\it trapped} if $r_{,V}<0$. It is {\it marginally trapped} if  $r_{,V}=0$. On a surface of constant $V$, the {\it apparent horizon} is the outermost (smallest $U$) marginally trapped 2-sphere. The apparent horizon of the full spacetime is defined to be the union of the apparent horizons of all surfaces of constant $V$. 

Eq.~(\ref{C1}) implies that if $r_{,V} \le 0$ on some 2-sphere then $r_{,V} \le 0$ along the outgoing null geodesics from that 2-sphere, hence $r$ is non-increasing along these geodesics so they cannot reach ${\cal I}^+$. Therefore the apparent horizon and any (marginally) trapped surfaces must lie in the black hole region of the spacetime. 

For a RN black hole solution, the apparent horizon coincides with the event horizon. For non-extreme RN, 2-spheres that lie between the inner (Cauchy) and outer (event) horizons are trapped. The (analytically extended) extreme RN solution has no trapped 2-spheres.

\subsection{Quasi-local mass}

In spherical symmetry, the quasi-local {\it Hawking mass} $m(U,V)$ is defined by
\begin{equation}
 1 - \frac{2m}{r} = g^{\mu\nu} \nabla_\mu r \nabla_\nu r = - 4 f^{-1} r_{,U} r_{,V}\ .
\end{equation}
In the charged case, it is convenient to follow Ref.~\cite{Poisson:1990eh} and introduce the {\it renormalized Hawking mass} $\varpi$ by replacing the LHS above with $1 - 2\varpi/r + Q^2/r^2$, which gives
\begin{equation}
 \varpi(U,V)=\frac{r}{2}\left(1+\frac{4r_{,U}r_{,V}}{f}+\frac{Q^2}{r^2}\right)\ .
\label{MHdef}
\end{equation}
The RN solution with mass $M$ and charge $Q$ has $\varpi(U,V)=M$ everywhere. Differentiating $\varpi$ with respect to $V$ and $U$ gives
\begin{equation}
\varpi_{,V}=-\frac{r^2r_{,U}}{2f}(\phi_{,V})^2\ ,\qquad
 \varpi_{,U}=-\frac{r^2r_{,V}}{2f}(\phi_{,U})^2\ ,
\label{dMPI}
\end{equation}
where we eliminated $r_{,UV}$, $r_{,VV}$ and $r_{,UU}$ using 
Eqs.~(\ref{rUV}), (\ref{C1}) and (\ref{C2}). As explained above, $r_{,U}<0$ everywhere so $\varpi_{,V} \ge 0$, i.e., $\varpi$ is non-decreasing along outgoing null geodesics. Outside any (marginally) trapped region, in particular outside ${\cal H}^+$, we have $r_{,V}>0$ and hence $\varpi_{,U} \le 0$ so $\varpi$ is non-increasing along ingoing null geodesics. 

Let $U=U_H$ be the event horizon. Outgoing null geodesics of constant $U<U_H$ will reach ${\cal I}^+$. For $U<U_H$ the {\it Bondi mass} is defined by
\begin{equation}
\label{bondidef}
 M_B(U) = \lim_{V \rightarrow \infty} \varpi(U,V)\ .
\end{equation}
The monotonicity of $\varpi$ implies that $M_B(U)$ is a non-increasing function of $U$. The {\it initial} Bondi mass is
\begin{equation}
 M_i \equiv M_B(U_0)\ .
\end{equation}
If the spacetime contains an apparent horizon then the Bondi mass will satisfy the BPS inequality
\begin{equation}
\label{bps}
 M_B(U) \ge |Q|\ .
\end{equation}
This can be proved as follows \cite{Dafermos:2012np}. Assume that the point $(U_1,V_1)$ lies on the apparent horizon. Since $r_{,V}=0$ on the apparent horizon we have $\varpi(U_1,V_1) = r/2+ Q^2/(2r) \ge |Q|$ (minimizing w.r.t.\ $r$). Since $\varpi_{,U}\le 0$ for $V=V_1$ outside the apparent horizon, it follows that $\varpi(U,V_1) \ge |Q|$ for $U<U_1$. Now $\varpi_{,V} \ge 0$ implies that $\varpi(U,V) \ge |Q|$ for $U<U_1$ and $V \ge V_1$. The result follows by taking $V \rightarrow \infty$.

\subsection{Construction of initial data}

The Ansatz~(\ref{anzatz}) has the residual gauge freedom
\begin{equation}
 U\to \tilde{U}(U)\ ,\quad
 V\to \tilde{V}(V)\ .
\label{resgauge}
\end{equation}
We fix this freedom as follows. First we set $Q=1$, which just fixes the scale. Then we choose 
\begin{equation}
U_0=-5.1, \qquad V_0=0
\end{equation}
and set
\begin{equation}
\label{fix}
 r(U,V)=\bar{r}(U,V) \qquad {\rm on} \; \Sigma
\end{equation}
where $\bar{r}(U,V)$ is the function $r(U,V)$ for the {\it extreme} RN spacetime given in Appendix \ref{RNDN}. On $\Sigma_1$, this implies that (Eq.~(\ref{rRN}) with $Q = 1$)
\begin{equation}
\label{rinit}
r(U,0)=1-U
\end{equation}
Note that $r \rightarrow 0$ as $U \rightarrow 1$ on $\Sigma_1$. Since $F^{\mu\nu} F_{\mu\nu} \propto 1/r^4$ it follows that there is a singularity at $U=1,V=V_0$ so we restrict to $U<1$. We will consider only initial data for which the singularity lies inside a black hole. For orientation, we note that the event horizon of the extreme RN spacetime is the surface $U=0$. Hence we expect the event horizon in a spacetime describing a small perturbation of extreme RN to be a surface $U=U_{EH}$ with $U_{EH} \approx 0$. 

\begin{figure}[htbp]
\centering
\includegraphics[width=8cm, clip]{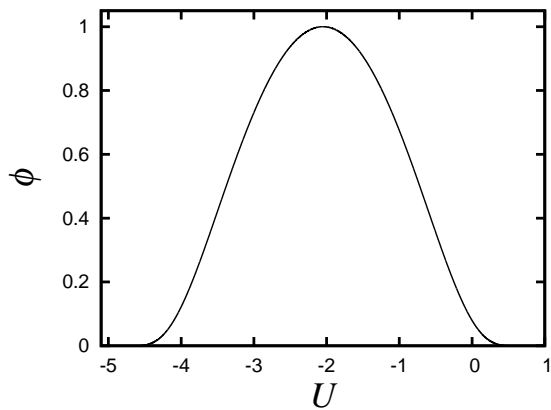}
\caption{$\phi(U,0)$ for $\epsilon=1$.}
\label{Fig:ID}
\end{figure}

On $\Sigma_2$, our choice (\ref{fix}) implies that (from Appendix \ref{RNDN})
\be
\label{dVr0}
 r_{,V}(U_0,V) = \frac{1}{2}\left(1-\frac{1}{r(U_0,V)} \right)^2\ .
\ee
We take the initial data for the scalar field to be trivial on $\Sigma_2$:
\begin{equation}
 \phi(U_0,V)=0\ .
\end{equation}
On $\Sigma_1$ we take initial data describing an outgoing wavepacket (see Fig.~\ref{initial})
\begin{equation}
 \phi(U,0)=
\begin{cases} 
\epsilon
  \exp\left[\alpha \left(\frac{1}{U-U_\textrm{in}}-\frac{1}{U-U_\textrm{out}}
+\frac{4}{U_\textrm{in}-U_\textrm{out}}
\right)\right] & (U_\textrm{out}<U<U_\textrm{in}) \\ 
0 & (\textrm{else}) 
\end{cases}
\ ,
\label{init}
\end{equation}
The right hand side of Eq.~(\ref{init}) is a smooth, compactly supported, function whose maximum value is $\epsilon$. 
In our numerical calculations, we fix the parameters as $U_\textrm{in}=0.9$, 
$U_\textrm{out}=-5$ and $\alpha=4$. This fixes the shape of the wavepacket. 
See Fig.~\ref{Fig:ID}. We will consider different values for the amplitude $\epsilon$. 

Note that this wavepacket is broad compared to the size of the black hole, extending to $r = 1-U_{\rm out} = 6$. Most of the wavepacket will lie outside the event horizon. The reason for this is that we are looking for an instability in which second $r$-derivatives of the scalar field become large. To claim an instability, we need to be sure that these derivatives are not large initially. We can make all derivatives of $\phi$ small initially by taking $\epsilon$ small enough. However, it is hard to maintain acceptable numerical accuracy for very small $\epsilon$ because the instability takes longer to develop. Therefore we ensure that $r$-derivatives of $\phi$ are small initially by taking a broad wavepacket. Since we have $U \le 1$ and $U_{EH} \approx 0$, this means that most of the wavepacket must lie outside the event horizon.

Having chosen initial data for $\phi$ and $r$, we can integrate the constraint equations $C_1$ and $C_2$ to determine $f$ on $\Sigma$:
\begin{equation}
  f(U,0) = f_0 \exp \left[ -\frac{1}{4} \int_{U_0}^U r(x,0) \phi_{,U}(x,0)^2 dx \right]
\end{equation}
\begin{equation}
 f(U_0,V) = f_0 \frac{r_{,V}(U_0,V)}{r_{,V}(U_0,0)} = f_0 \left( \frac{1-1/r(U_0,V)}{1-1/r(U_0,0)} \right)^2
\end{equation}
where $f_0\equiv f(U_0,0)$ is a constant of integration and we used (\ref{dVr0}). 

We can relate the constant $f_0$ to the initial Bondi mass $M_i$ as follows. Note that $\varpi$ is constant along
$\Sigma_2$ because of Eq.~(\ref{dMPI}) and $\phi=0$ on $\Sigma_2$. Thus, we have $M_i = M_B(U_0) = \varpi(U_0,0)$. Calculating the RHS using the definition of $\varpi$ gives
\begin{equation}
 M_i = \frac{r(U_0,0)}{2} \left[ 1 + \frac{1}{r(U_0,0)^2} - \frac{2}{f_0} \left( 1 - \frac{1}{r(U_0,0)} \right)^2 \right]\ .
\label{M_i}
\end{equation}
Hence the freedom to choose the parameter $f_0$ is equivalent to the freedom to choose $M_i$. So we have a 2-parameter family of initial data specified by $(\epsilon,M_i)$ or $(\epsilon,f_0)$. 

\begin{figure}
\centering
\includegraphics[width=8.0cm, clip]{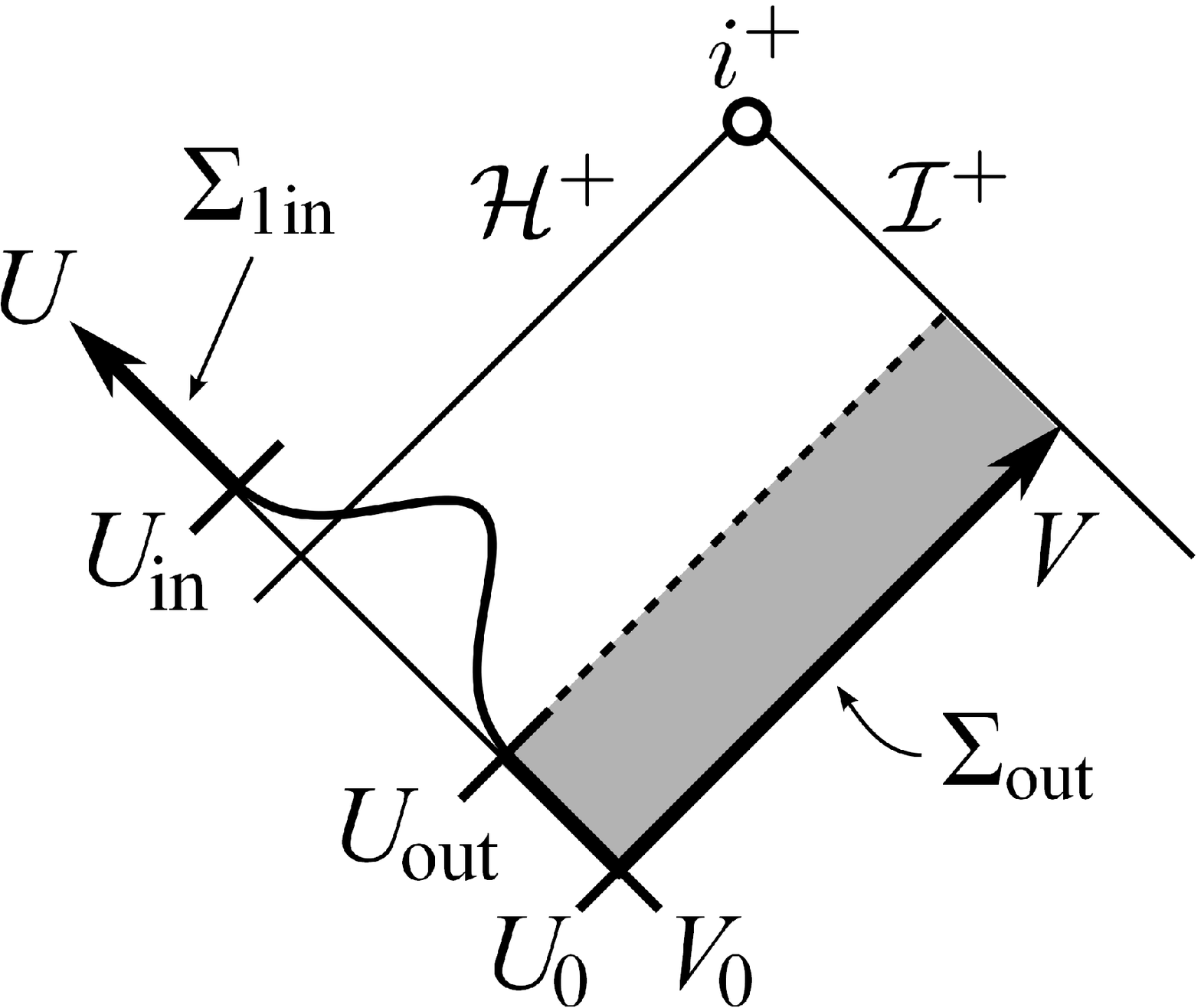}
\caption{Sketch of $\Sigma_\text{1in}$ and $\Sigma_\text{out}$. 
The spacetime in the future domain of dependence of $\Sigma_\text{out}$ (shaded 
region) is exactly RN.}
\label{Fig:Sigma}
\end{figure}

On those parts of $\Sigma$ where $\phi$ vanishes, we have initial data for Einstein-Maxwell theory, with spherical symmetry. The electrovac generalization of Birkhoff's theorem implies that such initial data must be Reissner-Nordstrom. Hence our initial data is RN on $\Sigma_{\rm out} \equiv \Sigma_2 \cup \{(U,V_0) \in \Sigma_1 : U < U_{\rm out} \}$ with mass parameter $M_i$, and the spacetime will be exactly RN in the future domain of dependence of $\Sigma_{\rm out}$ shown in Fig.~\ref{Fig:Sigma}. 

The initial data is also RN on $\Sigma_{1 {\rm in}} \equiv \{(U,V_0) \in \Sigma_1 : U > U_{\rm in} \}$ (see Fig.~\ref{Fig:Sigma}). Eq.~(\ref{dMPI}) implies that the renormalized Hawking mass $\varpi$ is constant on $\Sigma_{1 {\rm in}}$ but the data on $\Sigma_{1 {\rm in}}$ alone does not determine its value. Scattering of $\phi$ implies that the spacetime immediately to the future of $\Sigma_{1{ \rm in}}$ will not be RN in general.  

By integrating Eq.~(\ref{rUV}), we can determine $r_{,V}$ along $\Sigma_1$:
\be
\label{dVr}
 (r r_{,V})(U,0)=\frac{r(U_0,0)}{2} \left( 1 - \frac{1}{r(U_0,0)} \right)^2 - \frac{1}{4} \int_{U_0}^U \left( 1 - \frac{1}{r(x,0)^2} \right) f(x,0) dx\ .
\ee
The RHS is manifestly positive at $U=U_0$ so $r_{,V}>0$ at $(U_0,0)$. But as $U$ increases, it is possible that $r_{,V}$ becomes negative, in which case trapped 2-spheres are present on $\Sigma_1$. The apparent horizon, if present on $\Sigma_1$, is located at the smallest value of $U$ for which $r_{,V}=0$. 

\section{Nonlinear instability of extreme RN}
\label{NIERN}

\subsection{Choice of initial perturbation}
\label{Sec:ID}

Our initial data reduces to extreme RN initial data when $\epsilon=0$ and $M_i = 1$ ($f_0=2$). Hence to study perturbations of extreme RN we will consider small $\epsilon$ and small $M_i-1>0$ (the latter inequality comes from (\ref{bps}), recall we have set $Q=1$). Exploring a 2-parameter family of solutions numerically is difficult so we will consider 1-parameter subfamilies in which we set $M_i=M_i(\epsilon)$ with $M_i(\epsilon) \rightarrow 1+$ as $\epsilon \rightarrow 0$. We have considered two different 1-parameter families which are defined as follows. 

{\bf 1. Degenerate apparent horizon.} 
For fixed $\epsilon$ and large enough $M_i$, our initial data contains trapped 2-spheres in some region $U_+<U<U_-$ of $\Sigma_1$. The 2-spheres $U=U_{\pm}$ are marginally trapped and $U=U_+$ is the apparent horizon. If we now reduce $M_i$ then the trapped region shrinks until we reach a critical value $M_i(\epsilon)$ at which $U_+ = U_-$. In this case we have an apparent horizon but no trapped 2-spheres.  We will refer to this as a {\it degenerate apparent horizon}.

In this case, the function $r_{,V}$ on $\Sigma_1$ has a unique zero (at $U=U_+$) which
is also a global minimum and hence, $r_{,V} =
(r_{,V})_{,U} = 0$ at $U=U_+$. Eq.~(\ref{rUV}) reveals that $r=1$ at $U=U_+$ so (\ref{rinit}) implies that $U_+=0$. Vanishing of $r_{,V}$ at $U=0$ now determines (using (\ref{dVr}))
$f_0$, and hence $M_i$, in terms of $\epsilon$. 
For small $\epsilon$ we find that $f_0 = 2 + 0.740 \epsilon^2 + {\cal O}(\epsilon^4)$ and $M_i = 1+ 0.789 \epsilon^2 + {\cal O}(\epsilon^4)$. 
The latter implies that the solution in the future domain of dependence of $\Sigma_{\rm out}$ (the shaded region of Fig.~\ref{Fig:Sigma}) is non-extreme RN.

For this choice of initial data, the perturbation does not  change  the initial size of the apparent horizon, which remains at $r=1$. We have achieved this by ``adding negative energy'' behind the horizon to cancel the energy of the scalar field. The negative energy corresponds to reducing the value of $\varpi$ on $\Sigma_{1\rm in}$ below the value ($\varpi=1$) corresponding to extreme RN. Hence, very loosely, one might think of the initial data as being ``super-extreme'' RN on $\Sigma_{1\rm in}$. Although there is no {\it initial} perturbation to the size of the apparent horizon, we will see that the apparent horizon rapidly grows to a radius $r=1+{\cal O}(\epsilon)$.

This data has $M_i = 1 + {\cal O}(\epsilon^2)$, i.e., the perturbation to the mass is second order in $\epsilon$. Equivalently, $f_0 = 2 + {\cal O}(\epsilon^2)$, which implies that the metric (and Maxwell field) of a solution arising from this initial data differ from those of extreme RN only at second order in $\epsilon$. Hence the linearization of the solution gives vanishing metric and Maxwell field perturbations: it is just a scalar field evolving in a fixed extreme RN spacetime, precisely the situation considered by Aretakis. 

{\bf 2. First order mass perturbation.} A generic perturbation would give rise to a first order change in the mass: $M_i = 1 + {\cal O}(\epsilon)$ so it is interesting to consider initial data of this type. We make the choice $M_i = 1 + \epsilon$. Eq.~(\ref{M_i}), gives 
$f_0 = 2 + 0.938 \epsilon + 0.440\epsilon^2 + {\cal O}(\epsilon^3)$ for small $\epsilon$.
In this case, the linearization of the solution gives non-vanishing metric and Maxwell perturbations. These do not couple to the scalar field at linear order, so Birkhoff's theorem implies that the linearized metric and Maxwell perturbations correspond simply to a first order increase in the mass parameter of the RN solution. In the linearized solution, the scalar field evolves exactly as in the case studied by Aretakis. Of course, at the nonlinear level, the metric, Maxwell field, and scalar field perturbations are coupled.

For this initial data we find $U_{\pm} = -\sqrt2 \epsilon^{1/2}+{\cal O}(\epsilon)$, corresponding to marginally trapped 2-spheres of radius $r_{\pm} = 1 - U_{\pm}$ so the initial radius of the apparent horizon is $r_+ = 1 + \sqrt2 \epsilon^{1/2} + {\cal O}(\epsilon)$. The ${\cal O}(\epsilon^{1/2})$ term here comes from the linearized metric perturbation. The scalar field contributes only at ${\cal O}(\epsilon)$.

\subsection{Results: initial data with degenerate apparent horizon}
\label{Sec:degAH}

We have solved the equations of motion numerically following the method of Ref.~\cite{Burko:1997tb}. Appendix \ref{errors} contains a discussion of the numerical accuracy of our results.

We have considered $\epsilon=0.05,0.04,0.03,0.02, 0.01$.  In all cases we find that the solution on, and outside, the event horizon approaches a non-extreme RN solution at late time (large $V$). However, there is a long time for which the scalar field exhibits behaviour similar to that of a test field in extreme RN. In particular, there is an instability at the event horizon. We now describe our results in detail, starting with the properties of the apparent and event horizons.

\begin{figure}
  \centering
\subfigure[$U$ coordinate of horizons]
{\includegraphics[scale=0.5]{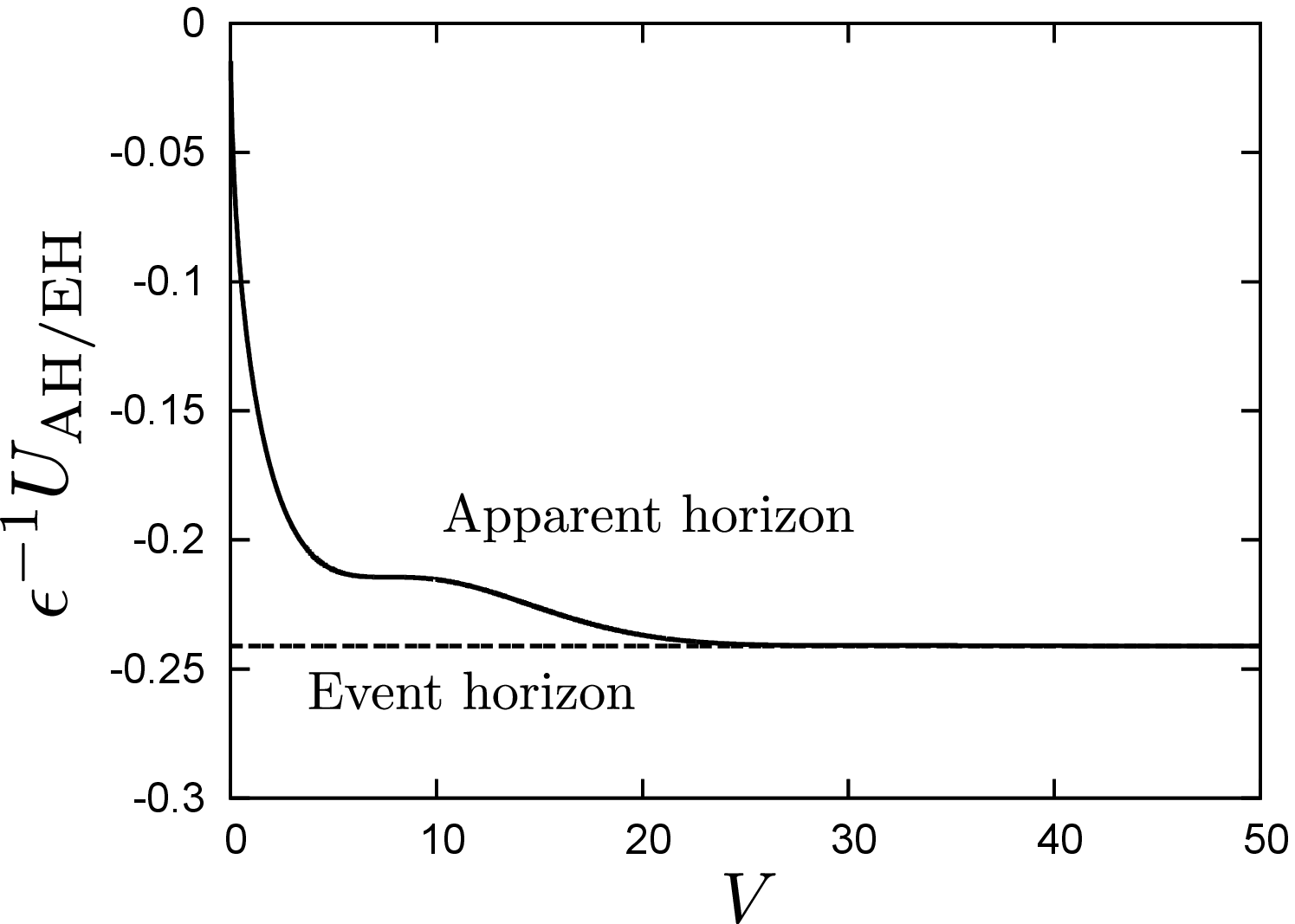}
\label{UAHEH}
}
\subfigure[radius of horizons]
{\includegraphics[scale=0.5]{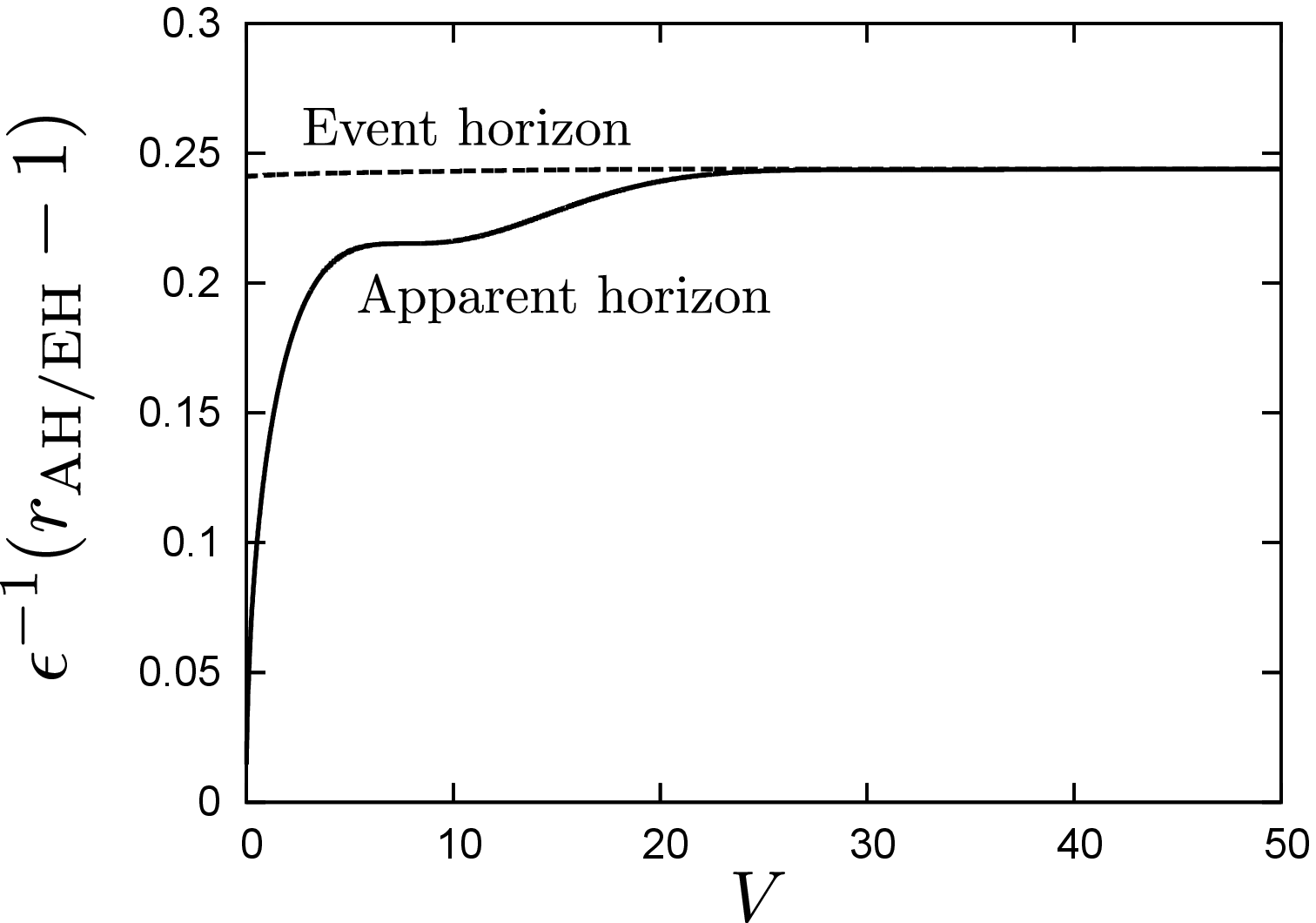}
\label{rAHEH}
}
\caption{
Panel (a): Time ($V$) dependence of the $U$-coordinate of the apparent horizon (solid curve) and event horizon (dashed curve) normalized by $\epsilon^{-1}$. Panel (b): Time dependence of radii of the apparent/event horizons. 
Differences of the radii from $1$ normalized by $\epsilon^{-1}$ are plotted in this panel. Note that the radius of the event horizon is almost constant. In both panels we show results for $\epsilon=0.02$. The results for $\epsilon=0.03,0.04$ are almost indistinguishable from these curves.  
 \label{bondi_r}
}
\end{figure}

On a surface of constant $V$, we denote the position of the apparent horizon by $U=U_{\rm AH}(V)$ and its radius by $r_{\rm AH}(V)\equiv r(U_{\rm AH}(V),V)$. These functions approach constant values at late time. Since we expect the apparent horizon to approach the event horizon at late time, we can determine the position of the event horizon as $U_{\rm EH} = \lim_{V \rightarrow \infty} U_{\rm AH}(V)$. Once we have found $U_{\rm EH}$ we can determine the radius of the event horizon $r_{\rm EH}(V) \equiv r(U_{\rm EH},V)$. In Fig.~\ref{bondi_r}, we show how these quantities change with time $V$ for initial data with different values of $\epsilon$. 

The radius of the apparent horizon grows rapidly before settling down to a constant value. Since $r>1$ for $V>0$ it follows that the apparent horizon is non-degenerate for $V>0$, i.e., trapped surfaces form immediately in the time evolution of this initial data. In contrast with the behaviour of the apparent horizon, we find that the radius of the event horizon is almost constant in time: 
$r_{\rm EH}(V) \approx 1 + 0.245 \epsilon$.
Hence, although this perturbation does not change the initial radius of the apparent horizon, it results in an ${\cal O}(\epsilon)$ change in the initial radius of the event horizon. 

\begin{figure}
  \centering
  \subfigure
  {\includegraphics[scale=0.5]{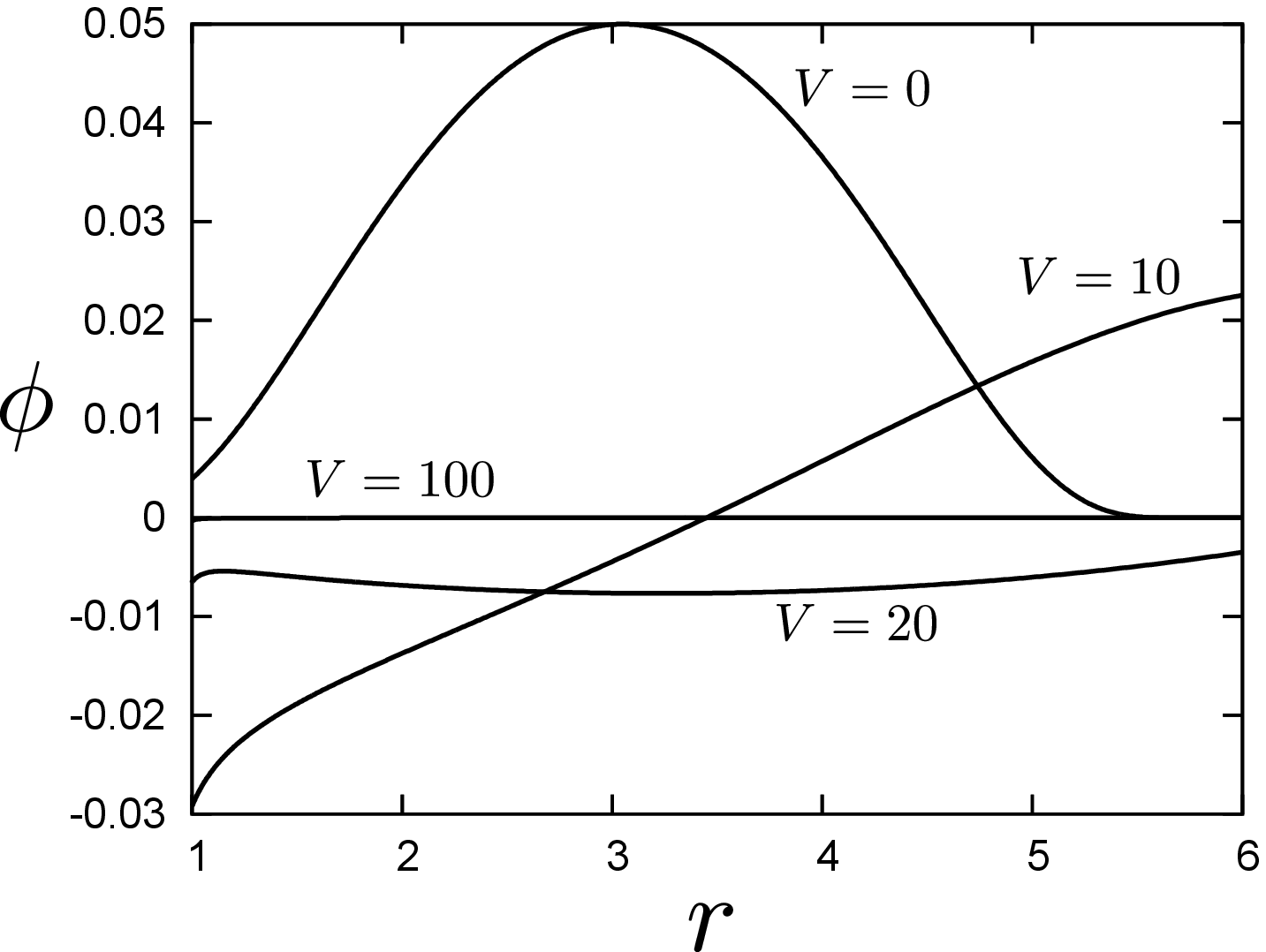}
\label{pev}
  }
  \subfigure
  {\includegraphics[scale=0.5]{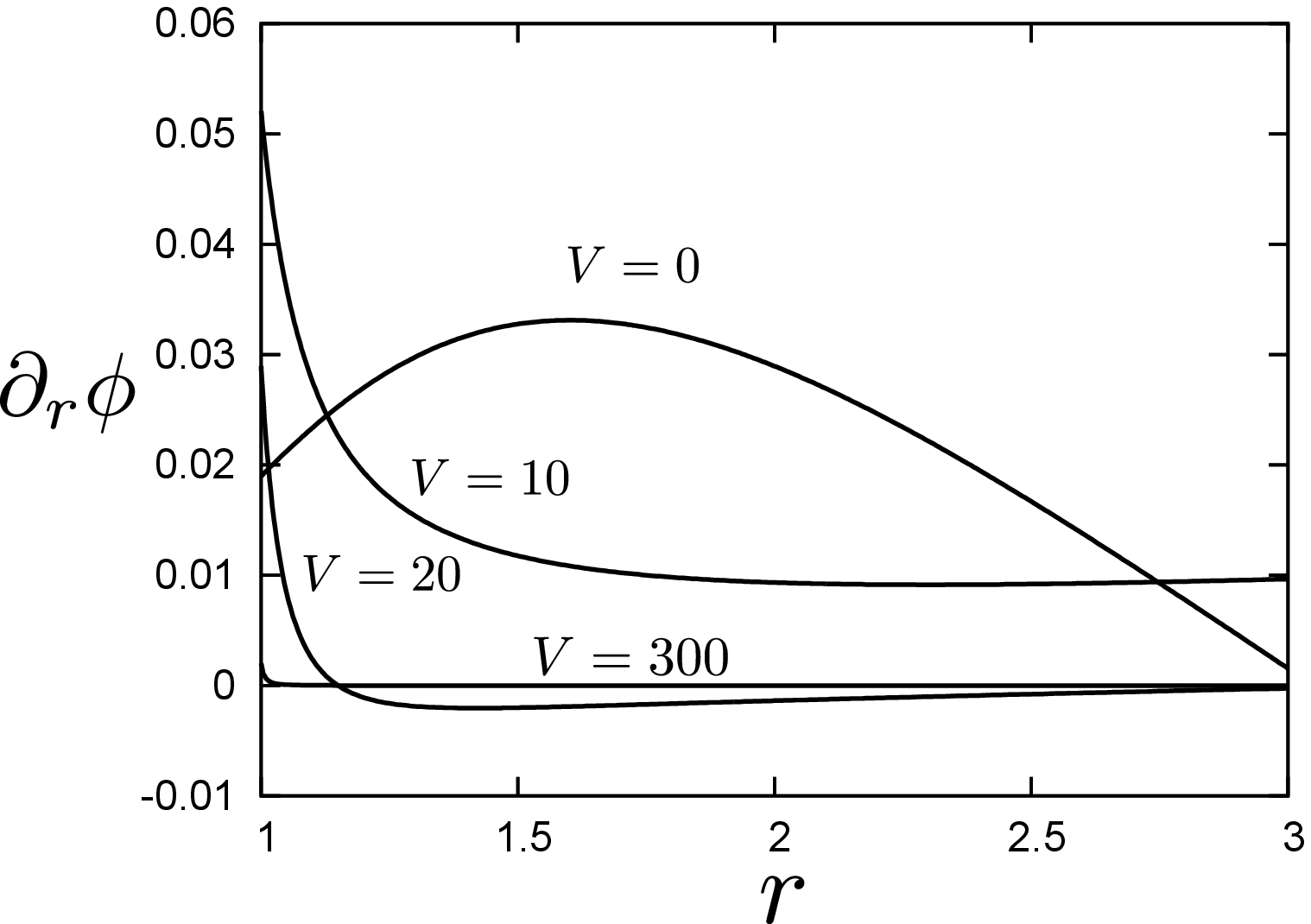} 
\label{dpev}
  }
  \caption{
$\phi$ and $\partial_r\phi$ as functions of $r$ on constant $V$ slices for $\epsilon=0.05$. The event horizon is at $r \approx 1.01$ (from Fig.~\ref{bondi_r}).  We can see that these functions decay as $V$ increases.
\label{pevs}
}
\end{figure}

Next we will describe the behaviour of the scalar field. In order to examine the behaviour near the horizon we can plot $\phi(U,V)$ against $r(U,V)$ at a fixed time $V$, viewing $U$ as a parameter.\footnote{Plotting $\phi$ against $U$ is less useful because lines of constant $U<U_{\rm EH}$ move outwards at the speed of light and hence are far from the horizon at large $V$.} Doing this for different values of $V$ gives us snapshots of the scalar field at different times. This is done in Fig.~\ref{pevs} for $\epsilon=0.05$. We also plot $\partial_r \phi \equiv \phi_{,U}/r_{,U}$.\footnote{
We calculate $\partial_r^n \phi$ ($n=1,2,3$) not by taking derivatives of $\phi(U,V)$ numerically but by obtaining and solving evolution equations for $\partial_r^n \phi$.
See Appendix~\ref{App:phir} for the details.
}
These plots reveal that $\phi$ and $\partial_r \phi$ both decay for large $V$ at fixed $r \ge 1$. In particular, they decay on, and outside, the apparent and event horizons. This implies that the energy-momentum tensor of the scalar field decays on, and outside, the apparent/event horizon. This strongly suggests that the solution approaches a RN solution at late time (because of the electrovac version of Birkhoff's theorem). The latter is uniquely determined by its horizon radius 
$r_+(\epsilon) = \lim_{V \rightarrow \infty} r_{EH}(V)=1+0.245 \epsilon$. 
Since $r_+(\epsilon)>1$, this RN solution is non-extreme. 

\begin{figure}[htbp]
\centering
\includegraphics[width=8cm,clip]{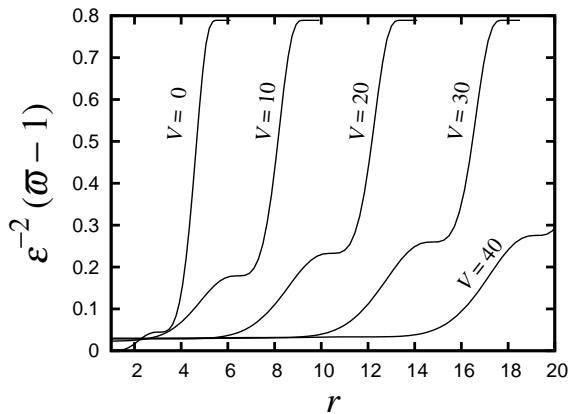}
\caption{
$\epsilon^{-2}(\varpi-1)$ for initial data with $\epsilon=0.02$ at $V=0, 10, 20, 30, 40$.
The curves for $\epsilon=0.03, 0.04$ almost completely coincide with the curves for $\epsilon=0.02$.
}
\label{Fig:deg_r-mP}
\end{figure}

A quantitative measure of closeness to a RN solution is given by the renormalized Hawking mass $\varpi$, which is constant for the RN solution. In Fig.~\ref{Fig:deg_r-mP}, we plot $\epsilon^{-2}(\varpi-1)$ against $r$ for different fixed values of $V$. From this figure one can read off the time it takes for the solution to settle down to the RN solution within a given radius of the horizon. For example, $\varpi$ settles down to a constant value in $r_\text{EH}< r <  6 r_\text{EH}$ within a time $V\sim 20$. 

\begin{figure}
\begin{center}
\includegraphics[scale=0.5]{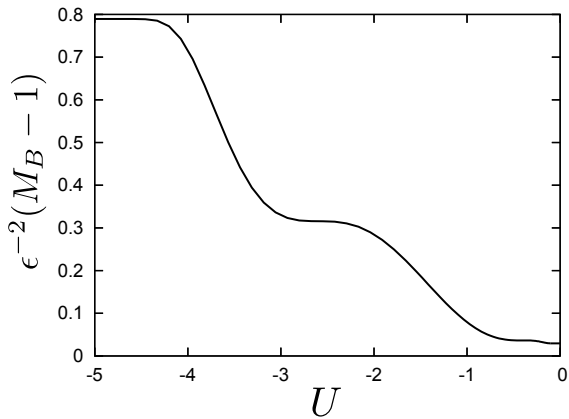}
\end{center}
\caption{
Time dependence of the Bondi mass $M_B(U)$.
The right edge of the figure corresponds to future timelike infinity with $U=U_{\rm EH} \approx 0$.
In the plot, $\epsilon^{-2}(M_B(U)-1)$ is shown for $\epsilon=0.02$.
The curves for $\epsilon=0.03,0.04$ almost completely coincide with the one shown.
}
 \label{MB}
\end{figure}

In Fig.~\ref{MB}, we plot the Bondi mass $M_B(U)$, defined by (\ref{bondidef}). To calculate this, we used the approximation $M_B(U)\simeq \varpi (U,V=150)$ which is valid because $\varpi$ does not vary significantly with $V$ for $V>150$. As explained above, $M_B(U)$ must be a non-increasing function of $U$ and respect the BPS bound: $M_B(U)\ge 1$. 

The behaviour of the Bondi mass can be understood from Eq.~(\ref{dMPI}), which shows that $\varpi_{,U}$ is proportional to $-r^2\phi_{,U}^2$. The profile of the wavepacket remains qualitatively the same as it propagates outwards. Hence at constant $V$, $\varpi$ is approximately constant at sufficiently early time ($U<-4$, corresponding to the outer edge of the wavepacket). $\varpi$ then decreases as $U$ approaches the value $U \approx -2$ where the wavepacket has its maximum. At this maximum, $\phi_{,U}$ vanishes so $\varpi$ is approximately constant again. For larger $U$, $\varpi$ decreases again. 

At late time ($U \rightarrow U_{\rm EH}$), the Bondi mass settles down to its final value $M_f(\epsilon)$. We have checked that this final value agrees, to within numerical accuracy, with the mass of a RN black hole of horizon radius $r_+(\epsilon)$. The Figure shows that most of the ``excess'' mass $M_i - 1$ of the spacetime is radiated to infinity instead of ending up in the final black hole: we find $(M_f-1)/(M_i-1)\approx 0.0375$. This is because most of our initial wavepacket lies outside the event horizon. 

We now return to the behaviour of the scalar field and the question of its stability. Recall that, for the test scalar field in extreme RN, Aretakis proved that $(\partial_r\phi)_V\equiv \phi_{,U}/r_{,U}$ does not decay on the event horizon, but generically approaches a non-zero constant value at late time. He proved also that $(\partial_r^2 \phi)_V$ blows up at late time on the event horizon.

\begin{figure}
  \centering
  \subfigure
  {\includegraphics[width=5cm]{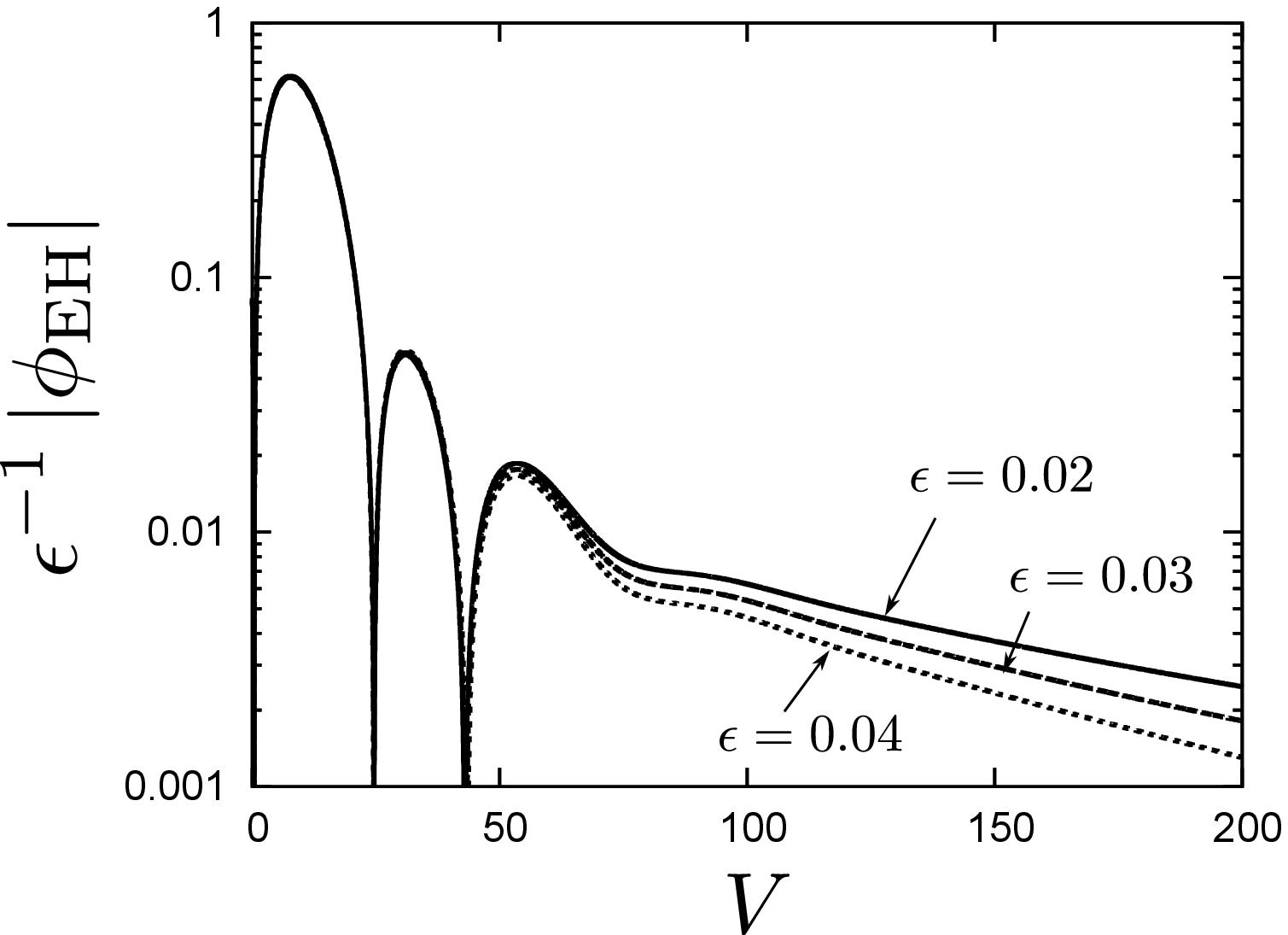}
\label{pH}
  }
  \subfigure
  {\includegraphics[width=5cm]{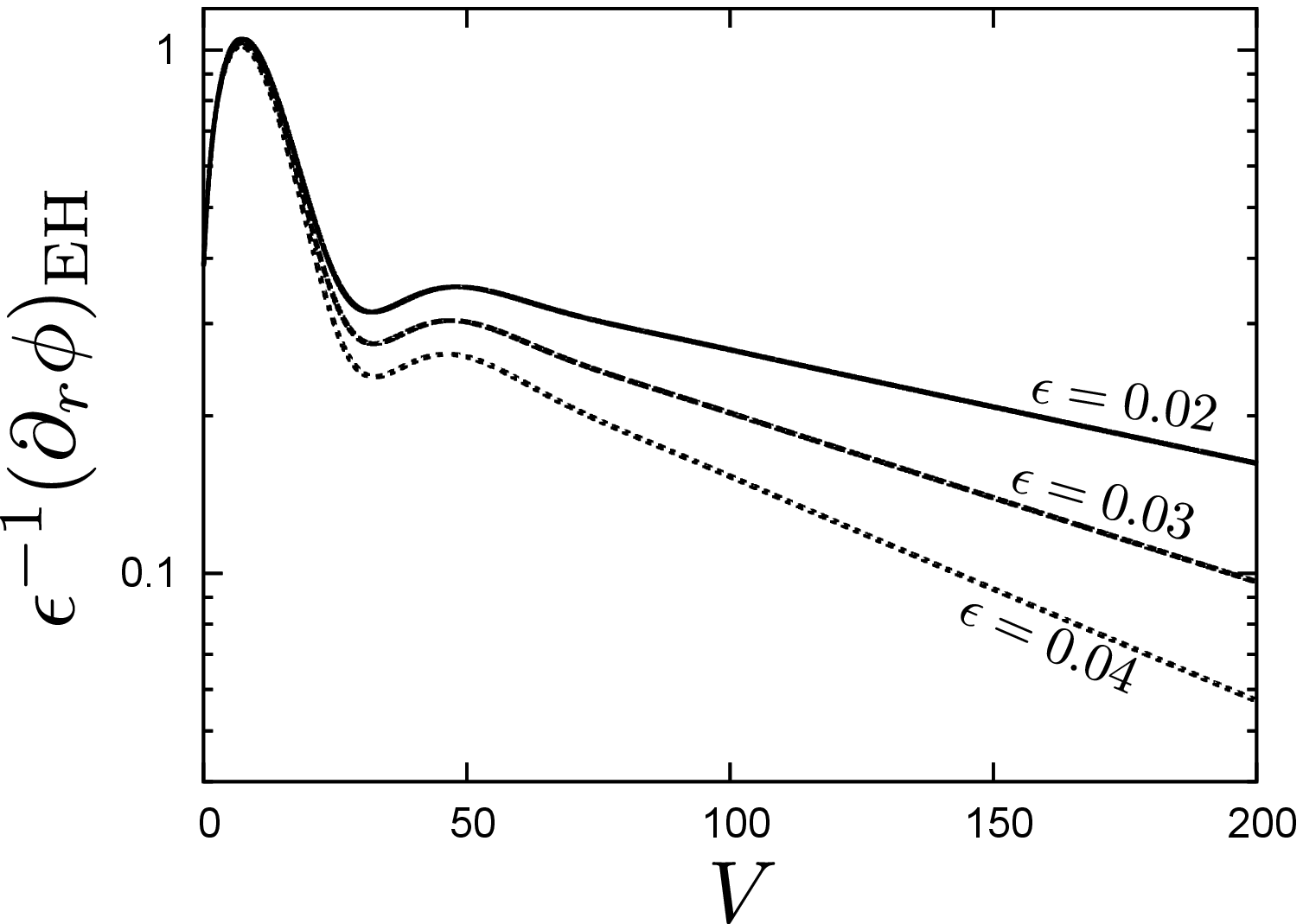} 
\label{dpH}
  }
  \subfigure
  {\includegraphics[width=5cm]{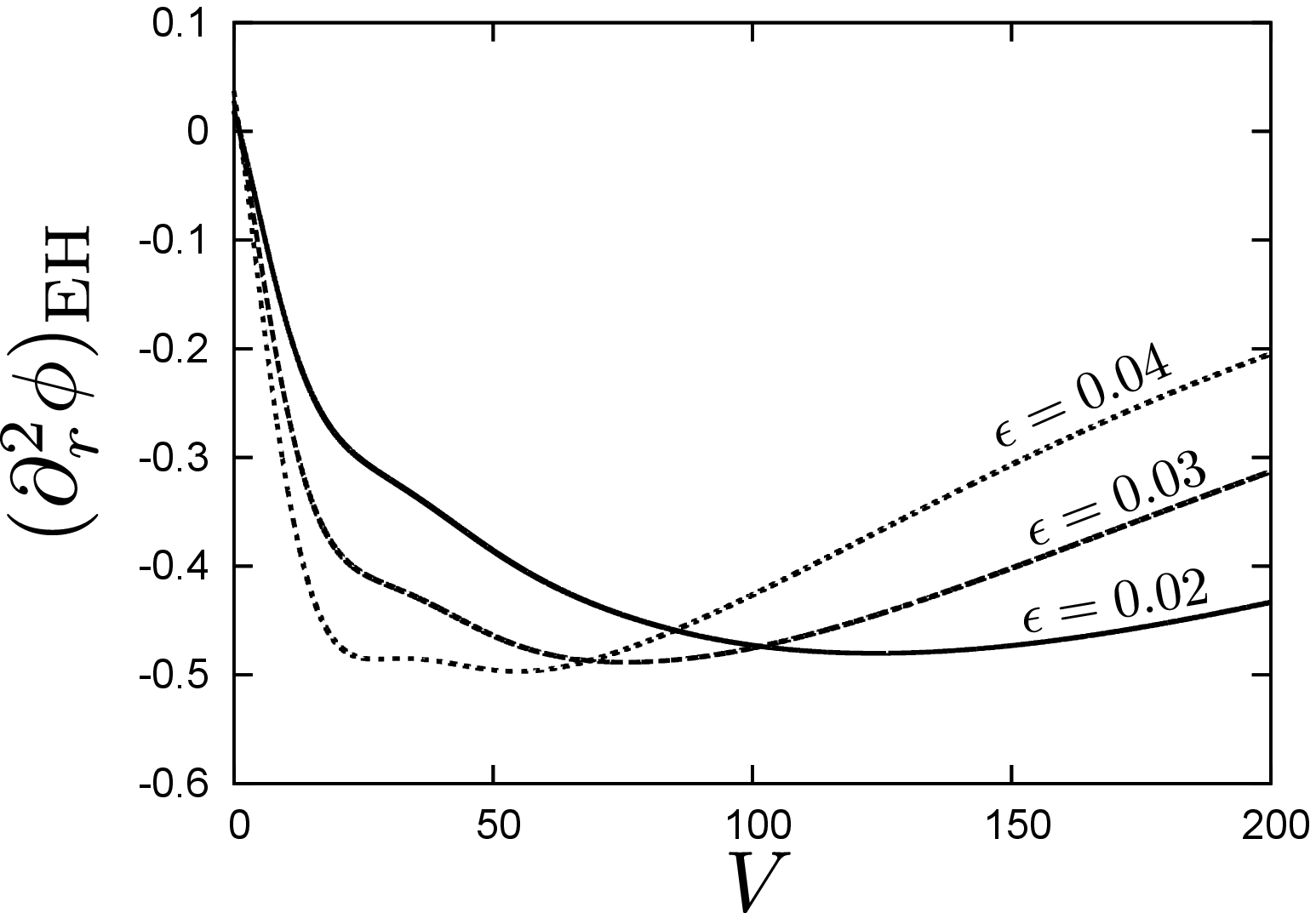} 
\label{ddpH}
  }
  \caption{
$\epsilon^{-1}\bigl|\phi_\textrm{EH}\bigr|$, 
$\epsilon^{-1} (\partial_r \phi)_\textrm{EH}$ and $(\partial_r^2 \phi)_\textrm{EH}$
against $V$ for 
$\epsilon=0.02,0.03,0.04$. $|(\partial_r^2 \phi)_\textrm{EH}|$ grows until it reaches a maximum $\sim 0.5$ and then decays to zero. The maximum value does not tend to zero as $\epsilon \rightarrow 0$. This is an instability.
\label{pHs}
}
\end{figure}

In Fig.~\ref{pHs}, we show our results for the time-dependence of $\phi$ and its first two $r$-derivatives (taken at fixed $V$) on the event horizon. 
The late time behaviour of $\partial_r \phi$ on the event horizon exhibits an important difference from the case of a test field in extreme RN. In the nonlinear solution, $(\partial_r \phi)_\text{EH}$ does not approach a constant value at late time. Instead Fig.~\ref{pHs} shows a slow decay. We can fit this decay to a function $\alpha e^{-\beta V}$ for $V \in [150,300]$ and find 
$\beta=9.78\times 10^{-3}$, $7.33\times 10^{-3}$, $4.87\times 10^{-3}$ for $\epsilon=0.04,0.03,0.02$ respectively. This value for $\beta$ is very close to the surface gravity $\kappa(\epsilon)$ of the final RN black hole (as determined from its horizon radius $r_+(\epsilon)$), which is $\kappa=9.54\times 10^{-3}$, $7.19\times 10^{-3}$, $4.82\times 10^{-3}$ respectively. 
We will explain this decay rate analytically below. Physically, the redshift effect of the final non-extreme black hole causes the radiation at the horizon to decay.\footnote{
We expect that scattering outside the event horizon will cause this exponential decay to transition to power-law decay at very late time.} Nevertheless, there is still an instability, as we now explain. 

\begin{figure}
\begin{center}
\includegraphics[scale=0.6]{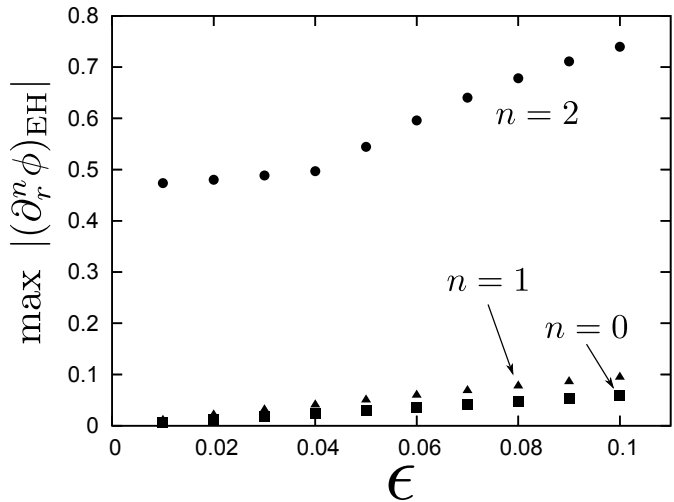}
\end{center}
\caption{
Maximum values of $\big|(\partial_r^n \phi)_\textrm{EH}\big|$ for $n=0,1,2$ against
$\epsilon$. The maximum value for $n=0,1$ tends to zero as $\epsilon \rightarrow 0$. However, for $n=2$ it tends to a non-zero limit as $\epsilon\to 0$, which demonstrates the existence of an instability.
}
 \label{ddpmax}
\end{figure}

We find that the magnitude of $\partial_r^2 \phi$ on the event horizon initially grows with $V$, reaches a maximum value and then decays. As $\epsilon$ is decreased, we find that $|\partial_r^2 \phi|$ grows for a longer time but at a slower rate. The maximum value of $|\partial_r^2 \phi|$ on the event horizon is about $0.5$ in all cases.\footnote{
For comparison, we note that the maximum value of $|\partial_r^2 \phi|$ on the initial surface $V=0$ is $1.27 \epsilon$.}  In particular, this maximum value does not tend to zero as $\epsilon \rightarrow 0$. We explore this in more detail in Fig.~\ref{ddpmax}, which gives the maximum value of $|\partial_r^n \phi|$ on the event horizon for $n=0,1,2$ as $\epsilon$ is decreased. Note that the maximum values of $|\phi|$ and $|\partial_r \phi|$ are ${\cal O}(\epsilon)$ and hence tend to zero as $\epsilon \rightarrow 0$. However, the maximum value of $|\partial_r^2 \phi|$ approaches a non-zero limit as $\epsilon \rightarrow 0$.

In summary, as the amplitude of the initial perturbation is taken to zero, an effect caused by the perturbation does not tend to zero. {\it This demonstrates an instability of extreme RN in the nonlinear theory.} The reason for this instability is that $\partial_r\phi$ decays much faster outside the event horizon than it does on the event horizon, so $\partial_r^2 \phi$ becomes large on the horizon. Our work reveals that, for any finite amplitude of perturbation, the instability eventually decays and the solution settles down to non-extreme RN. We will argue below that the time at which decay occurs diverges as $1/\epsilon$ as $\epsilon \rightarrow 0$. 

\subsection{Results: initial data with first order mass perturbation}
\label{Sec:O(e)pert}

In this section, we discuss briefly the results for the initial data with non-vanishing first order mass perturbation defined in section \ref{Sec:ID}. As above, the solution settles down to a non-extreme RN black hole at late time. This has surface gravity $\kappa={\cal O}(\epsilon^{1/2})$, whereas above we had $\kappa = {\cal O}(\epsilon)$. We will show below that $\kappa^{-1}$ is the time scale over which the instability develops and eventually decays. This implies that we can study smaller $\epsilon$ in this section than we did above. We have studied the cases $\epsilon = 10^{-1}, 10^{-2}, 10^{-3}, 10^{-4}$, for which the final black hole has surface gravity $\kappa=0.19, 0.11, 0.041, 0.014$ respectively.

\begin{figure}[htbp]
\centering
\subfigure[$\epsilon^{-1/2}\bigl|(\partial_r^2\phi)_\text{AH}\bigr|$]{\includegraphics[width=8cm, clip]{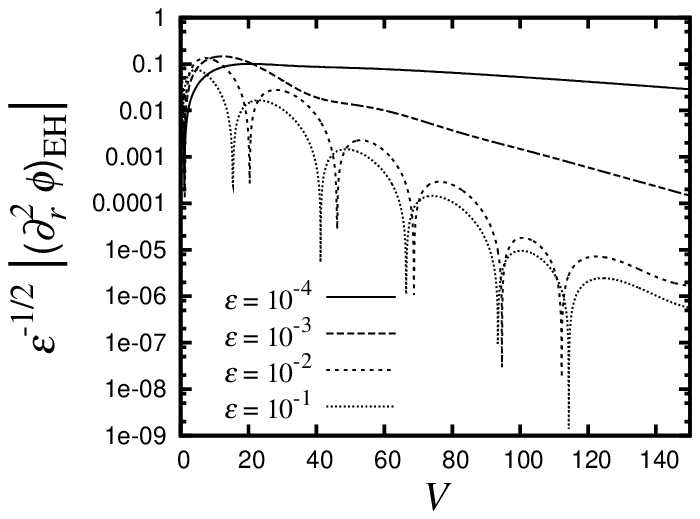} \label{Fig:ne(a)}}
\subfigure[$\bigl|(\partial_r^3\phi)_\text{EH}\bigr|$]{\includegraphics[width=8cm, clip]{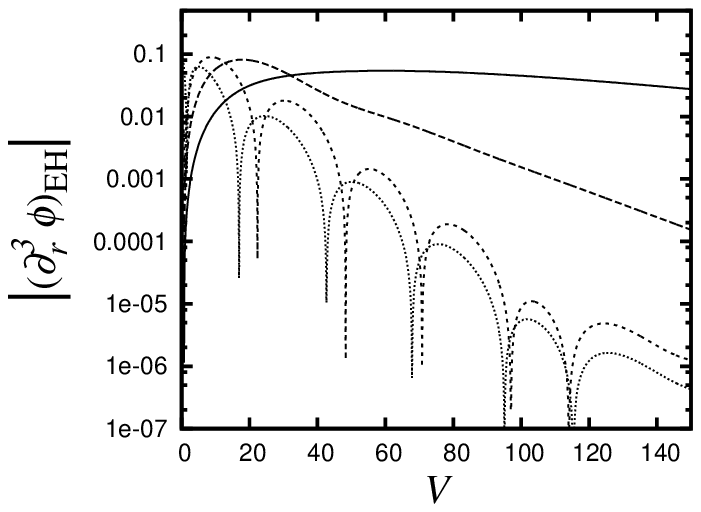} \label{Fig:ne(b)}}
\caption{$\epsilon^{-1/2}\bigl|(\partial_r^2\phi)_\text{EH}\bigr|$ and $\bigl|(\partial_r^3\phi)_\text{EH}\bigr|$ for the initial data with a first order mass perturbation and $\epsilon = 10^{-1}, 10^{-2}, 10^{-3}, 10^{-4}$. Damped oscillations ("quasinormal ringing") are apparent for the larger values of $\epsilon$. There is an instability because the maximum value of $\bigl|(\partial_r^3\phi)_\text{EH}\bigr|$ does not vanish as $\epsilon \rightarrow 0$.}
\label{Fig:ne}
\end{figure}

As above, we find that $\phi$, $\partial_r \phi$ decay on, and outside the event horizon. In contrast with the above case, we find that there is no instability associated to the behaviour of $\partial_r^2 \phi$: Fig.~\ref{Fig:ne}(a) shows that the maximum value of $|(\partial_r^2\phi)_{\rm EH}|$ is proportional to $\epsilon^{1/2}$ and hence vanishes as $\epsilon \rightarrow 0$. Hence this initial perturbation is more stable than the initial perturbation with a degenerate apparent horizon. However, an instability is still present. To see it,  consider $|(\partial_r^3\phi)_{\rm EH}|$ as shown in Fig.~\ref{Fig:ne}(b). The maximum value of this quantity does not vanish as $\epsilon \rightarrow 0$ and so there is an instability. We will reproduce this result analytically below. 

A similar instability occurs for initial data describing an extreme RN solution perturbed by an {\it ingoing} wavepacket. This is discussed in Appendix \ref{app:ingoing}. 

\subsection{Toy model}

In this section, we will explain how our numerical results can be understood by analytic calculations using a toy model. 
Our numerical results reveal that the instability occurs at late time, when the metric near the event horizon has stopped evolving significantly with time and has settled down to the metric of the final non-extreme RN black hole. This suggests that the solution near the horizon will be well-approximated by a test scalar field of initial amplitude $\epsilon$ evolving in a fixed non-extreme RN spacetime with surface gravity $\kappa={\cal O}(\epsilon)$ (for the initial data with a degenerate apparent horizon) or $\kappa = {\cal O}(\epsilon^{1/2})$ (for the initial data with a first order mass perturbation). 
We are interested in the behaviour of the solution as $\epsilon \rightarrow 0$. We can consider both cases simultaneously by allowing for general values of $\kappa$ and $\epsilon$. 

Consider the RN solution (Appendix \ref{RNDN}) in ingoing Eddington-Finkelstein coordinates:
\be
 ds^2 = -F(r) dv^2 + 2 dv dr + r^2 d\Omega^2 \qquad F(r) = \left( 1 - \frac{r_+}{r} \right) \left( 1 - \frac{r_-}{r} \right)
\ee
where $r_{\pm} = M \pm \sqrt{M^2 -1}$ (since $Q=1$). Lines of constant $v$ are ingoing radial null geodesics. The future event horizon ${\cal H}^+$ lies at $r=r_+$ and has surface gravity
\be
 \kappa = \frac{F'(r_+)}{2} = \frac{r_+ - r_-}{2r_+^2} < \frac{1}{2}\ .
\ee
We will label the black hole by its surface gravity $\kappa$. For small $\kappa$ we have
\be
 r_+ = 1+ \kappa + {\cal O}(\kappa^2)\ .
\ee
Note that $v$ is not quite the same as the coordinate $V$ used in our numerical simulations because our gauge choice (\ref{fix}) refers to the {\it extreme} RN spacetime. However, $v$ and $V$ will agree as $\kappa \rightarrow 0$, which is the limit we will eventually take. 

We consider a spherically symmetric massless scalar field $\phi(v,r)$ with initial data prescribed on intersecting null hypersurfaces as described above. For definiteness, we take the ``ingoing'' hypersurface ($\Sigma_1$) to be at $v=0$. The data on this hypersurface is an outgoing wavepacket of amplitude $\epsilon$: $\phi = \epsilon \bar{\phi}$ where $\bar{\phi}$ is independent of $\epsilon$. The initial data on the outgoing hypersurface ($\Sigma_2$) is assumed to be trivial as above. If we define
\be
 \Phi(v,r)  = r \phi(v,r)
\ee
then the spherically symmetric massless scalar wave equation in this geometry is
\be
\label{Phieq}
2   \Phi_{vr} + F \Phi_{rr} + F' \left(\Phi_r - r^{-1} \Phi \right) = 0
\ee
where subscripts denote derivatives. Following Aretakis' approach for extreme RN, let's evaluate this equation on ${\cal H}^+$. The result can be rearranged to give
\be
 \partial_v \left( e^{\kappa v} \Phi_r(v,r_+) \right) = \kappa e^{\kappa v} \Phi(v,r_+)
\ee
and hence
\be
\label{noncons}
 \Phi_r(v,r_+) = e^{-\kappa v} \Phi_r(0,r_+) + \kappa I_1 (v)
\ee
where we define (for positive integer $n$)
\be
I_n(v)= \int_0^v dx \; e^{-n\kappa(v-x)} \Phi(x,r_+) \ .
\ee
For extreme RN we have $\kappa=0$ and so we recover Aretakis' result that $\Phi_r(v,r_+)$ is constant on ${\cal H}^+$. For $\kappa>0$, we need to assume something about the behaviour of $\Phi$ on ${\cal H}^+$ to deduce how $\Phi_r$ behaves there. The assumption we make is:
\be
 |\Phi(v,r_+)| \le C \epsilon \max(1,v)^{-p}
\ee
for $v \ge 0$, where $C$ and $p$ are positive constants independent of $\epsilon$ and $\kappa$ (but could depend on the choice of profile $\bar{\phi}$, which we are holding fixed). This assumption is motivated by the known results for the decay of a scalar field on ${\cal H}^+$. For non-extreme RN, it has been proved that the result above applies for $p=3-\delta$ (for any $\delta>0$) but with a coefficient $C$ that depends on $\kappa$ (and $\delta$), and diverges as $\kappa \rightarrow 0$ \cite{Dafermos:2003yw}. For extreme RN, Aretakis proved that the above result holds with $p=3/5$ although the results of Ref.~\cite{Lucietti:2012xr,Bizon:2012we} suggest that this can be improved to $p=1$. In our case, none of these results apply strictly because we want a result with $C$ independent of $\kappa$. So we make the most conservative choice $p=3/5$, although in practice our numerical results indicate that the solution will decay much faster than this.

Using this assumption in (\ref{noncons}) one can bound the second term in (\ref{noncons}):
\bea
\label{Inbound}
 |I_n(v)| &\le& C \epsilon  \int_0^v dx \; e^{-n \kappa(v-x)} \max(1,x)^{-3/5} \nonumber \\
  &\le&  C\epsilon  \left[ e^{-n\kappa v} \frac{(e^{n \kappa}-1)}{n \kappa} + \frac{5}{2} (v^{2/5} - 1) \right]  \le   C\epsilon  \left(\frac{5}{2} v^{2/5} + D_n \right)
   \eea
where, in the second inequality we wrote $\int_0^v = \int_0^1 + \int_1^v$ and used $e^{-\kappa(v-x)} \le 1$, and in the third inequality we used $\kappa<1/2$ and $D_n$ is a positive constant independent of $\epsilon,\kappa$.

Let us now assume now that $\kappa=\kappa(\epsilon)$ such that $\kappa \rightarrow 0$ as $\epsilon \rightarrow 0$, as is the case for the choices of $\kappa(\epsilon)$ discussed at the start of this section. We also fix a time interval $v \in [0,N/\kappa]$ where $N$ is independent of $\epsilon$. Using $\Phi_r(0,r_+)=\epsilon (r\bar{\phi})_r(0,r_+)$, we see that the first term in (\ref{noncons}) is ${\cal O}(\epsilon)$. Over this time interval, the second term in (\ref{noncons}) is ${\cal O}(\epsilon \kappa^{3/5})$ (using (\ref{Inbound})), which can be made arbitrarily small compared to the first term by taking $\epsilon$ sufficiently small. Hence, for small enough $\epsilon$, the first term in (\ref{noncons}) dominates for $v \in [0,N/\kappa]$. This slow exponential decay of $\Phi_r$ at the event horizon, with exponent $\kappa$, is in good agreement with our numerical results for the full nonlinear system we studied in section \ref{Sec:degAH}.

Now we consider $\Phi_{rr}$. To investigate the behaviour of this quantity, we again follow Aretakis' approach. Taking an $r$-derivative of (\ref{Phieq}) gives
\be
 2\Phi_{vrr} + F \Phi_{rrr} + 2 F' \Phi_{rr} + \left( F'' - r^{-1} F'\right) \left( \Phi_r - r^{-1} \Phi \right) = 0
\ee
and evaluating this at $r=r_+$ gives
\be
\label{r2eq}
 \partial_v \left( e^{2\kappa v} \Phi_{rr} \right) = - k e^{2\kappa v}  \left( \Phi_r - r_+^{-1} \Phi \right)
\ee
where
\be
 k \equiv \frac{F''(r_+)}{2} - \frac{\kappa}{r_+}=\frac{5r_--3r_+}{2 r_+^3} = 1+{\cal O}(\kappa)\ .
 \ee
Using (\ref{noncons}) and integrating gives
\be
\label{blowup}
 \Phi_{rr}(v,r_+) = -\frac{k}{\kappa} \left(e^{-\kappa v}- e^{-2\kappa v} \right) \Phi_r (0,r_+)  - k \kappa J(v) + k r_+^{-1} I_2(v) + e^{-2\kappa v} \Phi_{rr}(0,r_+)
\ee 
where
\be
J(v)=  \int_0^v dx\; e^{-2\kappa(v-x)}I_1(v)\ . 
\ee
As before, assume $\kappa=\kappa(\epsilon)$ with $\kappa \rightarrow 0$ as $\epsilon \rightarrow 0$ and consider the time interval $v \in [0,N/\kappa]$ where $N$ is independent of $\epsilon$. The first term in (\ref{blowup}) is ${\cal O}(\epsilon \kappa^{-1})$. Using (\ref{Inbound}) we have
\be
 |J(v)| \le C \epsilon \int_0^v dx \left( \frac{5}{2} x^{2/5} + D_1 \right) = C\epsilon \left( \frac{25}{14} v^{7/5} + D_1 v \right)\ .
\ee
Hence the second term in (\ref{blowup}) is ${\cal O}(\epsilon \kappa^{-2/5})$. Eq.~(\ref{Inbound}) implies that the third term is also ${\cal O}(\epsilon \kappa^{-2/5})$. The final term is ${\cal O}(\epsilon)$. 

Now consider the case for which $\kappa = \kappa_0 \epsilon + {\cal O}(\epsilon^2)$ where $\kappa_0>0$ (as for our degenerate apparent horizon initial data). Using the results just obtained, it follows that the second, third and fourth terms of (\ref{blowup}) vanish in the limit $\epsilon \rightarrow 0$. However,  the first term is ${\cal O}(1)$ and hence survives this limit. To see this, set $v = \hat{v}/\kappa$ with $\hat{v} \in [0,N]$ independent of $\epsilon$. We then have
\be
 \lim_{\epsilon \rightarrow 0} \Phi_{rr}(v=\hat{v}/\kappa,r_+) = -\frac{1}{\kappa_0}  \left(e^{-\hat{v}}- e^{-2\hat{v}} \right)  (r\bar{\phi})_r(0,r_+)\ .
\ee
As explained above, the non-vanishing of this quantity demonstrates the existence of an instability. 
For small $\hat{v}$, the absolute value of the RHS grows linearly with $\hat{v}$, just as in the Aretakis instability. However, it reaches a maximum at $\hat{v}=\log 2$ ($v = \kappa^{-1} \log 2$) and decays exponentially thereafter. Using the values of $\kappa(\epsilon)$ obtained in section \ref{Sec:degAH} we have $\kappa^{-1} \log 2 \approx 72, 96, 144$ for $\epsilon=0.04,0.03,0.02$, so the position of the maximum determined analytically is in good agreement with the numerical result shown in Fig.~\ref{pHs}.\footnote{
Since $\phi$ and $\partial_r \phi$ decay on the event horizon and $r_+ = 1 + {\cal O}(\epsilon)$, we expect $\Phi_{rr}$ to agree with $\phi_{rr}$ at sufficiently late time.} 

This argument for instability fails when $\kappa = {\cal O}(\epsilon^{1/2})$ which corresponds to the case in which the full nonlinear problem has a non-zero first order metric perturbation. In this case, the first term in (\ref{blowup}) is ${\cal O}(\epsilon^{1/2})$, in agreement with the numerical results shown in Fig.~\ref{Fig:ne}. However, as explained above, an instability {\it is} present in this case, we just need an extra derivative to see it. Taking an $r$-derivative of (\ref{r2eq}) and evaluating at $r=r_+$ gives (using $\Phi,\Phi_r = {\cal O}(\epsilon)$)
\be
 \partial_v \left( e^{3\kappa v} \Phi_{rrr} \right) = -k' e^{3\kappa v} \Phi_{rr} + {\cal O}(\epsilon )
\ee
where
\be
 k' =\frac{3 F''(r_+)}{2} - \frac{\kappa}{r_+} =\frac{13 r_- - 7 r_+}{2 r_+^3} = 3 + {\cal O}(\kappa)
\ee
and we assumed $v \in [0,N/\kappa]$. Using (\ref{blowup}) and integrating
\be
 \Phi_{rrr}(v,r_+) = \frac{kk'}{2 \kappa^2} \left(e^{-\kappa v} - 2 e^{-2\kappa v} + e^{-3\kappa v} \right)  \Phi_r (0,r_+) + {\cal O}(\epsilon \kappa^{-7/5})\ .
\ee
If we now set $\kappa=\kappa_1 \epsilon^{1/2} + {\cal O}(\epsilon)$ ($\kappa_1>0$) then the correction term vanishes as $\epsilon \rightarrow 0$ and, for $v = \hat{v}/\kappa$ with $\hat{v} \in [0,N]$, we have
\be
\lim_{\epsilon \rightarrow 0} \Phi_{rrr}(v=\hat{v}/\kappa,r_+) =  \frac{kk'}{2 \kappa_1^2} e^{-\hat{v}}\left(1-e^{-\hat{v}}  \right)^2  (r \bar{\phi})_r (0,r_+)
\ee
which is ${\cal O}(1)$, demonstrating instability. The above expression has a maximum at $\hat{v}=\log 3$ ($v = \kappa^{-1} \log 3$) and decays exponentially at larger $\hat{v}$. Using the values of $\kappa(\epsilon)$ obtained in section \ref{Sec:O(e)pert} we have $\kappa^{-1} \log 3 \approx 27, 73$ for $\epsilon=10^{-3}, 10^{-4}$, so the position of the maximum determined analytically is in reasonable agreement with the numerical results shown in Fig.~\ref{Fig:ne}. 

In summary, subject to a reasonable assumption concerning the decay of the scalar field along the event horizon, we have shown that this linear toy model provides a good explanation of our numerical results for the full nonlinear evolution. In particular, it explains the existence of an instability in the second or third transverse derivative of the scalar field at the event horizon. The instability develops over a time of order $\kappa^{-1}$ where $\kappa$ is the surface gravity of the final black hole. It is followed by a period in which the scalar field and its derivatives decay as $e^{-\kappa v}$ on the event horizon. This is the horizon redshift effect. At times much later than $\kappa^{-1}$ we would expect this exponential decay to be replaced by a power law tail arising from scattering outside the black hole. 

\section{Dynamical extreme black holes}

\label{extreme}

\subsection{Introduction}

\label{extremeintro}

We have seen that a generic perturbation of an extreme RN black hole leads to an instability but eventually the spacetime settles down to non-extreme RN. In this section we will show that it is possible to fine-tune the initial perturbation so that the solution settles down to extreme RN {\it outside} the event horizon. Because of the third law of black hole mechanics (``a non-extreme black hole cannot become extreme''), it is natural to regard such a black hole as extreme for all time, so it is a time-dependent extreme black hole. To make the discussion more precise, we need a definition of extremality that encompasses time-dependent black holes. 

One could define extremality in terms of the late-time behaviour of the event horizon. For example we could define the black hole to be extreme if $r \rightarrow 1$ along the event horizon as $V \rightarrow \infty$ (we continue to set $Q=1$ in this section). Alternatively we could define the black hole to be extreme if $\varpi \rightarrow 1$ along the event horizon as $V \rightarrow \infty$. Another definition might be $M_B \rightarrow 1$ at late time. However, we will adopt a version of Israel's definition \cite{israel} and say that our black hole spacetime is extreme if there is no trapped symmetry 2-sphere in the future domain of dependence of $\Sigma$.  In fact, the black holes that we will discuss in this section are extreme with respect to all of these definitions, as is extreme RN.

This definition apparently requires us to monitor the entire black hole interior for trapped 2-spheres. Fortunately, this is not the case. The LHS of Eq.~(\ref{rUV}) can be written $(rr_{,V})_{,U}$ so the unique stationary point of $r r_{,V}$  on a surface of constant advanced time $V$ is located at $r=Q=1$. (It is unique because $r_{,U}<0$ so $r$ decreases monotonically along this surface.) Furthermore, this corresponds to a global minimum because $r_{,U}<0$ and $(rr_{,V})_{,U}$ change sign from negative to positive as $r$ decreases through $r=1$. If trapped 2-spheres are present on this surface of constant $V$ then $r r_{,V}$ is negative somewhere on this surface, so its minimum value must be negative. Hence trapped 2-spheres are present on this surface of constant $V$ if, and only if, $r_{,V}<0$ at the value of $U$ for which $r=1$. So trapped 2-spheres are absent on a surface of constant $V$ if $r_{,V} \ge 0$ where $r=1$ this surface. 

If a surface of constant $V$ has no trapped 2-spheres but does have an apparent horizon then, since $r_{,V}=0$ at the apparent horizon but $r_{,V}>0$ elsewhere on the surface, the apparent horizon corresponds to the minimum of $r_{,V}$ and hence has radius $r=1$. This is what we called a degenerate apparent horizon above. We found that trapped 2-spheres formed immediately when such data was evolved in time. So the spacetime arising from such initial data does not describe an extreme black hole.\footnote{It is tempting to say that the initial data describes an extreme black hole but this becomes non-extreme when evolved. However, we will use the word ``extreme'' only in reference to the full spacetime.}

If we reduce $M_i$ slightly below the value which gives initial data with a degenerate apparent horizon then the initial data contains no apparent horizon. However, when this data is evolved, we find that  trapped 2-spheres form after a non-zero advanced time ($V$). Eventually it settles down to a non-extreme RN solution. So again, such data does not produce an extreme black hole.

If we reduce $M_i$ still further then there will be a critical value $M_*(\epsilon)$ such that the spacetime describes a black hole when $M_i>M_*(\epsilon)$ but has no event horizon when $M_i<M_*(\epsilon)$ (e.g.\ a naked singularity). We will now argue that the spacetime with $M_i = M_*(\epsilon)$ describes an extreme black hole.

The argument is based on Cauchy stability of the equations of motion. This states that, in a compact subset of the future domain of dependence of $\Sigma$, the solution depends continuously on the initial data prescribed on $\Sigma$.

Consider a solution with $M_i$ slightly greater than $M_*(\epsilon)$. This settles down to a non-extreme RN black hole with horizon radius $r_+ \le 2$ (say). It follows that the event horizon radius is always less than $2$. Since $r_{,U}<0$, we have $r \le 2$ inside the black hole too. Numerically we find that the horizon of the black hole is always close to $U=0$ for small $\epsilon$. In particular, the region $U\ge 0.5$ lies inside the black hole and therefore has $r \le 2$. It follows that, for $M_i$ sufficiently close to (but greater than) $M_*(\epsilon)$, we have $r(U,V) \le 2$ for $U \in [0.5,0.6]$ (say) and $V \in [0,V_1]$ for any $V_1 > 0$. Cauchy stability now implies that the solution with $M_i = M_*(\epsilon)$ also must have $r(U,V) \le 2$ in the compact region $[0.5,0.6] \times [0,V_1]$ But this holds for any $V_1 > 0$.\footnote{
Our numerics shows that the solution with $M_i = M_*(\epsilon)$ exists throughout $[0.5,0.6] \times [0,V_1]$ for any $V_1>0$.} Hence, for the solution with $M_i = M_*(\epsilon)$, outgoing null geodesics with $U \in [0.5,0.6]$ do not reach infinity, so this region of the spacetime must lie inside a black hole. 

To prove that this black hole is extreme, assume the converse. Then the spacetime with $M_i=M_*(\epsilon)$ has a trapped 2-sphere, i.e., $r_{,V}<0$ at some point $(U_1,V_1)$. But then we can apply Cauchy stability to deduce that there exists $M_i<M_*(\epsilon)$ for which the spacetime has a trapped 2-sphere at $(U_1,V_1)$. But, as explained in section \ref{trappedapparent}, a trapped 2-sphere must lie inside a black hole, which contradicts the fact that a spacetime with $M_i <M_*(\epsilon)$ does not have a black hole region. Hence the spacetime with $M_i = M_*(\epsilon)$ describes a black hole without trapped 2-spheres: an  extreme black hole. 

For given $\epsilon$, we determine $M_*(\epsilon)$ as follows. As mentioned above, the minimum value of $r r_{,V}$  on a surface of constant $V$ is located at $r=1$. For $M_i>M_*(\epsilon)$, the black hole is non-extreme at late time so this minimum value is negative. For $M_i<M_*(\epsilon)$ the minimum value must be positive because there are no marginally trapped 2-spheres. So $M_*(\epsilon)$ is determined by tuning $M_i$ so that the value of $rr_{,V}|_{r=1}$ on a surface of constant $V$ approaches $0$ at late time (large $V$). In practice, it is easier to tune $f_0$, which is related to $M_i$ by (\ref{M_i}). 

\subsection{Results}

We were led to consider dynamical extreme black holes by the question of stability of extreme RN. When investigating stability, we considered small values of $\epsilon$. However, now it seems more interesting to consider larger $\epsilon$ since this gives an extreme black hole solution which is not particularly close to extreme RN initially. For the rest of this section we will report results for $\epsilon=0.5,0.1$. Smaller values of $\epsilon$ give qualitatively similar results. We will state results for $\epsilon=0.5$ and give the results for $\epsilon=0.1$ in parentheses. 

We find that 
$M_*(\epsilon)-1 = 1.8\times 10^{-1}$ ($7.6\times 10^{-3}$) for $\epsilon=0.5$ ($0.1$)
and the corresponding critical value of 
$f_0-2$ is $1.8\times 10^{-1}$ ($7.1\times 10^{-3}$).
We have determined the solution up to 
$V=500$. 

\begin{figure}
 \centering
 \subfigure[$|\phi|$ at $r=1.5$]
  {\includegraphics[width=8cm,clip]{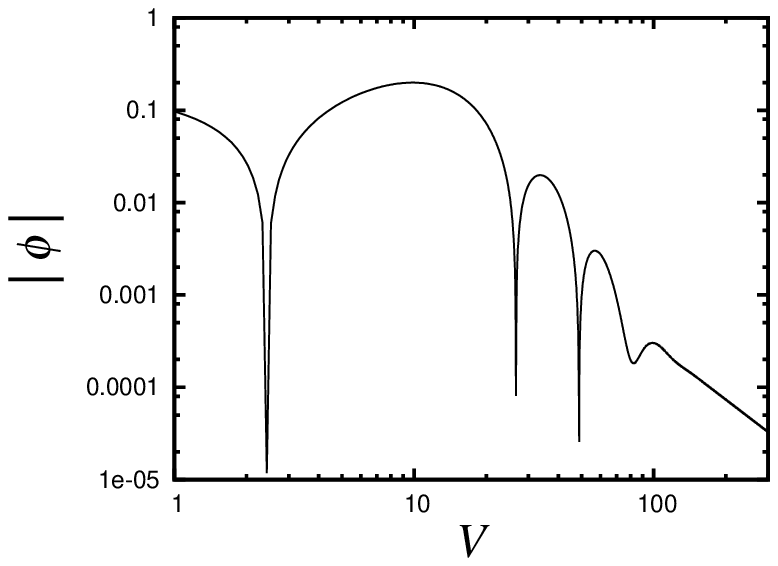} 
\label{fixed_r}}
\subfigure[$r(U,V)$ at fixed $U$]{
\includegraphics[width=8cm,clip]{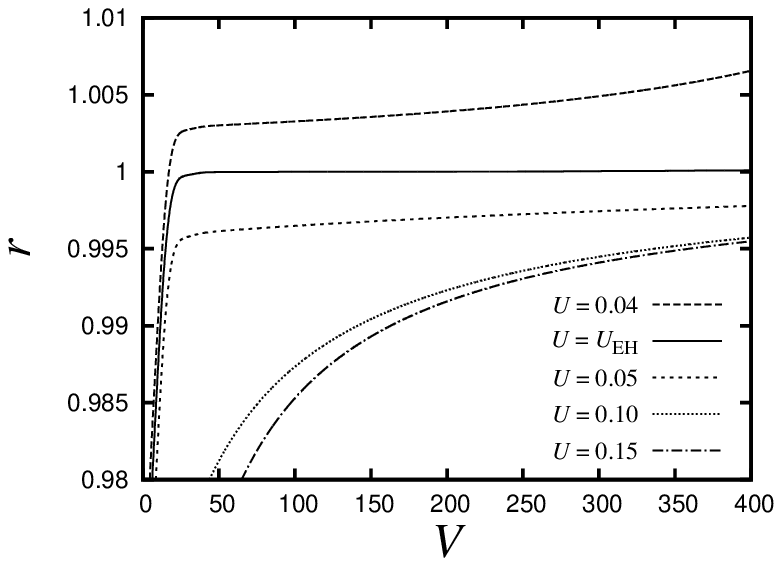}
\label{rEH_deg}
}
  \caption{Results for dynamical extreme black hole with $\epsilon=0.5$. (a) Decay of scalar field at $r=1.5$. Damped oscillations ("quasinormal ringing") are followed by power law decay. (b) Radius of outgoing null geodesics (lines of constant $U$).}
 \end{figure}

For $M_i>M_*(\epsilon)$, the solution settles down to non-extreme RN at late time outside the horizon. For $M_i = M_*(\epsilon)$, we find that the solution settles down to {\it extreme} RN at late time outside the horizon. The scalar field decays for $r>1$, as shown for $r=1.5$ in Fig.~\ref{fixed_r}. At late time, the decay is consistent with a power law tail: fitting the data for $r=1.5$, $V \in [150,300]$ to a power $V^{-a}$ gives $a=2.0$ ($2.0$) which agrees with results for a test field in extreme RN \cite{Lucietti:2012xr,Ori:2013iua}. 

When we studied solutions settling down to non-extreme RN, we determined the location of the event horizon from the late time location of the apparent horizon. The latter is determined as the boundary of the region containing trapped 2-spheres. This does not work here because, as we have explained, a dynamical extreme black hole does not have trapped 2-spheres. In principle, the location of the event horizon can be determined by identifying the value $U_{EH}$ for which $r \rightarrow \infty$ as $V \rightarrow \infty$ along outgoing null geodesics with $U<U_{EH}$ but $r$ is bounded as $V \rightarrow \infty$ for outgoing null geodesics with $U \ge U_{EH}$. However, it is difficult to do this with high accuracy.\footnote{The reason can be understood by considering the extreme RN solution, for which $U_{EH}=0$ and, for small $U$, $r$ becomes large only when $V \gtrsim -2/U$ (see Appendix \ref{RNDN}). Since we evolve only up to $V=500$ we would not expect to determine $U_{EH}$ to an accuracy of better than $0.004$ this way.}

We will adopt a less rigorous way of identifying the location of the event horizon. Since the solution settles down to extreme RN for $r>1$, we assume that the late time limit of the radius of the event horizon coincides with the extreme RN value $r=1$. This is equivalent to assuming that the position of the event horizon behaves continuously as $M_i \rightarrow M_*(\epsilon)$. 

To justify this assumption, we can examine in more detail how quickly the solution settles down to extreme RN. A plot of the renormalized Hawking mass $\varpi$ looks just like Fig.~\ref{Fig:deg_r-mP} except that the late time value is now $\varpi=1$. This shows $\varpi$ has settled down to the extreme RN value $\varpi=1$ in the region $1 < r \le 8$ by time $V=30$. This indicates that the metric settles down to extreme RN near the horizon at early time in our numerical evolution, which extends to $V=500$. Hence our assumption that the late time horizon radius is the same as that of extreme RN seems very reasonable.

Fig.~\ref{rEH_deg} shows that indeed there exists a value $U_{EH}$ such that the outgoing null geodesics $U=U_{EH}$ have $r \rightarrow 1$ at large $V$.\footnote{
In practice we determine $U_{EH}$ by the condition $r(U_{EH},V)=1$ for some large value of $V$. We used $V=100$.}
 We find $U_{\rm EH} = 0.044$ ($0.0015$). As a check on our assumption that $U=U_{EH}$ is the event horizon, Fig.~\ref{rEH_deg} shows that an outgoing radial null geodesic with $U=0.040$ expands at large $V$ so $U=0.040$ must be outside the event horizon.
  
The radius of the event horizon as a function of $V$ is given by $r_\textrm{EH}(V)=r(U_\textrm{EH},V)$. This can be seen in Fig.~\ref{rEH_deg}. The initial event horizon radius is $0.96$ ($0.998$) so the event horizon radius does not vary much with time, just as for the degenerate apparent horizon initial data with much smaller $\epsilon$ (Fig.~\ref{rAHEH}). This is because most of the initial wavepacket lies outside the horizon and propagates to infinity instead of falling into the black hole. 

\begin{figure}
\centering
 \subfigure[Bondi mass]
 { \includegraphics[width=8cm,clip]{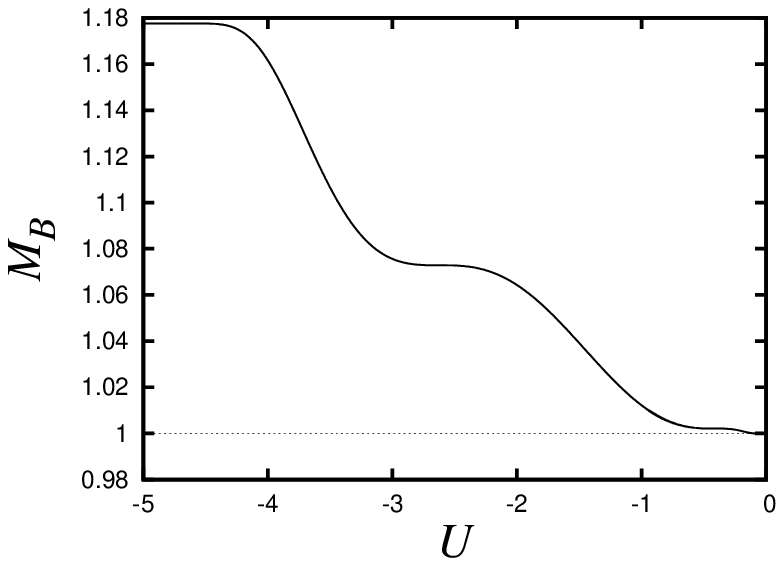}
  \label{fig:bondi_extreme}}
\subfigure[$\varpi(U,V)$ at fixed $U$]{
\includegraphics[width=8cm,clip]{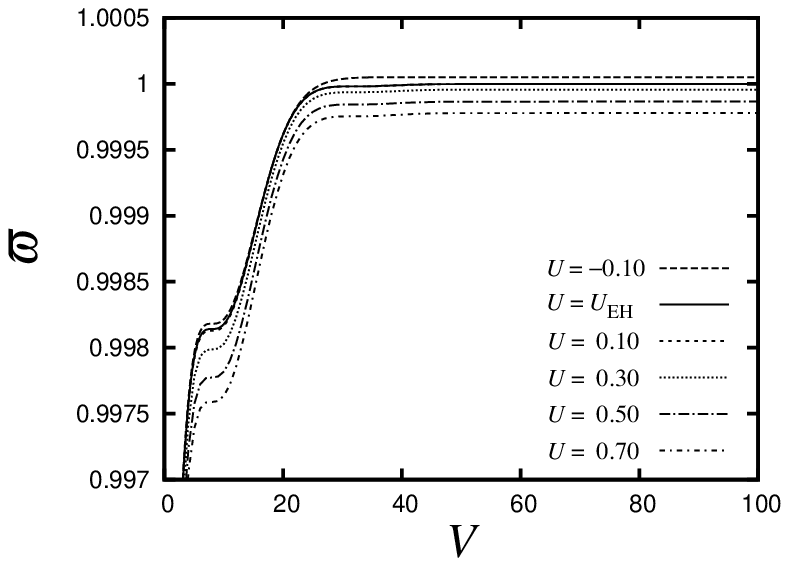}
\label{Fig:V-varpi_eps=0.5}}
\caption{
Bondi mass and renormalized Hawking mass for $\epsilon =0.5$.
}
\end{figure}

The  Bondi mass is shown in Fig.~\ref{fig:bondi_extreme}. It approaches the BPS value $M_B=1$ at late time ($U \rightarrow U_{EH}$).

Fig.~\ref{Fig:V-varpi_eps=0.5} shows the behaviour of $\varpi$ along outgoing null geodesics. For $U=U_{EH}$ we see that $\varpi \rightarrow 1$, indicating that the metric settles down to extreme RN on the event horizon. Note that $\varpi$ has settled down to its asymptotic value by time $V=30$. Since the metric is settling down to extreme RN both on and outside the event horizon, we would expect the late-time evolution of the scalar field to resemble a test field in extreme RN, in which case it must exhibit the Aretakis instability and therefore the solution as a whole does {\it not} approach extreme RN on the horizon. This is indeed the case.

\begin{figure}
  \centering
  \subfigure[$\bigl|\phi|_\text{EH}\bigr|$]
  {\includegraphics[width=5cm]{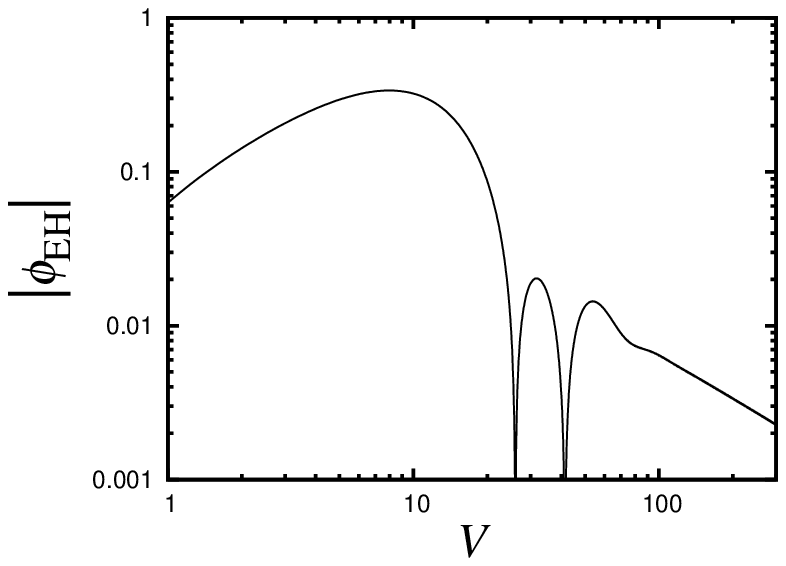}
\label{peh_ex}
  }
  \subfigure[$(\partial_r\phi)_\text{EH}$]
  {\includegraphics[width=5cm]{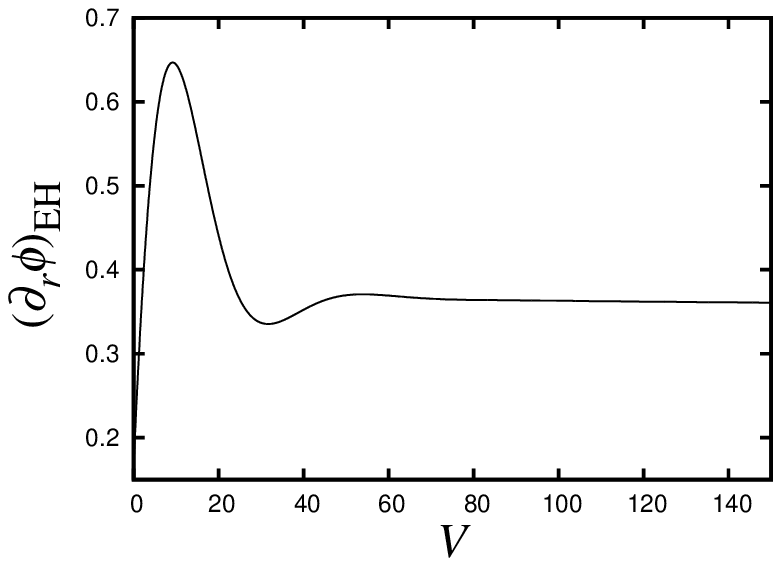} 
\label{dpeh_ex}
  }
  \subfigure[$(\partial_r^2\phi)_\text{EH}$]
  {\includegraphics[width=5cm]{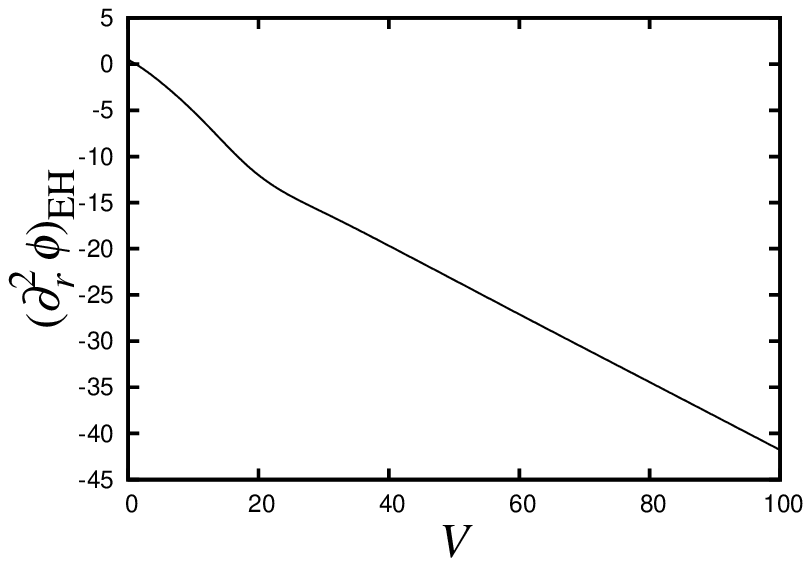} 
\label{ddpeh_ex}
  }
  \caption{
$\phi$ and its first two $r$-derivatives at the event horizon for $\epsilon=0.5$. $(\partial_r \phi)_\textrm{EH}$ tends to a non-zero constant and $(\partial_r^2 \phi)_\textrm{EH}$ blows up linearly as $V \rightarrow \infty$. This is qualitatively the same as the evolution of a test field in extreme RN. 
\label{peh_exs}
}
\end{figure}

First, Fig.~\ref{peh_ex} shows that the scalar field decays on the event horizon. Fitting the solution for $V \in [150, 300]$ to a power law $V^{-a}$ we find $a = 0.95$ ($0.95$)
which agrees (up to numerical error) with the result $\phi \sim 1/V$ for a test field in extreme RN \cite{Lucietti:2012xr,Bizon:2012we}. 

Fig.~\ref{dpeh_ex} shows the behaviour of $(\partial_r \phi)_V \equiv \partial_U \phi/\partial_U r$ at the event horizon. 
At late time this quantity approaches a non-zero constant 
$H_0=0.36$ ($0.047$)
just like a test field in extreme RN. However, outside the horizon $\partial_r \phi$ decays. This implies that $\partial_r^2 \phi$ blows up at late time on the horizon, as shown in Fig.~\ref{ddpeh_ex}. The blow-up is linear in time, just as in the Aretakis instability of a test field in extreme RN. So our results demonstrate that dynamical extreme black holes exhibit a nonlinear version of the Aretakis instability. 

These results show that, at late time, the solution on the horizon is not extreme RN: there is an additional parameter $H_0$, which depends on the initial data e.g., it depends on $\epsilon$. Hence there is ``hair on the horizon'' of a dynamical extreme black hole. This ``hair'' has an effect that can be detected outside the horizon: at fixed $r>1$, the scalar field decays as $V^{-2}$ when $H_0 \ne 0$ but as $V^{-3}$ when $H_0 = 0$ \cite{Lucietti:2012xr,Bizon:2012we}.

These results are for $\epsilon=0.5$ ($0.1$), which is not very small. However, we find qualitatively similar behaviour for smaller $\epsilon$, corresponding to a small perturbation of extreme RN. This shows that there exist non-generic initial perturbations of extreme RN for which the Aretakis instability never decays. 


Ref.~\cite{Ori:2013iua} made a prediction for the late time behaviour of a test scalar field outside the event horizon of extreme RN. In our coordinates, this is
\begin{equation}
\phi \sim -\frac{4H_0}{(r-1)V^2}
\label{Ori}
\end{equation}
for $V \gg |r_*(r)|$ where $r_*$ is defined by (\ref{tortoise}). Since our field is behaving like a test field at late time, it should match this prediction. We have performed numerical fits of $(r-1)V^2 \phi/(-4H_0)$ for $V \in [200,500]$ using a function $a + b/V + c/V^2$ for $r=1.5,3.0,5.0$. In all cases we find that $a$ lies between $1.01$ and $1.02$,\footnote{
The fit is slightly improved if we allow the subleading terms to include $(\log V)/V$.} so the prediction of Ref.~\cite{Ori:2013iua} is in good agreement with our results.

Since we are considering ``critical'' solutions obtained by tuning a 1-parameter family of initial data, it is natural to ask whether there is any analogue of black hole critical phenomena \cite{Choptuik:1992jv} in this model. Is there universal behaviour as $M_i \rightarrow M_*(\epsilon)$? Here, ``universal'' means ``independent of initial data'', i.e., independent of $\epsilon$. Hence if there is universality then the behaviour for general $\epsilon$ must be the same as for $\epsilon=0$, i.e., for the RN solution. We find that this is indeed the case: the final Bondi mass $M_f$ obeys $M_f -1 \propto M_i-M_*(\epsilon)$ as $M_i \rightarrow M_*(\epsilon)$ for $\epsilon=0.5,0.1$, in agreement with the trivial case $\epsilon=0$. This implies that the surface gravity of the final black hole is $\kappa \propto (M_i - M_*(\epsilon))^{1/2}$. Given our results in the previous section, it follows that the ``decay time'' for the solution to settle down to non-extreme RN (on and outside the horizon) scales as $V \propto (M_i - M_*(\epsilon))^{-1/2}$.  

\subsection{Black hole interior}

We will now describe the interior of a dynamical extreme black hole. Our main interest is what happens as $V \rightarrow \infty$ inside the black hole. In the non-extreme case, it has been proved that one can introduce a new coordinate $\hat{V}(V)$ with $\hat{V}(\infty)= \hat{V}_*$ finite, such that the metric, Maxwell field and scalar field can be continuously extended across the null surface $\hat{V} = \hat{V}_*$ \cite{Dafermos:2003wr}. This surface is a Cauchy horizon in the extended spacetime (denoted ${\cal CH}$ in Fig.~\ref{initial}). However, one finds that $\partial_{\hat{V}} \phi$, $\partial_{\hat{V}} r$ and $\varpi$ diverge at the Cauchy horizon. This implies that one cannot extend the fields so that they are $C^1$ at the Cauchy horizon. In particular, one cannot extend the fields so that they satisfy the equations of motion, even weakly, at the Cauchy horizon.

At the level of test fields, the divergence of $\partial_{\hat{V}} \phi$ at the Cauchy horizon can be understood as follows \cite{chandra}. One can argue that $\partial_V \phi \sim V^{-p}$ as $V \rightarrow \infty$ for some $p>0$. For a test field, the metric is exactly non-extreme RN, which can be extended across the Cauchy horizon (at $r=r_-$) by defining $\hat{V} = - e^{-\kappa_- V}$ where $\kappa_-$ is the surface gravity of the Cauchy horizon. But then we have $\partial_{\hat{V}} \phi \sim e^{\kappa_- V} V^{-p}$, which diverges as $V \rightarrow \infty$. But note that for {\it extreme} RN we have $\hat{V} \sim -V^{-1}$ so {\it if} we had $\partial_V \phi \sim V^{-p}$ then we would get $\partial_{\hat{V}} \phi \sim \hat{V}^{p-1}$ which is continuous at $\hat{V}=0$ for $p \ge 1$. This suggests that, when one includes backreaction, the Cauchy horizon of a dynamical extreme black hole might be smoother than that of a non-extreme black hole. We will now argue that this is indeed true.

First, it is easy to see that $\varpi$ is bounded in an extreme black hole. Consider a black hole which has no trapped surfaces on the surface $V=V_1$. Pick a point $(U_1,V_1)$ inside the black hole. Since the black hole is extreme we have $r_{,V}(U,V_1) \ge 0$ and so (from (\ref{dMPI})) $\varpi_{,U}<0$. Hence $\varpi(U_1,V_1) \le \varpi(U_2,V_1)$ for any $U_2<U_1$. Take $U_2<U_{\rm EH}$ so $(U_2,V_1)$ lies outside the event horizon. We then have $\varpi(U_2,V_1) \le M_B(U_2)$ using $\varpi_{,V} \ge 0$. Hence $\varpi(U_1,V_1) \le M_B(U_2)$. In particular, $\varpi(U_1,V_1) \le M_f$ where $M_f = \lim_{U \rightarrow U_{\rm EH}} M_B(U)$ is the ``final'' Bondi mass. If the black hole is extreme then this holds for all $V_1$ so we have $\varpi \le M_f$ throughout the black hole interior. 

Our numerics indicate that $M_f = 1$ for our $M_i = M_*(\epsilon)$ solutions. Hence we must have $\varpi \le 1$ everywhere inside the black hole.\footnote{ 
This implies $g^{\mu\nu}\nabla_\mu r \nabla_\nu r = 1 -2\varpi/r+1/r^2  \ge (1-1/r)^2 >0$ (since $r<1$ inside the black hole). Therefore surfaces of constant $r$ are timelike inside a dynamical extreme black hole, just as for extreme RN.} In particular $\varpi$ cannot diverge as $V \rightarrow \infty$. 

Fig.~\ref{Fig:V-varpi_eps=0.5} shows how $\varpi$ behaves at large $V$ along lines of constant $U$ inside the black hole. In all cases, $\lim_{V\rightarrow \infty} \varpi(U,V)<1$. $\varpi$ is close to $1$ inside the black hole, indicating that the solution there is close to the extreme RN interior, even though $\epsilon=0.5$ is quite large. This is because only a small part of our initial wavepacket (Fig.~\ref{Fig:ID}) lies inside the black hole. Most of it lies outside the black hole and propagates to infinity. By considering different initial data (or larger $\epsilon$) it should be possible to construct dynamical extreme black holes for which $\varpi$ is not so close to $1$ behind the event horizon. 

\begin{figure}[htbp]
\includegraphics[width=7.5cm,clip]{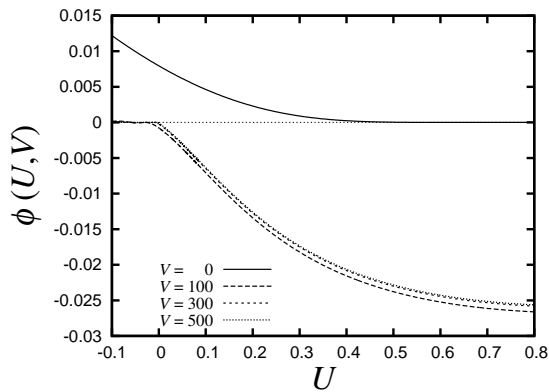}
\centering
\caption{$\phi(U,V)$ at $V=0, 100, 300, 500$ for the dynamical extreme black hole with $\epsilon=0.5$.}
\label{Fig:U-phi_dex}
\end{figure}

For a non-extreme black hole, the fact that $\phi$ admits a $C^0$ extension to the Cauchy horizon implies that $\phi(U,V)$ approaches a fixed profile as $V \rightarrow \infty$. We find that the same is true for a dynamical extreme black hole. This is shown in Fig.~\ref{Fig:U-phi_dex}. Note the apparent discontinuity in $\partial_U \phi$ at the event horizon that develops at large $V$. The same occurs for a non-extreme black hole. In the non-extreme case, this is a coordinate effect arising because $f \sim e^{\kappa V}$ on the event horizon (see Appendix \ref{RNDN}), implying that a small fixed interval of $U$ centred on the event horizon corresponds to an increasingly large region of spacetime as $V \rightarrow \infty$. 

In the extreme case, the interpretation is different and can be understood by thinking about a test field in extreme RN. In this case $f \rightarrow 2$ along the horizon at large $V$. From the work of Aretakis, we know that $\partial_r \phi$ decays outside the horizon but not {\it on} the horizon. We also have $\partial_U r = -1$ on the horizon (see Appendix \ref{RNDN}). Hence $\partial_U \phi$ decays outside, but not on, the event horizon and so $\partial_U^2 \phi$ becomes large on the event horizon at late time. So the late-time apparent discontinuity in Fig.~\ref{Fig:U-phi_dex} is the (nonlinear version of the) Aretakis instability discussed above.

Next consider the behaviour of $f$ as $V \rightarrow \infty$. It is convenient to recall how $f$ behaves for extreme RN (Appendix \ref{RNDN}). The event horizon of extreme RN is at $U=0$. For large $V$ and small $U$ one has $r \approx 1-2U/(2+UV)$ which implies $f \approx 8/(2+UV)^2$. Hence on the horizon we have $f \rightarrow 2$ as $V\rightarrow \infty$ and behind the horizon, for small $U$, we have $f \rightarrow 0$ as $V \rightarrow \infty$.

\begin{figure}[htbp]
\centering
\subfigure[$f(U,V)$]{\includegraphics[width=7.5cm,clip]{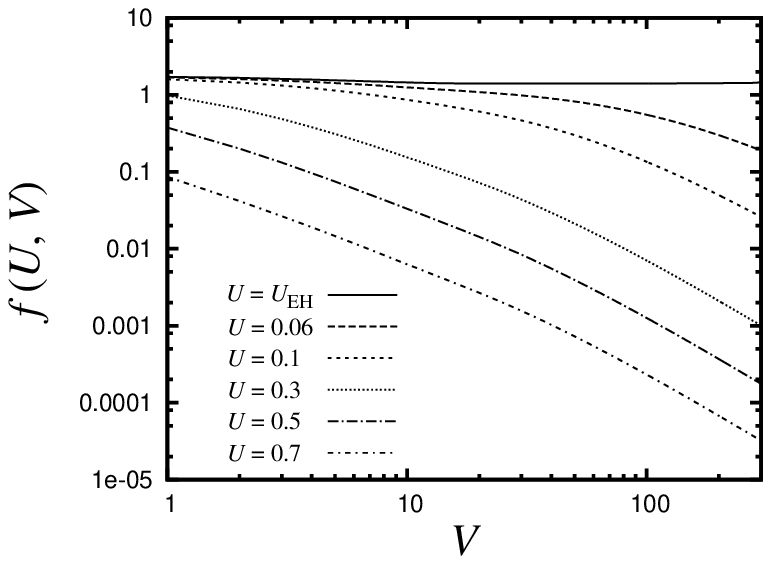}}
\subfigure[$\hat f(U,V)$]{\includegraphics[width=7.5cm,clip]{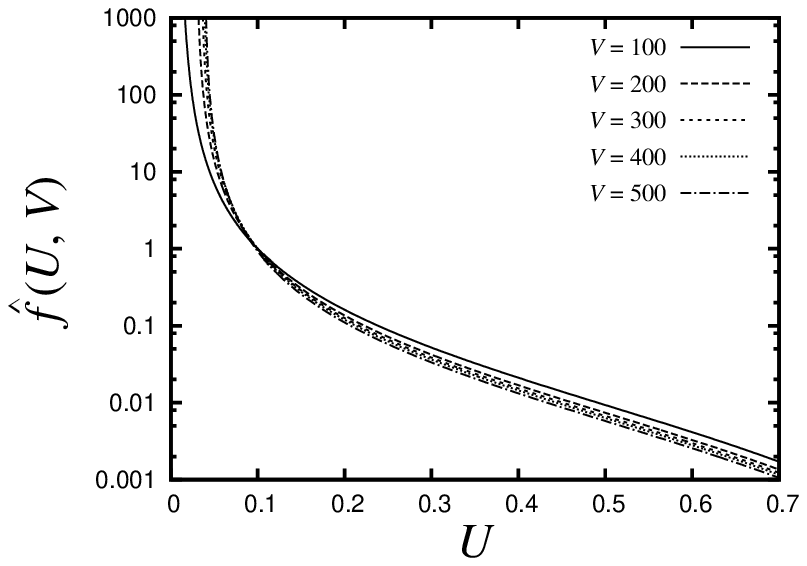}}
\caption{(a): $f(U,V)$ for the dynamical extreme black hole with $\epsilon=0.5$.
(b): $\hat f(U,V) = f(U,V)/f(0.1,V)$.}
\label{Fig:V-f_eps=0.1}
\end{figure}

We find qualitatively the same behaviour for $f$ in a dynamical extreme black hole. Along the event horizon we find that $f$ approaches a constant value $f(U_{EH},V) \rightarrow 1.41$ ($1.98$)
as $V \rightarrow \infty$ for $\epsilon=0.5$ ($0.1$).\footnote{This limiting value differs from the extreme RN value which appears to contradict our statement that the metric settles down to extreme RN along the event horizon. However, there is no contradiction because $f$ is not a gauge invariant quantity. Recall that $\varpi \rightarrow 1$ along the event horizon, and $\varpi$ is gauge invariant.} Inside the horizon, we find that $f$ decays at large $V$. This is shown in Fig.~\ref{Fig:V-f_eps=0.1}(a). Fitting to a power law, we find that $f(0.1,V)$ decays as $V^{-a}$ as $V \rightarrow \infty$ with $a \approx 1.8$ ($1.9$). Note that $a=2$ for extreme RN.

We now define $\hat{V}$ as follows. Pick $U_*>U_{\rm EH}$ (e.g.\ $U_*=0.1$) and define 
\be
 \hat{f}(U,V) = \frac{f(U,V)}{f(U_*,V)}
\ee
and
\be
 \hat{V} = \int_0^V f(U_*,V') dV'
\ee
The numerical results just mentioned show that this integral converges: $\hat{V} \rightarrow \hat{V}_*$ as $V \rightarrow \infty$. In $(U,\hat{V})$ coordinates, the metric becomes
\be
 ds^2 = - \hat{f} dU d\hat{V} + r^2 d\Omega^2
\ee 
Panel (b) of 
Fig.~\ref{Fig:V-f_eps=0.1}
shows how $\hat{f}(U,V)$ behaves at large $V$ for $U_*=0.1$. This plot is consistent with $\hat{f}(U,V)$ approaching a finite limit as $V \rightarrow \infty$ although the convergence to this limit is slower than for $\phi(U,V)$. However, the rate of convergence is exactly as for extreme RN: a plot of $\hat{f}$ for extreme RN is almost identical to our plot. So our results indicate that $\hat{f}$ extends continuously to the Cauchy horizon at $\hat{V} = \hat{V}_*$.

\begin{figure}[htbp]
\centering
\subfigure[$r(U,V)$]{\includegraphics[width=7.5cm,clip]{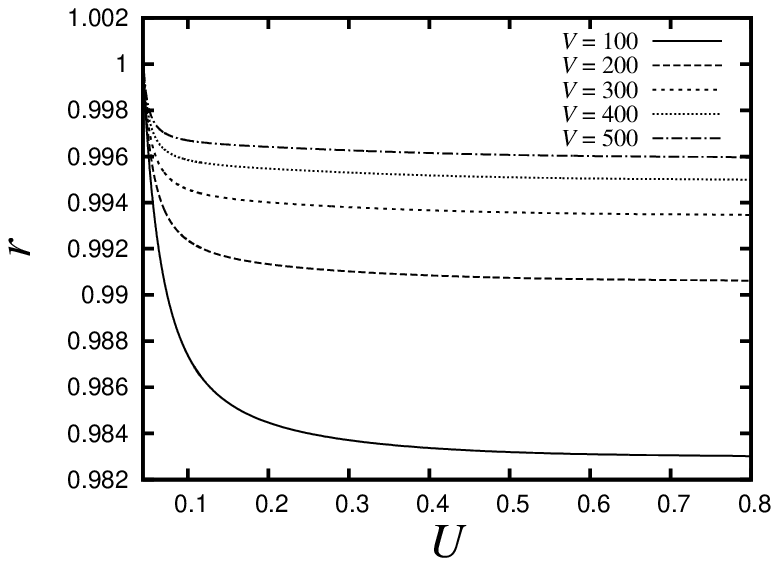}}
\subfigure[$-r_{,U}/\hat f$]{\includegraphics[width=7.5cm,clip]{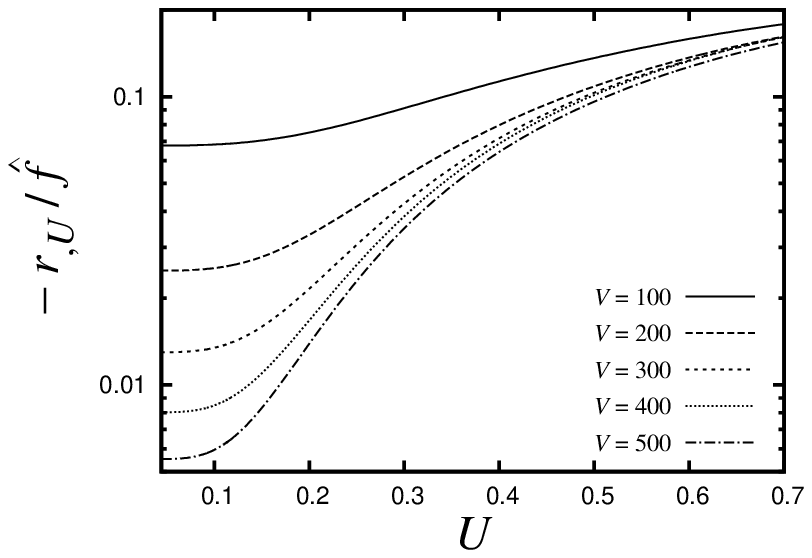}}
\caption{Panel (a), (b):
$r$ and $r_{,U}/\hat f$ of the dynamical extreme black hole with $\epsilon=0.5$.
$r_{,U}(U,V)$ converges into a nontrivial profile at late time, and the convergence appear to occur earlier at larger $U$. This implies that the limiting value of $r(U,V)$ for $V\to \infty$ is not constant.
}
\label{Fig:U-r_eps=0.1}
\end{figure}

The behaviour of $r(U,V)$ is shown in Fig.~\ref{Fig:U-r_eps=0.1}(a). The solution is close to the corresponding solution for extreme RN which has $r \approx 1-2/V$ for small $U$ with $V \gg 2/U$. However, unlike extreme RN, the limiting value of $r$ as $V \rightarrow \infty$ cannot be constant.  To see this, note that the equations of motion are gauge-invariant with respect to choice of null coordinates. Hence we can insert hats on $V$ and $f$ in these equations. The ``hatted version'' of (\ref{C2}) implies that $r_{,U} \ne 0$ at $\hat{V} = \hat{V}_*$. Since $r_{,U}<0$ for $\hat{V}<\hat{V}_*$ we expect $r_{,U}<0$ at $\hat{V} = \hat{V}_*$: the Cauchy horizon contracts in response to the scalar field energy crossing it, just as in the non-extreme case. Fig.~\ref{Fig:U-r_eps=0.1}(b) shows $r_{,U}/\hat{f}$, which is the rate of change of $r$ with respect to an affine parameter along a line of constant $V$, so the limit as $V \rightarrow \infty$ gives the rate of change of $r$ w.r.t.\ an affine parameter along the Cauchy horizon. 

\begin{figure}[htbp]
\centering
\includegraphics[width=8cm,clip]{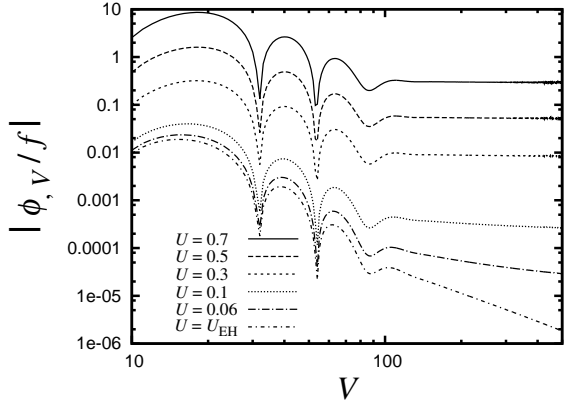}
\caption{$\phi_{,\hat V}/\hat f=\phi_{,V}/f$ for the dynamical extreme black hole with $\epsilon=0.5$. Damped oscillations occur early in the evolution. For any $U>U_\text{EH}$, $\phi_{,\hat V} / \hat f$ converges into a constant at late time (the convergence is slower for $U$ close to $U_{EH}$).}
\label{Fig:V-phihatV=0.1}
\end{figure}

So far, this is exactly as for a non-extreme black hole: the fields can be continuously extended to the Cauchy horizon. But for a non-extreme black hole, it is known that the extension is not $C^1$, as discussed above. We will now argue that the extension {\it is} $C^1$ for a dynamical extreme black hole. Fig.~\ref{Fig:V-phihatV=0.1} shows that $\phi_{,V}/f = \phi_{,\hat{V}}/\hat{f}$ converges to a finite limit as $V \rightarrow \infty$. (Recall that this quantity diverges exponentially with $V$ in the non-extreme case.) It follows that $\phi_{,\hat{V}}$ extends continuously to the Cauchy horizon. Now the ``hatted version'' of (\ref{C1}) implies that $ r_{,\hat{V}}/\hat{f}$ is $C^1$ at the Cauchy horizon and hence $r_{,\hat{V}}$ must be $C^0$ there. Finally, let $X$ denote the RHS of the ``hatted version'' of (\ref{fUV}). The results just obtained imply that $X$ extends continuously to the Cauchy horizon.\footnote{
Actually we also need that $\phi_{,U}$ and $r_{,U}$ extend continuously to ${\cal CH}$. For $\phi_{,U}$ this is apparent from Fig. \ref{Fig:U-phi_dex}. For $r_{,U}$ it follows by writing the LHS of the hatted version of (\ref{rUV}) as $(rr_{,U})_{,\hat{V}}$ and integrating w.r.t $\hat{V}$. Similarly, integrating the hatted version of (\ref{fUV}) w.r.t. $\hat{V}$ shows that $\hat{f}_{,U}$ extends continuously to ${\cal CH}$.}
 Integrating with respect to $U$ gives
\be
 \left( \log \hat{f}(U,\hat{V}) \right)_{,\hat{V}} = \int_{U_*}^U X(U',\hat{V}) dU'
\ee
where we used $\hat{f}(U_*,V) = 1$. This shows that $\hat{f}_{,\hat{V}}$ extends continuously to the Cauchy horizon. 

\begin{figure}[htbp]
\centering
\includegraphics[width=6cm, clip]{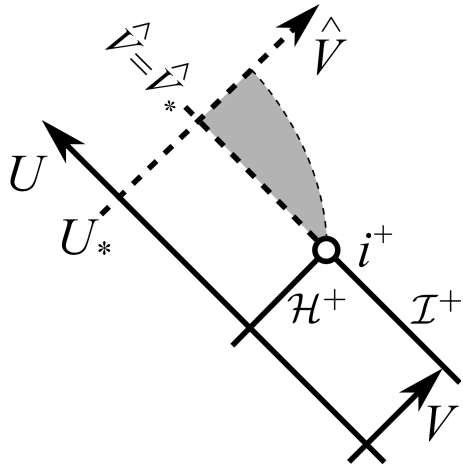}
\caption{Extension across the Cauchy horizon.}
\label{Fig:cauchy}
\end{figure}

We have shown that the fields can be extended to the Cauchy horizon in a $C^1$ manner. An extension beyond the Cauchy horizon can be constructed using a standard argument. View the Cauchy horizon $\hat{V} = \hat{V}_*$ as an ingoing null hypersurface, and pick an outgoing null hypersurface e.g.\ $U = U_*$, $ \hat{V}>\hat{V}_*$ (see Fig.~\ref{Fig:cauchy}).  Take initial data on $\hat{V} = \hat{V}_*$ to be the data just obtained. Take any smooth data on the outgoing null hypersurface that is smooth for $ \hat{V}>\hat{V}_*$ and $C^1$ at $ \hat{V}=\hat{V}_*$. Now one can solve the equations of motion in the region $U<U_*$, $V> \hat{V}_*$. A solution will exist at least in a neighbourhood of the hypersurfaces (the shaded region of Fig.~\ref{Fig:cauchy}).\footnote{Note that this is ``solving in a spacelike direction''. This is legitimate because we are solving wave equations in $1+1$ dimensions. In $1+1$ dimensions, reversing the sign of the metric interchanges space and time.} This construction gives a $C^1$ extension of the fields across the Cauchy horizon. The equations of motion will be satisfied everywhere except perhaps on the Cauchy horizon.

On the Cauchy horizon, the fields are at least a {\it weak} solution of the equations of motion, which is defined as follows. Multiply each (hatted) equation of motion by an arbitrary smooth test function of compact support and integrate by parts to eliminate the second derivatives. A set of fields is said to be a weak solution if the resulting equations are satisfied for arbitrary test functions. To see that our $C^1$ extension is a weak solution, split each integral into two parts, one on each side of the Cauchy horizon. Now integrate by parts to recover the second derivatives. The bulk terms vanish because the equations of motion are satisfied on either side of the Cauchy horizon. The surface terms from the two sides cancel because the fields are $C^1$ at the Cauchy horizon.

It would be interesting to know whether the extension just constructed is actually $C^2$ at the Cauchy horizon, and therefore a solution, rather than just a weak solution, of the equations of motion. However, we have not been able to determine the limiting behaviour of second derivatives (e.g.\ $\phi_{,\hat{V}\hat{V}}$) with sufficient numerical accuracy to answer this.

The above discussion applies to the Cauchy horizon denoted ${\cal CH}$ in Fig.~\ref{initial}. There is another Cauchy horizon ${\cal CH}'$ in Fig.~\ref{initial}. This arises because we cannot maintain acceptable numerical accuracy close to the singularity at $U=1$ on the initial data surface. Therefore we restricted our numerical domain to the region $U \le U_{\rm max}<1$, with ${\cal CH}'$ the surface $U=U_{\rm max}$. But it still makes sense to ask about the nature of the spacetime we would obtain with $U_{\rm max}=1$ even if we can't determine this numerically. In particular, what is the nature of the future boundary of the future domain of dependence of $\Sigma$? {\it A priori}, this boundary might have a singular spacelike component but Refs.~\cite{Dafermos:2003wr,Kommemi:2011wh} showed that $r \rightarrow 0$ on such a component. This is not possible here because we have $r_{,V}>0$ in the future domain of dependence of $\Sigma$. Hence the boundary must be null, composed of ${\cal CH}$ and ${\cal CH}'$, with the latter now at $U=1$, i.e., emanating from the singularity on $\Sigma$ in Fig.~\ref{initial}. We have argued that there exists a $C^1$ extension of the solution across (at least the early part of) ${\cal CH}$. It would be interesting to know what happens at ${\cal CH}'$: can the solution be extended across ${\cal CH}'$ or is this a null singularity?

\subsection{Observer crossing the event horizon at late time}

Marolf and Ori \cite{Marolf:2010nd,Marolf:2011dj} have suggested that a freely falling observer who crosses the event horizon of an extreme black hole {\it at late time} will experience a singularity there. (See also Ref. \cite{Garfinkle:2011pj}.) We will now review their argument, as applied to the Einstein-Maxwell-scalar field theory.

Marolf and Ori considered an observer following a timelike geodesic. However, to simplify the discussion we will examine the geometry probed by late time {\it null} ingoing geodesics. Consider a black hole which settles down to non-extreme RN at late time. Far from the black hole, $f \rightarrow 2$ and $(U,V)$ are affine parameters along ingoing and outgoing null geodesics. Consider an ingoing geodesic with affine parameter $\lambda$ normalized so that the geodesic has unit ``energy''. Such a geodesic is labelled by advanced time $V$ which is constant along the geodesic. Along such a geodesic we have
\be
 \frac{dU}{d\lambda} = \frac{2}{f}
\ee 
hence the rate of change of the scalar field along the geodesic is
\be
 \frac{d\phi}{d\lambda} = 2 \frac{\phi_{,U}}{f}\ .
\ee
But inside the black hole we know that $f \rightarrow 0$ and $ \phi_{,U}$ tends to a non-zero limit as $V \rightarrow \infty$ at fixed $U$. So, in the limit $V \rightarrow \infty$ (at fixed $U$), $d\phi/d\lambda$ diverges inside the black hole. This is because the affine time it takes such a geodesic to traverse a fixed $U$-interval inside the black hole tends to zero as $V \rightarrow \infty$. Marolf and Ori showed that the same applies to ingoing timelike geodesics of fixed energy/unit mass where $V$ labels the time at which the geodesic crosses the event horizon. Hence an observer who falls freely into the black hole at late time experiences very large gradients inside the black hole. We emphasize that the interior geometry is perfectly smooth, but the observer traverses this geometry ``at increasingly great speed'' as $V \rightarrow \infty$. 

For a non-extreme black hole, this large gradient occur a non-zero affine time after the late-time geodesic crosses the horizon. The affine time for our null geodesic (labelled by $V$) to travel from the event horizon $U=U_{EH}$ to some other value of $U>U_{EH}$ is 
\be
 \lambda(U,V) \equiv \frac{1}{2} \int_{U_{EH}}^{U} f(U',V) dU'
\ee
Now $f \rightarrow 0$ as $V \rightarrow \infty$ {\it inside} the black hole. However, {\it on} the event horizon $f$ diverges as $e^{\kappa V}$ and one finds that $\lambda(U,V)$ approaches a non-zero limit as $V \rightarrow \infty$. For example, in the case of a non-extreme RN solution, $\lambda(U,V)=r_+-r(U,V) \rightarrow r_+ - r_-$ as $V \rightarrow \infty$ where $r_-$ is the inner horizon radius.

By extrapolating from the non-extreme case, Marolf and Ori suggested that an observer of fixed energy/unit mass who crosses the event horizon of an extreme black hole at time $V$ will experience very large gradients within a proper time that vanishes as $V \rightarrow \infty$. In other words, a very late time observer experiences a singularity as soon as s/he crosses the horizon.

\begin{figure}[htbp]
\centering
\includegraphics[width=7.5cm,clip]{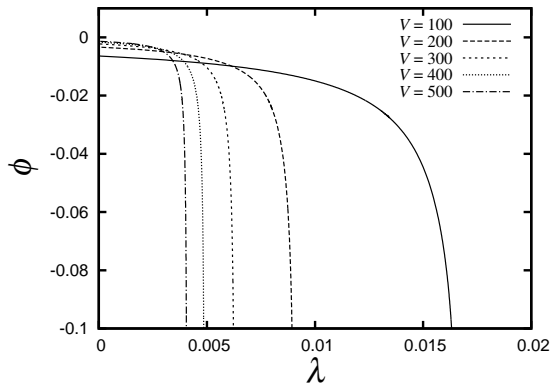}
\caption{
Dependence of $\phi$ on affine parameter $\lambda$ along ingoing null geodesics $V={\rm const}$. ($\epsilon=0.5$). 
}
\label{Fig:normalizedlambda_eps=0.5}
\end{figure}

Marolf and Ori could only conjecture about the extreme case because there were no results for the interior of a dynamical extreme black hole. But now we can use our results to confirm this conjecture. We have shown that, inside the black hole, $\phi_{,U}$ approaches a non-zero limit and $f$ vanishes as $V \rightarrow \infty$. Hence $d\phi/d\lambda$ diverges as $V \rightarrow \infty$ at fixed $U$ inside the black hole, just as in the non-extreme case. But in the extreme case, $f$ approaches a finite limit on the event horizon as $V\rightarrow \infty$. This implies that $\lambda(U,V) \rightarrow 0$ as $V \rightarrow \infty$. So the affine time it takes for the null geodesic to travel from the event horizon to a place where $d\phi/d\lambda$ is becoming large tends to zero as $V \rightarrow \infty$. Hence a very late time ingoing null geodesic of unit energy experiences large gradients immediately behind the horizon. This is shown in Fig.~\ref{Fig:normalizedlambda_eps=0.5}. The affine time at which the gradient $d\phi/d\lambda$ becomes large is well-approximated by $\lambda \approx 2/V$. This can be understood by recalling that, for our solutions, the metric is close to that of extreme RN, for which $\lambda(U,V) \approx 2/V$ for $V \gg 2/U$.  

These results confirm that an ingoing null geodesic of fixed energy that crosses the event horizon at late time will experience large gradients almost immediately inside the horizon. A similar result will apply to ingoing timelike geodesics of fixed energy, in agreement with the conjecture of Marolf and Ori. However, this result applies only in the limit $V \rightarrow \infty$. An observer who crosses the horizon at any finite value of $V$ will live for a non-zero time ($\sim 2/V$) before experiencing large gradients. 


\section{Discussion}
\label{discussion}

In this paper we have demonstrated that the instability of extreme RN discovered by Aretakis persists when gravitational backreaction is included. Generically, the endpoint of the instability is a non-extreme RN black hole. However, by fine-tuning the initial perturbation one can produce a dynamical extreme black hole, which settles down to extreme RN outside the event horizon, for which the instability never decays on the horizon. An observer who falls freely across the event horizon at late time will experience large gradients almost immediately, as conjectured by Marolf and Ori.  We have studied the interior of such a black hole and found strong evidence that it can be extended as a weak solution across an inner Cauchy horizon, in contrast with the non-extreme case. 

Ref.~\cite{Dain:2011kb} proved an inequality relating the area of a marginally trapped 2-surface to its quasilocal electric and magnetic charges: $A \ge 4 \pi (Q_E^2 + Q_M^2)$. where $Q_E$ and $Q_M$ are the integrals of $\star F/(4\pi)$ and $F/(4\pi)$ over the surface. Applying this to the apparent horizon in our model, this result reduces to the simple inequality $r_{AH} \ge |Q|$.\footnote{This is easy to prove directly: in section \ref{extremeintro} we showed that if there is a (marginally) trapped symmetry 2-sphere on a surface of constant $V$ then the 2-sphere at $r=|Q|$ is (marginally) trapped and so the apparent horizon must have $r \ge |Q|$.} This is, of course, respected by all of our solutions. However, it is interesting to note that the area of the event horizon need {\it not} satisfy this inequality.\footnote{We are grateful to Sergio Dain for a discussion of this point.} {\it If} an apparent horizon is present on a surface of constant $V$ then we must have $r_{EH} \ge r_{AH} \ge |Q|$ on this surface. However, our dynamical extreme black holes have $r_{EH}<|Q|$. This is possible because they do not have an apparent horizon. We can also construct solutions which settle down to non-extreme RN at late time but for which an apparent horizon is absent at early time (small $V$). Such solutions can also have $r_{EH}<|Q|$ at early time. Hence it is incorrect to interpret the results of Ref.~\cite{Dain:2011kb} as providing a firm lower bound on the size of a black hole with given charge.

We have considered the simplest theory for which the gravitational backreaction of the Aretakis instability can be studied. In this model, one can study evolution of non-equilibrium black hole spacetimes. However, since the model does not contain charged matter, one cannot study {\it formation} of charged black holes by gravitational collapse. Formation of an extreme RN black hole in spherically symmetric collapse of a thin charged shell was studied long ago (see e.g.\ Ref.~\cite{boulware}). In this case, (electrovac) Birkhoff's theorem implies that the solution is exactly extreme RN outside the shell so there is no possibility of an Aretakis instability. Therefore it would be very interesting to study a model in which the dynamics is non-trivial. One could consider gravitational collapse of a charged scalar field. Or a collapsing thin charged shell together with a massless scalar field. How do dynamical extreme black holes behave in such models?

The Aretakis instability involves growth of second and higher derivatives of the scalar field on the event horizon. Therefore, although they can be arbitrarily large compared to the string scale, our dynamical extreme black holes may be sensitive to higher-derivative corrections to the classical equations of motion. Could such corrections eliminate the Aretakis instability?  Higher derivative terms are usually treated perturbatively e.g.\ the leading order correction to the classical solution is sourced by the leading higher derivative terms evaluated on the classical solution. But for this to be consistent, the corrected solution must remain close to the classical solution. This suggests that higher derivative corrections could not eliminate the instability. However, it is conceivable that the corrected solution always remains close to {\it some} classical solution, but not the same classical solution for all time. For example, maybe such corrections lead to a slow decay of the ``hair'' (the constant $H_0$) on the horizon of a dynamical extreme black hole. It would be interesting to consider this further.

\medskip

\subsection*{Acknowledgments}

We are grateful to Mihalis Dafermos for helpful discussions. 
NT acknowledges hospitality and fruitful discussions at Yukawa Institute for Theoretical Physics, where a part of this work was accomplished, and thanks Tomohiro Harada, Hideki Maeda and Tsuyoshi Houri for useful comments. 
The work of KM is supported in part by JSPS Grant-in-Aid for Scientific
Research No.24$\cdot$2337.
HSR is supported by a Royal Society University Research Fellowship and by European Research Council Grant No.\ ERC-2011-StG 279363-HiDGR. 
The work of NT is supported in part by World Premier International Research Center Initiative (WPI Initiative), MEXT, Japan, JSPS Grant-in-Aid for Scientific Research 25$\cdot$755 and the DOE Grant DE-FG03-91ER40674. 
A part of numerical computation in this work was carried out at the Yukawa Institute Computer Facility.

\appendix 

\section{Reissner-Nordstrom black holes in double null coordinates}
\label{RNDN}

In this section, we will write the RN solution in double null
coordinates regular at the future event horizon ${\cal H}^+$. 
A familiar way of doing this is to use Kruskal-like coordinates. 
However, since we wish to consider both the non-extreme and extreme
cases, we will use slightly different coordinates. We start from the RN metric in static coordinates:
\begin{equation}
 ds^2=-F(r)dt^2+\frac{dr^2}{F(r)}+r^2d\Omega^2\quad 
F(r)=1-\frac{2M}{r}+\frac{Q^2}{r^2}\
\end{equation}
where $M$ is the mass. For the non-extreme case, $M>|Q|$, the outer and inner horizons are located at
$r=r_\pm\equiv M\pm\sqrt{M^2-Q^2}$. The horizons are coincident in the extreme case $M=|Q|$. Now define the tortoise coordinate for the non-extreme and extreme cases, respectively:
\begin{equation}
 r_\ast(r)=\left\{
\begin{array}{ll}
r-r_+ +
  \frac{1}{2\kappa_+}\log \left|\frac{r-r_+}{r_+}\right|
- \frac{1}{2\kappa_-}\log \left|\frac{r-r_-}{r_+}\right| & (r_+>r_-)\\
r-r_+ + 2r_+ \log \left|\frac{r-r_+}{r_+}\right|-\frac{r_+^2}{r-r_+} & (r_+=r_-)
\end{array}
\right.
\label{tortoise}
\end{equation}
where, in the non-extreme case, $\kappa_\pm=(r_+-r_-)/(2r_\pm^2)$ are the (positive) surface gravities of the outer and inner horizons respectively. Defining retarded and advanced time coordinates $u=t-r_\ast$ and $V=t+r_\ast$ we obtain
\begin{equation}
 ds^2=-F(r(u,V))dudV+r^2(u,V)d\Omega^2\ .
\end{equation}
where $r(u,V)$ is determined by solving $r_\ast(r)=(V-u)/2$. These coordinates cover only the black hole exterior $r>r_+$. Define a new retarded time coordinate $U(u)<0$ by
\begin{equation}
 \frac{u}{2}=-r_\ast(r_+-U)\ .
\end{equation}
Note that $U=0$ on the future event horizon ${\cal H}^+$. From the definition of the tortoise coordinate~(\ref{tortoise}), we have $du=2dU/F(r_+-U)$. Hence in $(U,V)$-coordinates the metric becomes
\begin{equation}
 ds^2=-f(U,V) dUdV+r^2(U,V)d\Omega^2 \qquad f(U,V) = \frac{2F(r(U,V))}{F(r_+-U)}
\label{RNUV}
\end{equation}
where $r(U,V)$ is determined by solving $r_\ast(r)=(V-u(U))/2$. It can be seen that $r$ is analytic in
$(U,V)$ at $U=0$: we can expand $r$ for small $U$ as
\begin{equation}
 r(U,V)=
\left\{
\begin{array}{ll}
r_+-e^{\kappa_+ V} U -\frac{(r_+-2r_-)(e^{2\kappa_+ V}-e^{\kappa_+
  V})}{r_+(r_+-r_-)}U^2 + \cdots & (r_+>r_-)\\
r_+-U+\frac{V}{2r_+^2}U^2+\cdots & (r_+=r_-)
\end{array}
\right.
\label{rRN}
\end{equation}
hence we can extend the definition of $r(U,V)$ to $U \ge 0$ by analyticity. It follows that $F(r)/F(r_+-U)=e^{\kappa_+ V} + \mathcal{O}(U)$ for small $U$ (in both cases). This implies that the metric~(\ref{RNUV}) is analytic at $U=0$ so we can analytically continue to $U>0$. This gives a double null coordinate system regular across the future event horizon $\mathcal{H}^+$ at $U=0$. 

Note that simplifications occur when $V=0$ (in both cases):
\begin{equation}
\label{zeroVRNdata}
 r(U,0) = r_+ - U, \qquad f(U,0) = 2
\end{equation}

Eq.~(\ref{rRN}) shows that outgoing radial null geodesics (lines of constant $U$) diverge exponentially in $V$ near the horizon ($U=0$) in the non-extreme case. However, in the extreme case, these geodesics do not diverge to first order: they remain close to the horizon. This is closely related to the absence of a red-shift effect in the extreme case. It is interesting to consider this in more detail. Note that
\be
 r_*(r) = -U + 2r_+ \log |U| + \frac{r_+^2}{U} + \frac{V}{2}
\ee
Consider an outgoing radial null geodesic (line of constant $U$) at large $V$. If such a geodesic remains close to the black hole, i.e., $r/r_+ = {\cal O}(1)$, then the large positive term $V/2$ above must be almost cancelled by a large negative term $r_+^2/U$, i.e., $U \approx -2 r_+^2/V$. Since these two terms are both large, if we make a small change in $U$ so that exact cancellation no longer occurs then $r_*(r)$ will be large: it will be large and negative if 
$U\gtrsim -2r_+^2/V$ 
but large and positive if 
$U \lesssim -2r_+^2/V$. 
It follows that, at large fixed $V$, $r(U,V)$ varies slowly with $r(U,V) \sim r_+$ for 
$U\gtrsim -2r_+^2/V$ 
but then varies rapidly at $U\approx -2r_+^2/V$ and becomes large for 
$U\lesssim-2r_+^2/V$. 
Note that $r_+ - U$ is the ``initial'' value of $r$ on the outgoing null geodesics, so outgoing radial null geodesics which start within a distance $2r_+^2/V$ of the horizon are still near the horizon at time $V$. This can be contrasted with the behaviour for a non-extreme black hole, for which the ``critical'' value of $U$ is $-r_+ e^{-\kappa_+ V}$ and so the set of geodesics which is near the horizon at time $V$ is exponentially smaller than in the extreme case.

\section{Results: initial data with ingoing wave}

\label{app:ingoing}

In the main text, we studied initial data with an outgoing wavepacket localized near ${\cal H}^+$. In this Appendix, we study the non-linear time evolution of initial data with an {\it ingoing} wavepacket. In other words, we will determine what happens when a wavepacket falls into an extreme RN black hole. 

\begin{figure}
\centering
\includegraphics[width=7cm, clip]{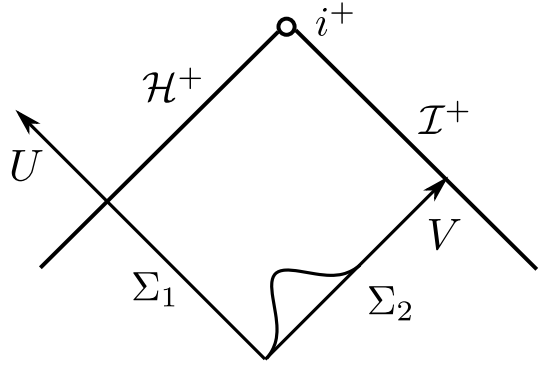}
\caption{Schematic plot for ingoing wave inital data. 
}
\label{Fig:inital_ing}
\end{figure}

On $\Sigma_2$ we take initial data describing an ingoing wavepacket (see Fig.~\ref{Fig:inital_ing})
\begin{equation}
 \phi(U_0,V)=
\begin{cases} 
\epsilon
  \exp\left[\alpha \left(\frac{1}{V-V_\textrm{ini}}-\frac{1}{V-V_\textrm{fin}}
+\frac{4}{V_\textrm{in}-V_\textrm{out}}
\right)\right] & (V_\textrm{ini}<V<V_\textrm{fin}) \\ 
0 & (\textrm{else}) 
\end{cases}
\ .
\label{init_ing}
\end{equation}
In our numerical calculations, we will fix the parameters as 
$\alpha=4$, $V_\textrm{ini}=0$ and $V_\textrm{fin}=5.9$.
On $\Sigma_1$, $\phi$ is set to be zero. 
In the same way as the outgoing wave initial data case, we fix the residual gauge freedom by choosing 
$r$ on $\Sigma$ as Eq.~(\ref{fix}). 

The final piece of data is the value of $f$ on $\Sigma_1$, which is constant because of Eq.~(\ref{C2}) and $r_{,U}=-1$ on $\Sigma_1$. We choose $f=2$ on $\Sigma_1$ (this is the extreme RN value of  $f$). On $\Sigma_2$, we determine $f$ by solving the
constraint~(\ref{C1}).

\begin{figure}
  \centering
  \subfigure[Apparent and event horizons]
  {\includegraphics[scale=0.5]{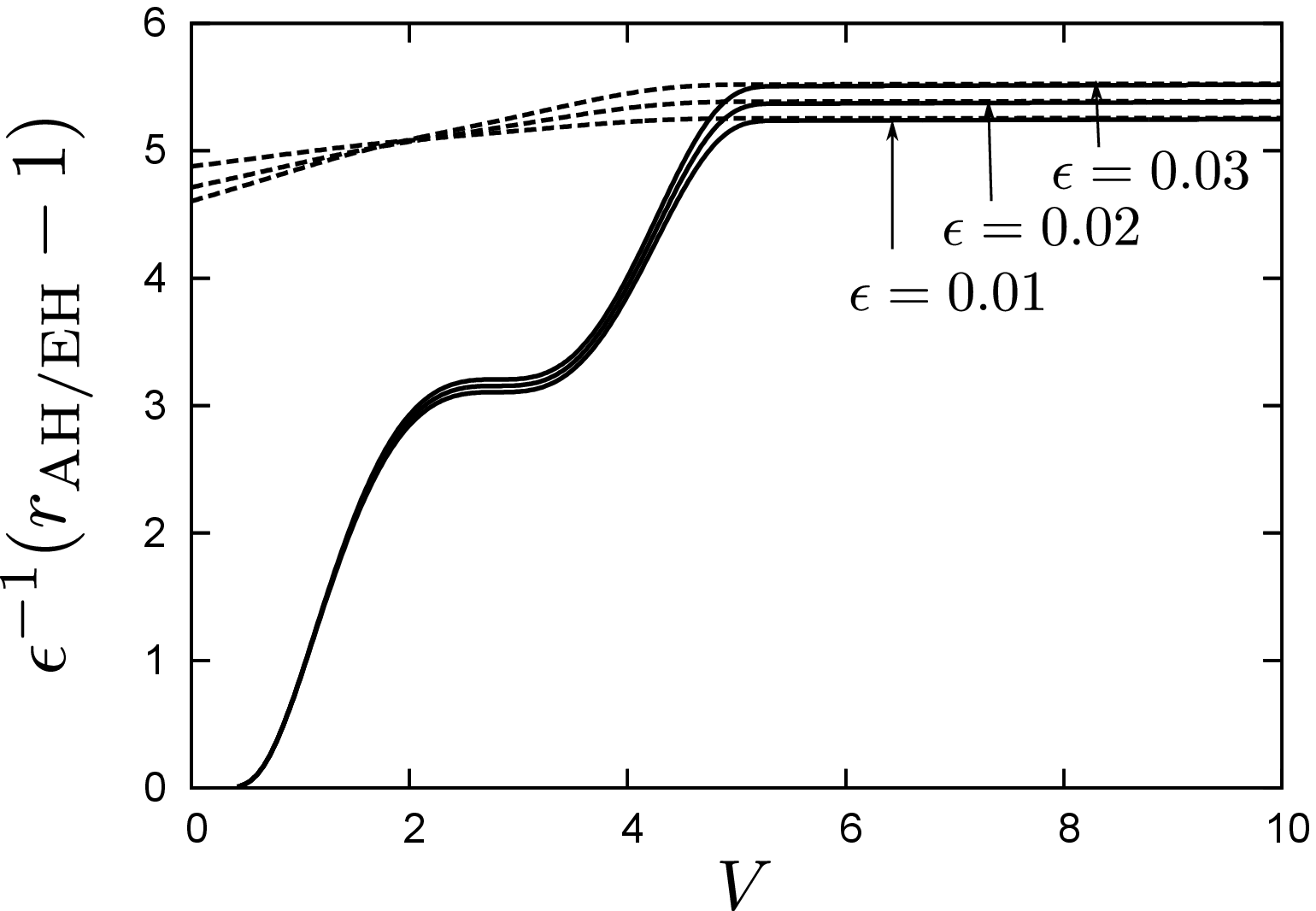} 
  }
  \subfigure[Bondi mass]
  {\includegraphics[scale=0.5]{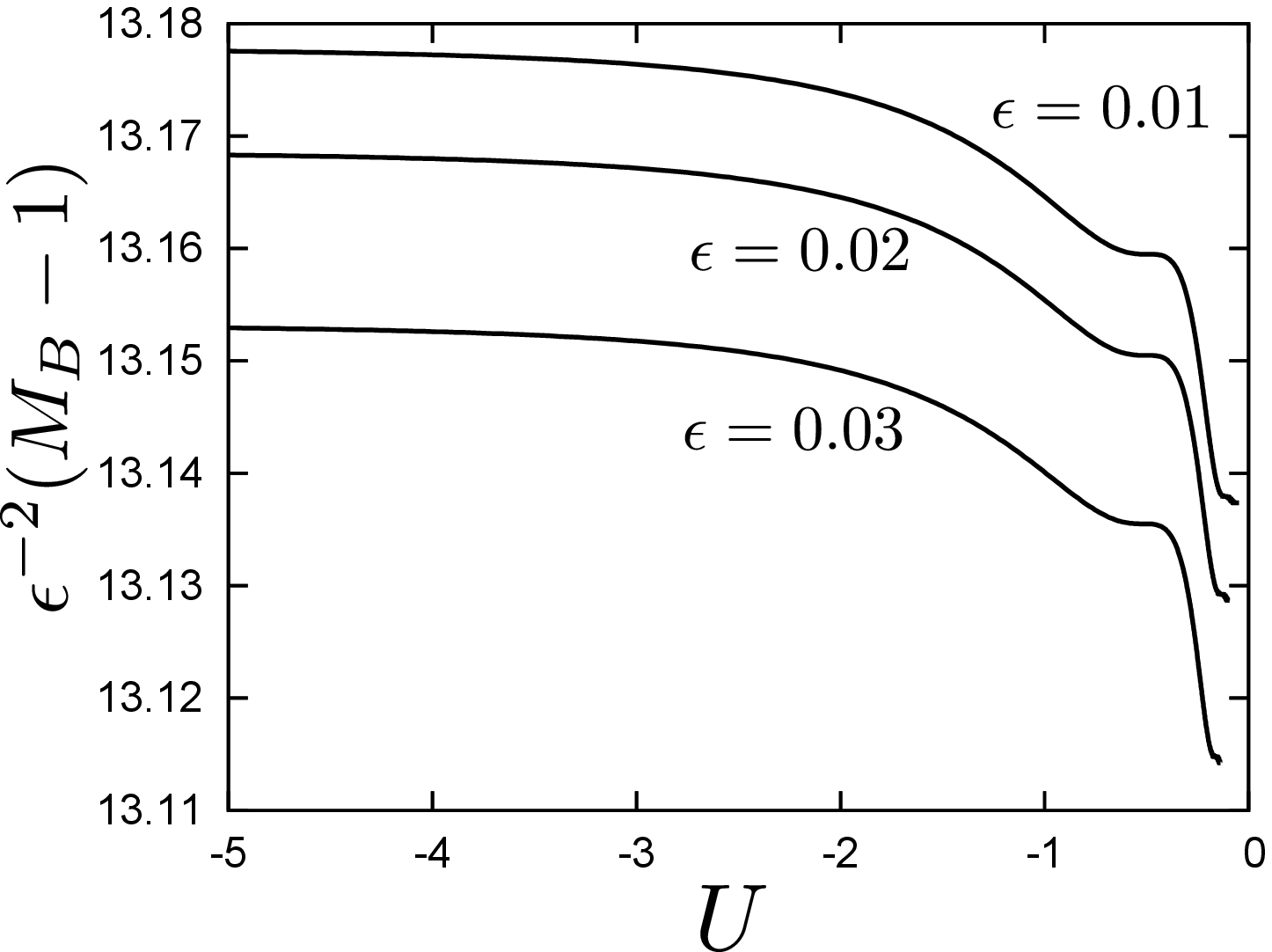}
  }
  \caption{
The left figure shows the time dependence of the radius of
 the apparent horizon (solid curves) and event horizon (dashed curves)
for ingoing wave initial data. 
They are almost constants in $V\gtrsim 5.9$.
The right figure shows the 
Bondi mass $M_B(U)$.
The right edges of the curves correspond to future timelike infinity.
\label{rUhoring}
}
\end{figure}

We solve the time evolution for $\epsilon=0.03,0.02,0.01$ using the same
numerical method as for an outgoing wave. 
In Fig.~\ref{rUhoring}(a), we show the time dependence of the apparent and event
horizons with different values of $\epsilon$. 
These functions change significantly in the region of $0\leq V \lesssim 5.9$
where the ingoing wave packet is supported. For $V\gtrsim 5.9$, the functions are almost independent of 
$V$. These results indicate that the backscatter of the wave
packet is negligible and the geometry is well-approximated by a non-extreme RN
solution in $V\gtrsim 5.9$.
In Fig.~\ref{rUhoring}(b), we show the $U$-dependence of the Bondi mass. 
We see that the Bondi mass is almost constant ($(M_f-1)/(M_i-1)\sim 10^{-3}$).
This confirms that most of the energy in the wavepacket enters the black hole rather than being scattered to infinity. 

\begin{figure}
  \centering
  \subfigure
  {\includegraphics[scale=0.45]{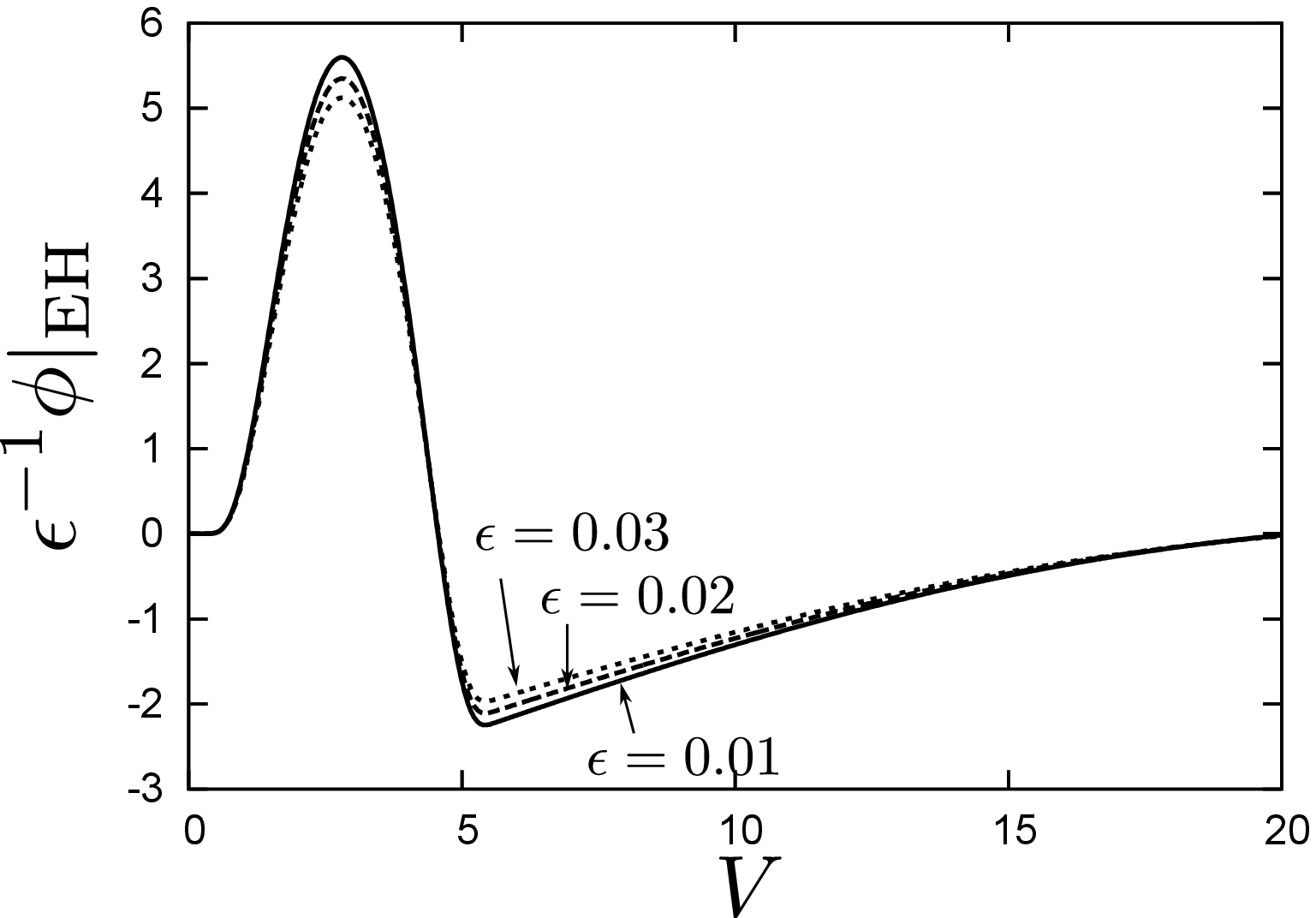}
  }
  \subfigure
  {\includegraphics[scale=0.45]{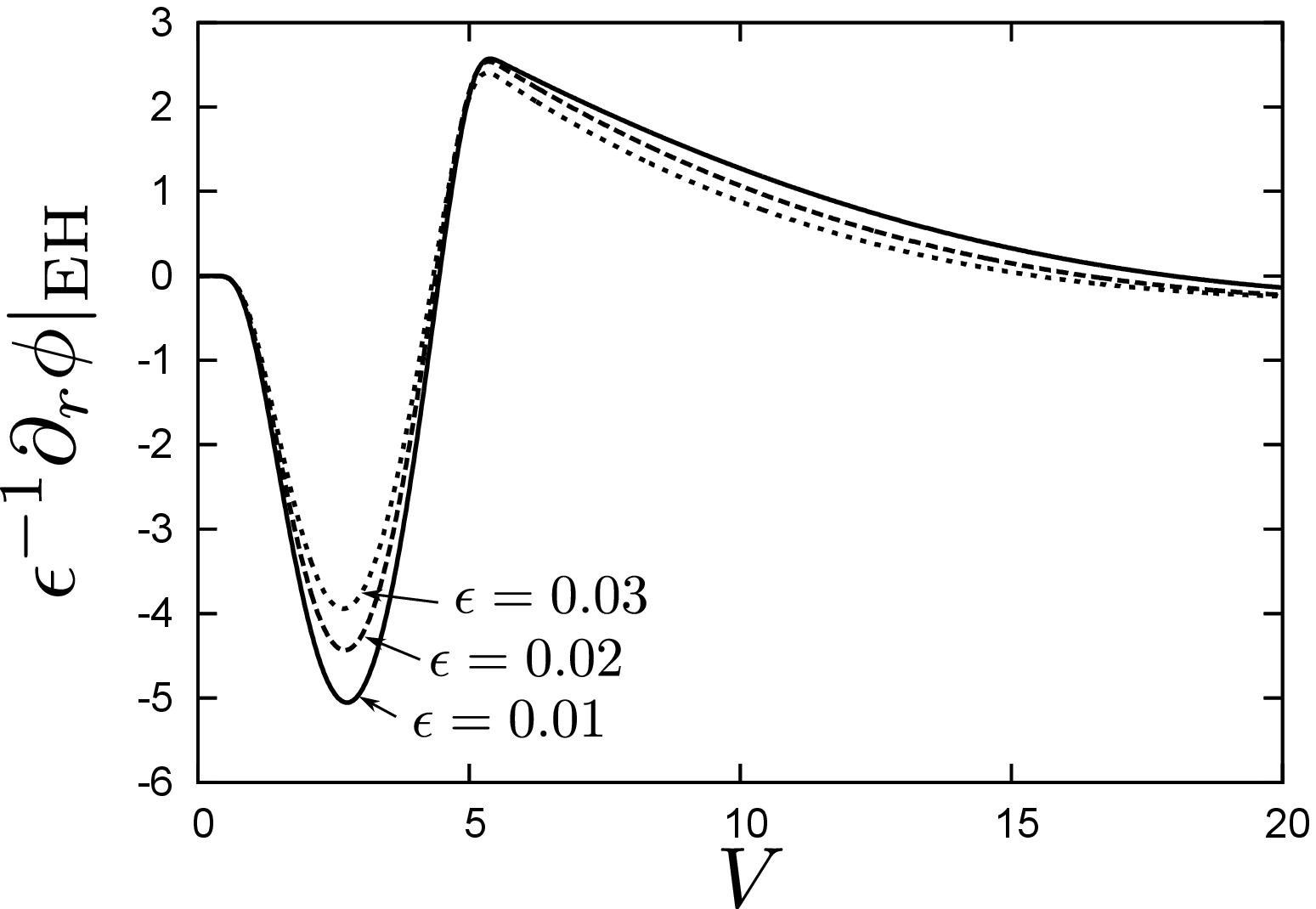} 
  }
  \subfigure
  {\includegraphics[scale=0.45]{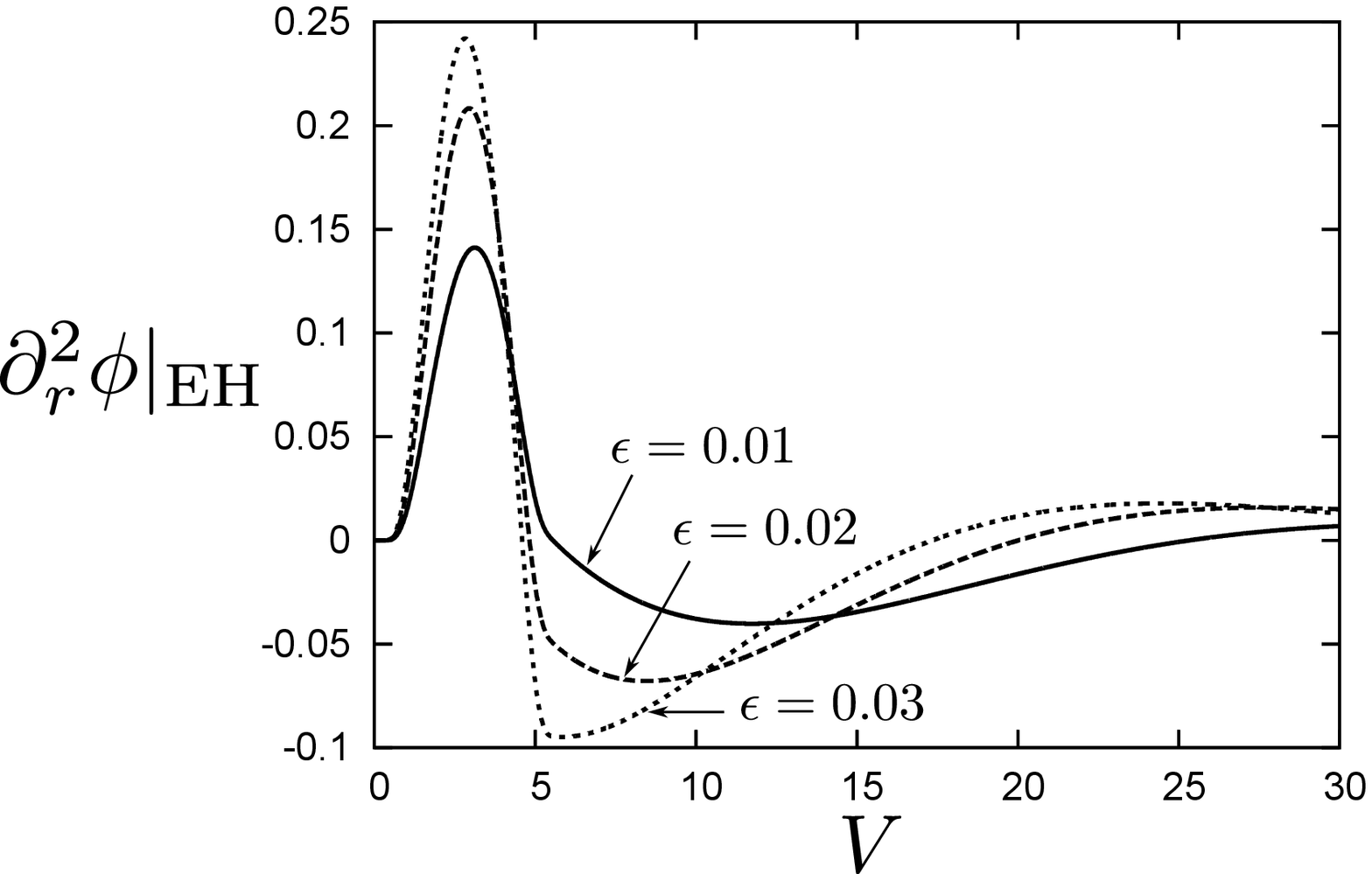} 
  }
  \subfigure
  {\includegraphics[scale=0.45]{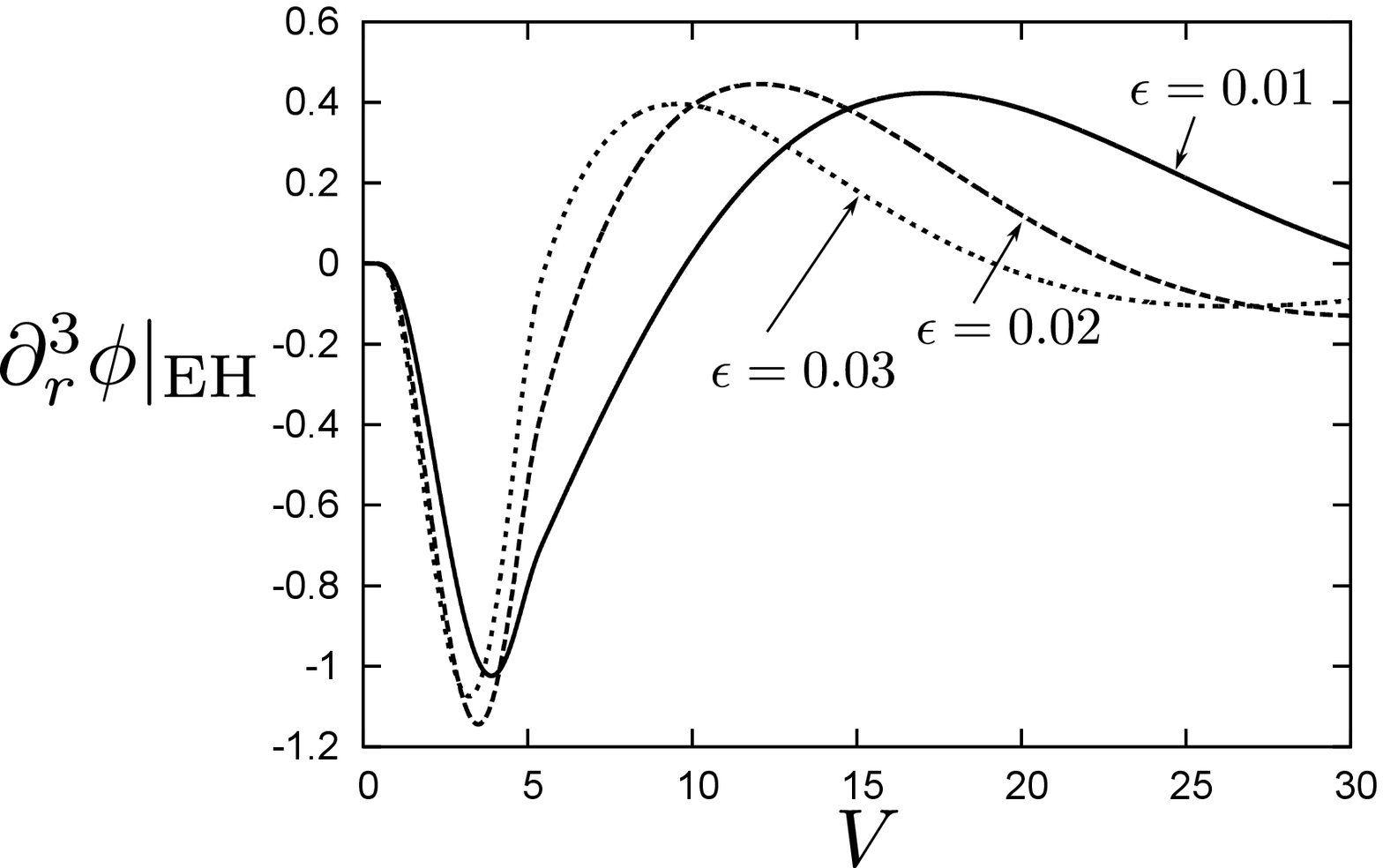} 
  }
  \caption{
$\partial_r^n \phi|_\textrm{EH}$ ($n=0,1,2,3$)
against $V$ for ingoing wave initial data. 
For $n=0,1,2$, they becomes small as $\epsilon$ decreases.
(Note that we factorize out $\epsilon$ for $n=0,1$.) 
For $n=3$, there are extrema at $V=17.2,12.0,9.5$ 
for $\epsilon=0.01,0.02,0.03$, respectively.
They appear to be almost independent on
$\epsilon$. 
\label{phiEH_ing}
}
\end{figure}

In Fig.~\ref{phiEH_ing}, we plot time dependence of scalar
field and its radial derivatives on the event
horizon. They decay at late time.
We also checked that scalar
field and its radial derivatives decay outside of the event horizon.
Thus, we can conclude that the end point of the time evolution is
non-extreme RN. The surface gravity of the final black hole is given by 
$\kappa=0.0526, 0.108, 0.166$ for
$\epsilon=0.01, 0.02, 0.03$, respectively.

Note that the maximum value of $|\partial_r^2\phi|_\textrm{EH}|$ decreases as $\epsilon \rightarrow 0$. 
Thus, we cannot find any evidence of an instability in the second derivative of the scalar field. However, 
in the plot for $\partial_r^3\phi|_\textrm{EH}$, 
there is a local maximum at $V=17.2,12.0,9.5$ 
for $\epsilon=0.01,0.02,0.03$, respectively.
We can see that the value of $\partial_r^3\phi|_\textrm{EH}$ at this local maximum appears to be almost independent of $\epsilon$. This implies that there is an instability: the maximum value of $\partial_r^3\phi|_\textrm{EH}$ is ${\cal O}(1)$ as $\epsilon \rightarrow 0$. 

In summary, for initial data describing an ingoing wave, there is an instability in the third transverse derivative of the scalar field at the event horizon, just as for a test scalar field \cite{Lucietti:2012xr,Bizon:2012we,Aretakis:2012bm}. The endpoint of the instability is a non-extreme RN black hole. 

\section{Error analysis}
\label{errors}

Our numerical calculation is based on Burko and Ori's method~\cite{Burko:1997tb}. 
In their method, an adaptive mesh refinement is used and the number of
grid points for the $U$-coordinate depends on $V$ although the time step
$\delta V$ is fixed. We refer to the initial number of grid points as
$N$ hereafter and see how our numerical errors depend on $N$.

\subsection{Constraint violation}

\begin{figure}
  \centering
    \subfigure
  {\includegraphics[scale=0.5]{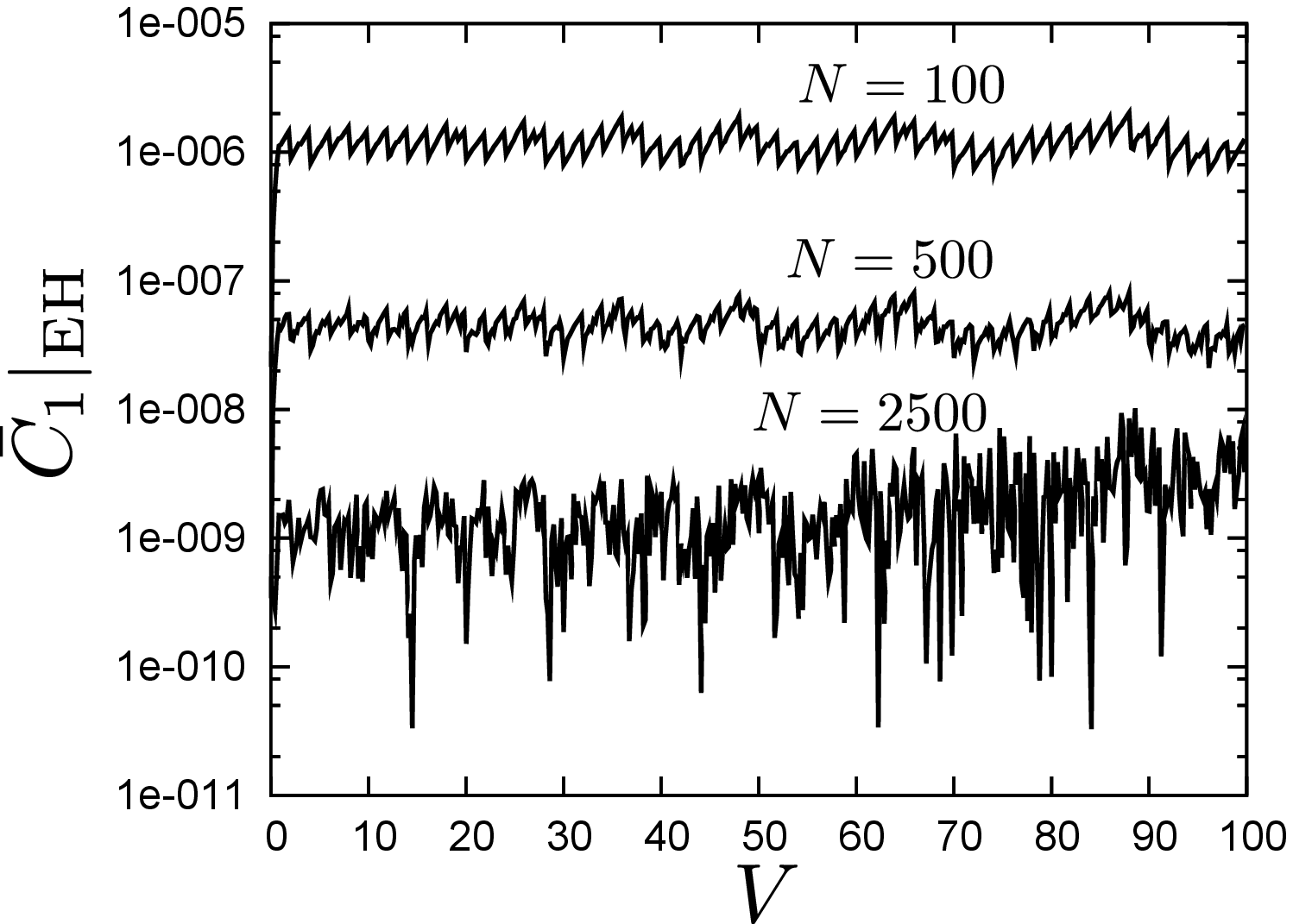}
  }
 \subfigure
  {\includegraphics[scale=0.5]{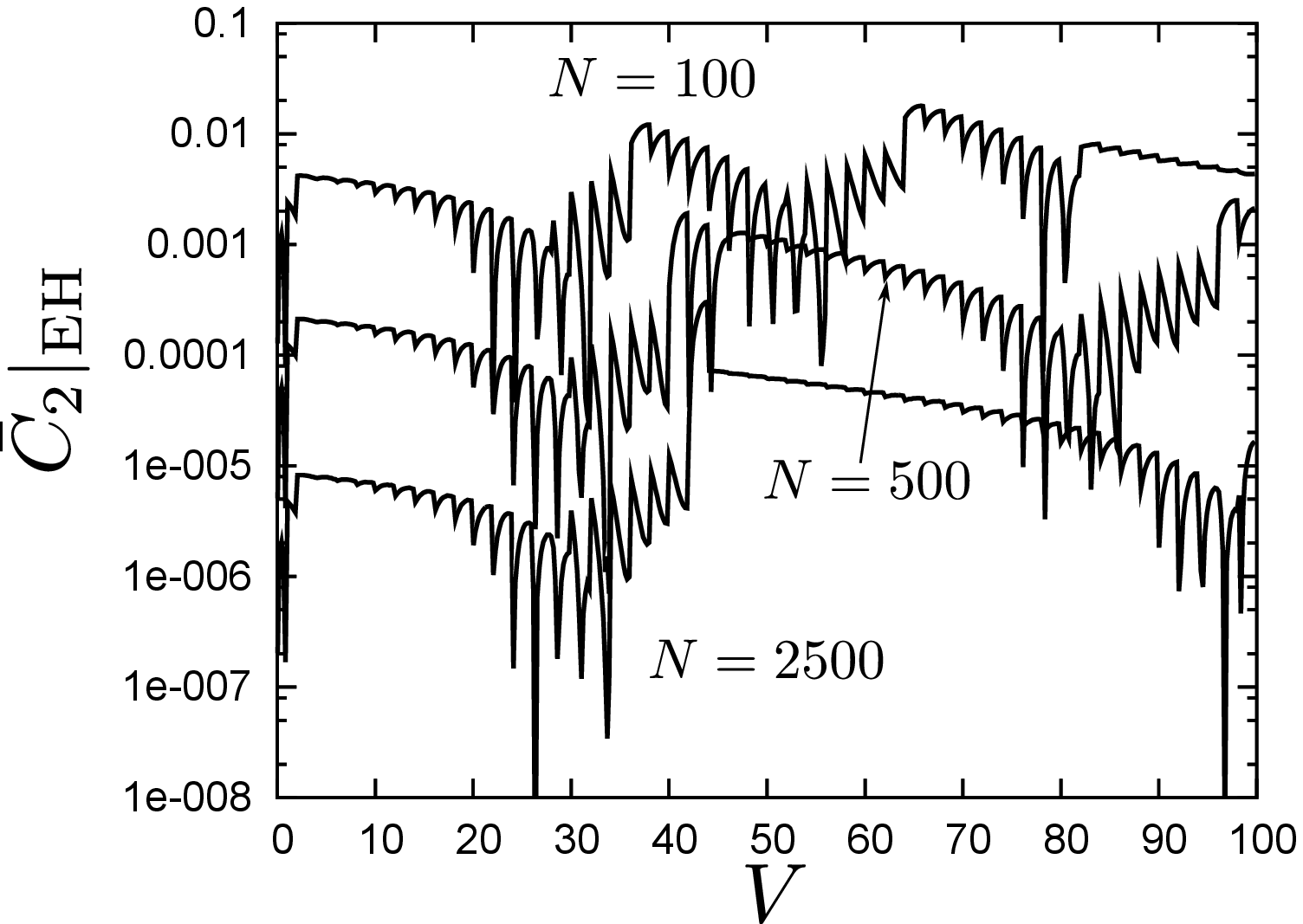}
  }
 \subfigure
  {\includegraphics[scale=0.5]{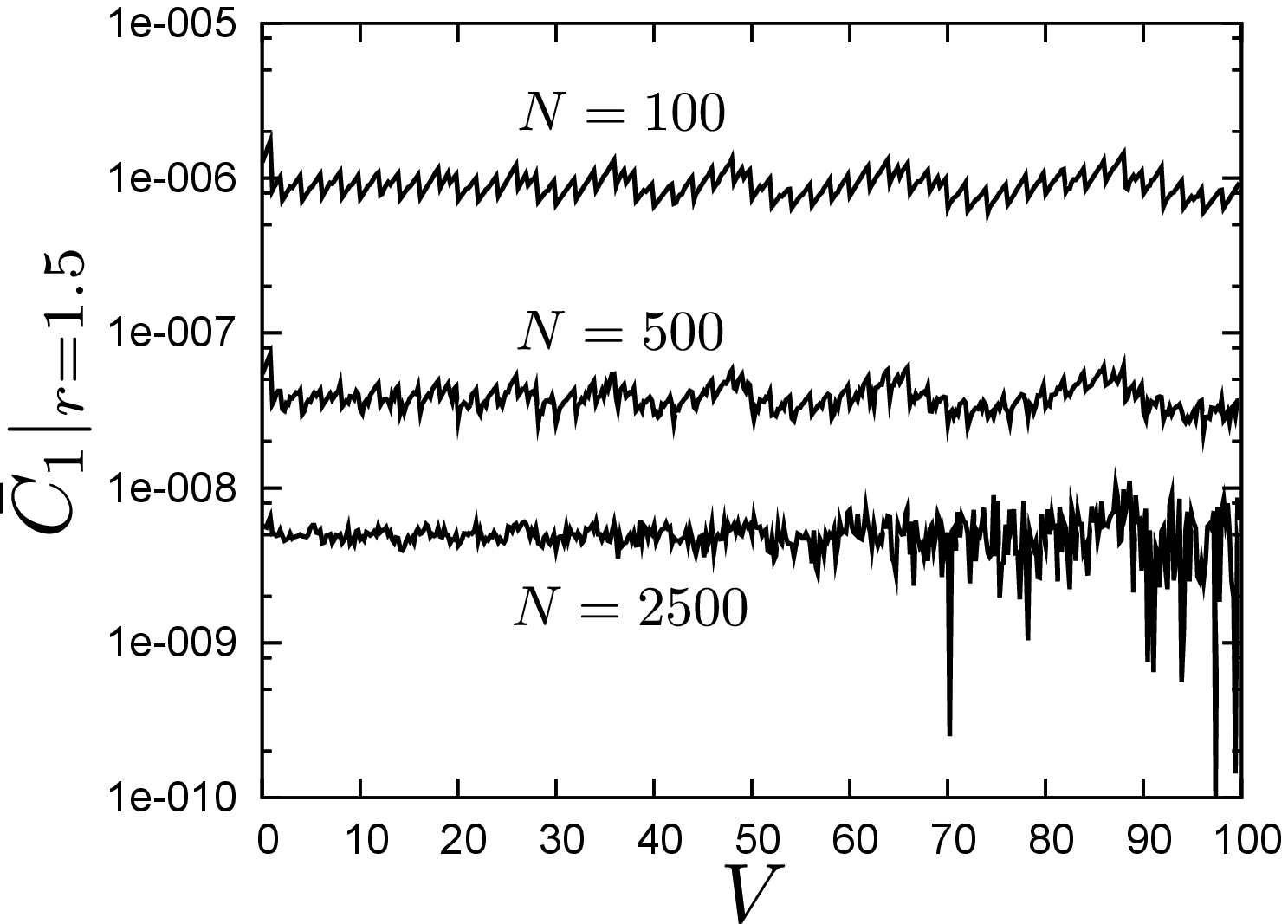}
  }
 \subfigure
  {\includegraphics[scale=0.5]{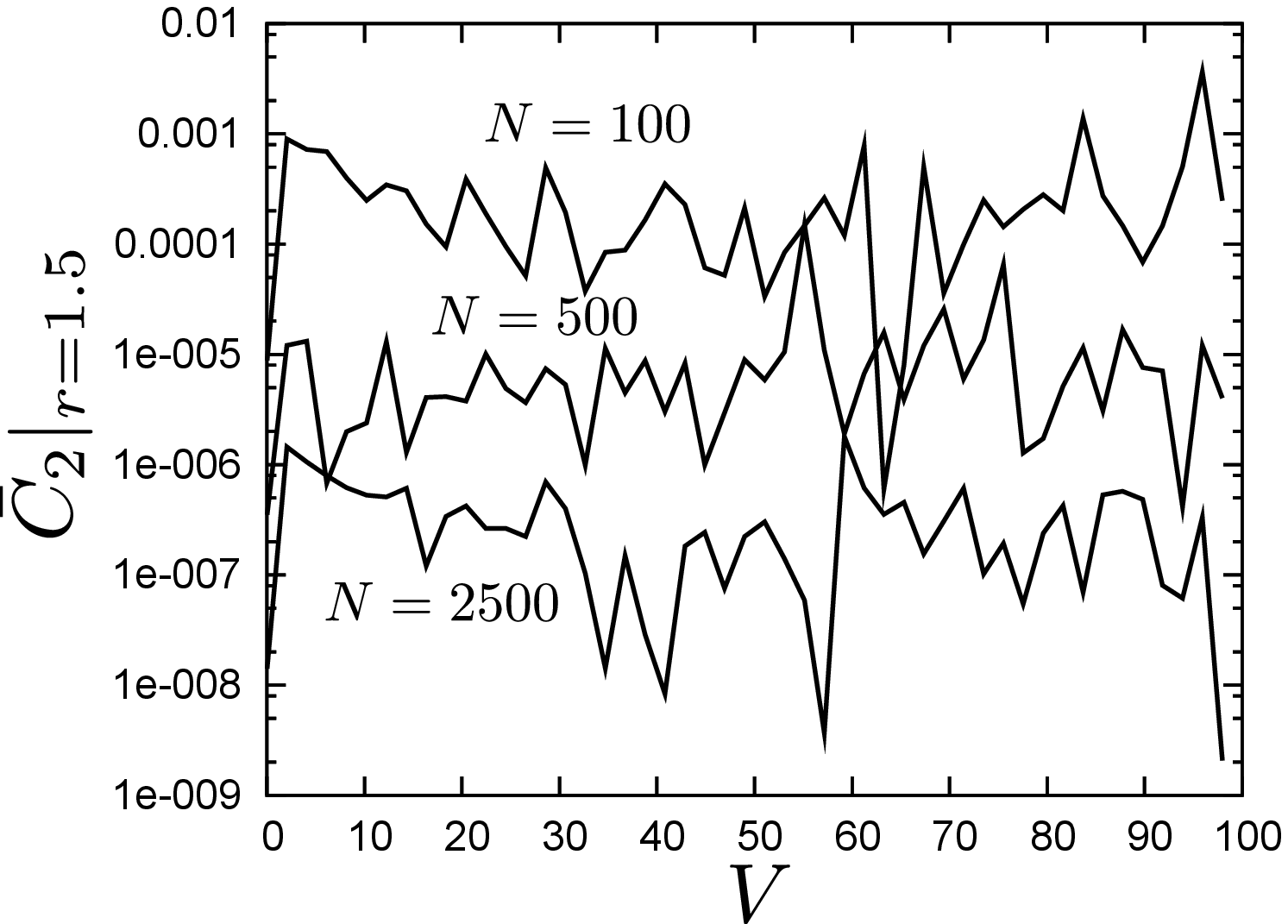}
  }
  \caption{
The gauge invariant constraints at $r=r_\textrm{EH}$ and $r=1.5$.
We vary the initial grid number for $U$-coordinate for
 $N=100,500,2500$ and fix the time step as $\delta V=0.02$.
They are roughly independent of $V$ and decrease as $N$ increases. 
This demonstrates the numerical stability
of the simulations. 
\label{C1C2}
}
\end{figure}

First, we check the constraint violation to monitor the numerical errors. 
Eqs.~(\ref{C1}) and (\ref{C2}) are not appropriate to
check the numerical errors since they are 
not invariant under the residual gauge
transformations~(\ref{resgauge}). 
Thus, we define gauge invariant constraints as
\begin{equation}
 \bar{C}_1=\frac{r_{,U}^2}{f^2}\left(r_{,VV}-\frac{r_{,V}f_{,V}}{f}+\frac{r\phi_{,V}^2}{4}\right)\
  ,\quad
 \bar{C}_2=\frac{1}{r_{,U}^2}\left(r_{,UU}-\frac{r_{,U}f_{,U}}{f}+\frac{r\phi_{,U}^2}{4}\right)\ .
\end{equation}
In Fig.~\ref{C1C2}, we plot the gauge invariant constraints evaluated at
the event horizon and at $r=1.5$. 
We took the initial data with a degenerate apparent horizon for
$\epsilon=0.05$. 
The time step is fixed as 
$\delta V=0.02$ and the initial grid number for $U$-direction is varied as 
$N=100,500,2500$. 
We find that the constraints are roughly constants and do not grow as $V$
increases. This demonstrates the numerical stability
of the simulations. 
The constraints decrease as $N$ increases. 
Imposing $\bar{C}_1,\bar{C}_2 \lesssim 10^{-2}$, we
obtain a condition for $N$: $N\gtrsim \textrm{few}\times 10^2$.
However, as we will explain in the next subsection, we need larger $N$ to resolve the apparent horizon. 
We also studied the $\delta V$ dependence of the constraints: 
We varied the time step in the range of $0.01\leq\delta V\leq 0.1$ for fixed $N=500$.
However, the constraints did not depend on $\delta V$ much in that range. 
This result indicates that $\delta V=0.02$ is small enough to keep
the constraints small.

\subsection{The drift of the apparent horizon}

\begin{figure}
  \centering
    \subfigure
  {\includegraphics[scale=0.5]{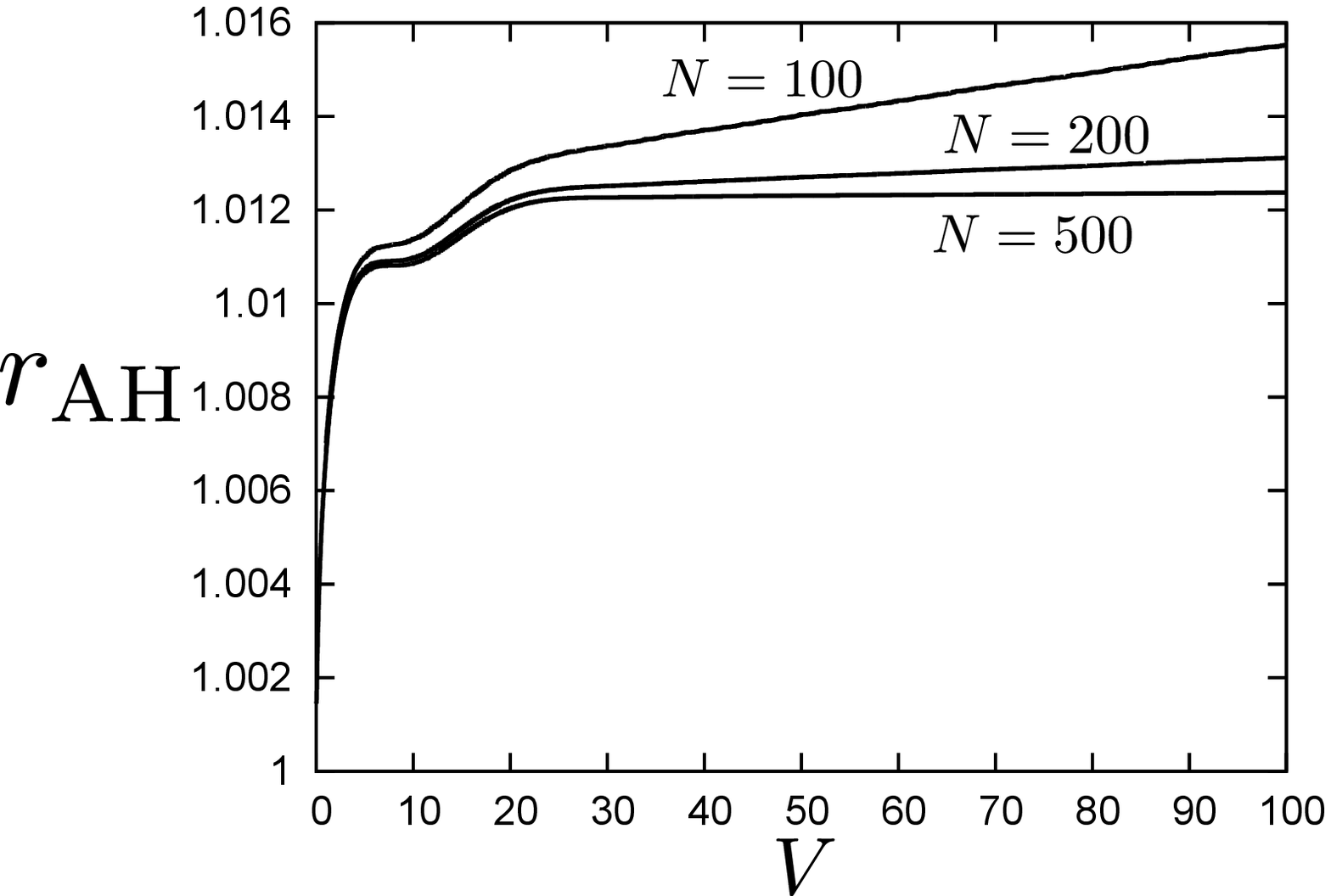}
  }
 \subfigure
  {\includegraphics[scale=0.5]{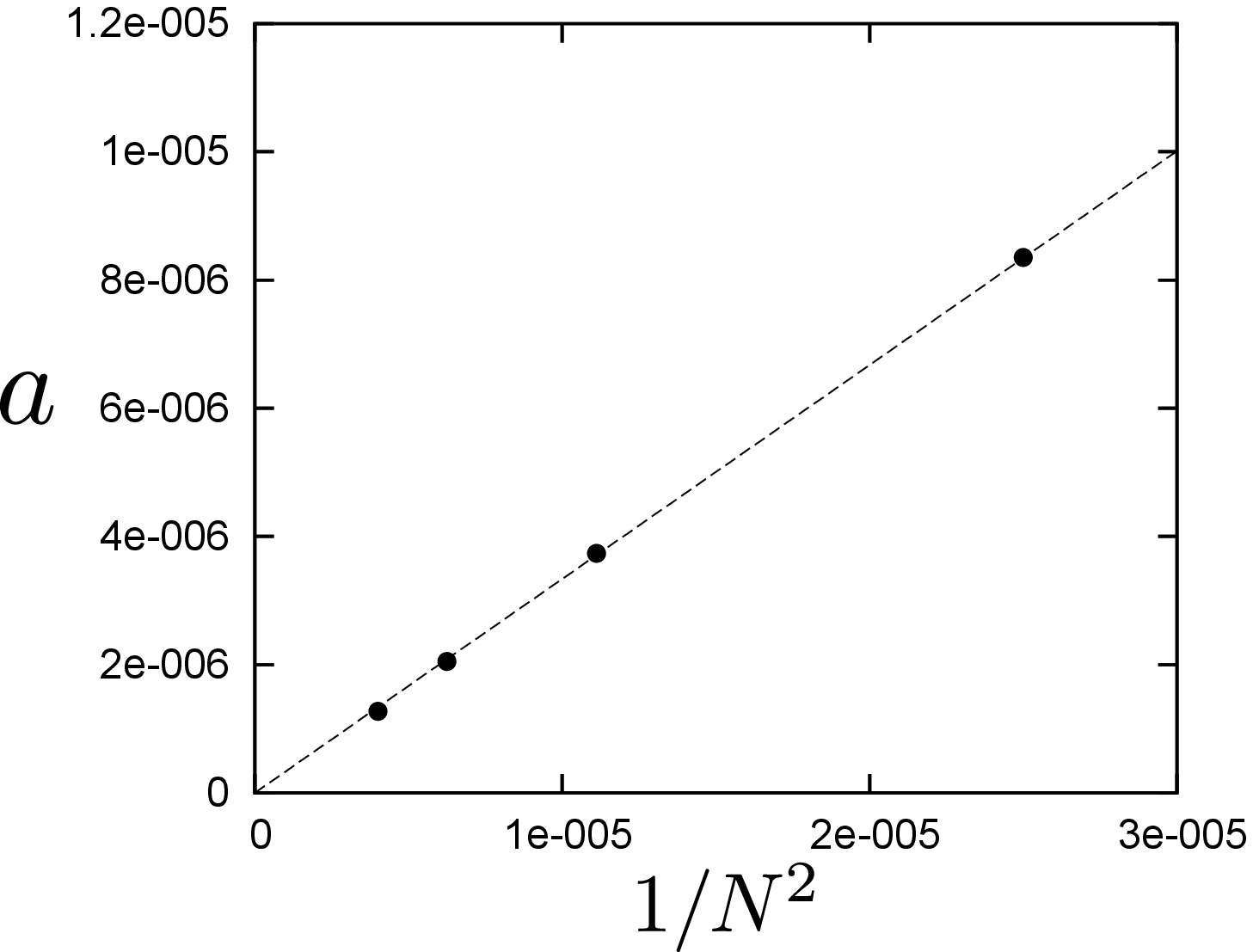}
  }
  \caption{
The left figure shows the radius of the apparent horizon for
 $N=100,200,500$. The step size is fixed as $\delta V=0.02$.
The apparent horizon
depends on $V$ linearly at late time.
Fitting the curve to $aV+b$ for $50\leq V\leq 100$, 
we plot $a$ against
 $1/N^2$ in the right figure. 
\label{drift}
}
\end{figure}

In our numerical simulations, the location of the apparent horizon is
relatively sensitive to the resolution. Thus, in this subsection, we
study the numerical error in the apparent horizon. In Fig.~\ref{drift} (left),
we plot the radius of the apparent horizon for the initial data with
degenerate apparent horizon with $\epsilon=0.05$. 
We take the initial grid number as $N=100,200,500$ and fix the step size as $\delta V=0.02$. 
At late time, the apparent horizon $r_\textrm{AH}$ should approach a
constant since the end point of the time evolution is a non-extreme RN black hole. 
However, in the actual numerical calculation, the apparent horizon
depends on $V$ linearly at the late time because of the numerical error.

We fit $r_\textrm{AH}$ to a linear function $aV+b$ for 
$50\leq V\leq 100$ and find that the $a$ depends on $N$ as in Fig.~\ref{drift} (right).
From the figure, we obtain $a\simeq 0.3/N^2$. Thus, the numerical error
in $r_\textrm{AH}$ can be estimated as $\delta r_\textrm{AH}\sim V/N^2$.
Now, we impose a condition that the relative error in $r_\textrm{AH}$ is
less than $1\%$: $\delta r_\textrm{AH}/(r_\textrm{AH}-1)\lesssim 10^{-2}$.
From the condition, we obtain $N^2 \gtrsim 10^2\times V/(r_\textrm{AH}-1)$.
In our numerical calculations, typically the parameters are
$V\sim 10^3$ and $r_\textrm{AH}-1\sim 10^{-3}$. 
Thus, we need a initial grid number with $N\gtrsim 10^4$. 
In this paper, we typically choose $N=4\times 10^4$.
We also checked the $\delta V$ dependence of the apparent horizon
and found that, in the range of $0.01\leq\delta V\leq 0.1$, 
the apparent horizon radius is almost independent of $\delta V$. This result
indicates that 
the step size $\delta V=0.02$ is small enough to determine the apparent
horizon accurately. Thus, in the most part of this paper, we set $\delta V=0.02$.

\section{Evolution equations for radial derivatives of the scalar field}
\label{App:phir}

In section \ref{NIERN}, we evaluated radial derivatives of
the scalar field: $\partial_r^n\phi$ ($n=1,2,3$). 
The direct numerical differentiations tend to lose accuracy as $n$
increases. Thus, instead, we determine the derivatives by obtaining and solving
evolution equations for $\phi^{(n)}\equiv \partial_r^n\phi$.
These equations are:
{\allowdisplaybreaks
\begin{align}
&\phi^{(1)}_{,UV}=\frac{1}{16r^4r_{,U}}
\big[
 r^2 (r^2-Q^2) r_{,U} f(\phi^{(1)})^3
\notag\\
&\hspace{3cm}
+4 r_{,U} (4 r^2 r_{,V} r_{,U}-r^2 f+3Q^2f)\phi^{(1)} 
+4r(r^2-Q^2)f\phi^{(1)}_{,U}
+32 r^2 r_{,U}^2 \phi_{,V}
\big]
\\
&
\phi^{(2)}_{,UV}=\frac{1}{64r^5 r_{,U}}\big[
-32r^2(-r^2f-2r^2r_{,V}r_{,U}+Q^2f)\phi^{(2)}_{,U}
\notag\\
&\hspace{1.5cm}
-4rr_{,U}\{
16r^2r_{,V}r_{,U}
+16r^2f
- 40 Q^2 f
-5r^2 f(r^2-Q^2) (\phi^{(1)})^2
\}
\phi^{(2)}
\notag\\
&\hspace{3cm}
+r^4r_{,U}f(r^2-Q^2)(\phi^{(1)})^5
-4r^2 r_{,U} f(r^2-5Q^2)(\phi^{(1)})^3
\notag\\
&\hspace{4.5cm}
-16r_{,U}(20r^2r_{,V}r_{,U}-r^2f+11Q^2f)\phi^{(1)}
-384r^2 r_{,U}^2 \phi_{,V}
\big]
\\
&\phi^{(3)}_{,UV}
=\frac{1}{256 r^6 r_{,U}}\big[
-64r^3(-3 r^2 f-8 r^2 r_{,V} r_{,U}+3Q^2 f)\phi^{(3)}_{,U}
\notag\\
&\hspace{1cm}
-64r^2 r_{,U}\{20r^2 r_{,V} r_{,U} - 2 r^4 f (\phi^{(1)})^2  
+ 2 Q^2 r^2 f (\phi^{(1)})^2 + 9 r^2 f - 21 Q^2 f\}\phi^{(3)}
\notag\\
&\hspace{1cm}
+160 r^4 (r^2-Q^2) r_{,U} f \phi^{(1)}(\phi^{(2)})^2
-8 r r_{,U} \{64 r^2 r_{,V} r_{,U}-5 r^6 f (\phi^{(1)})^4 +14
 r^4f(\phi^{(1)})^2 
\notag\\
&\hspace{1cm}
+5 Q^2 r^4 f (\phi^{(1)})^4-80 r^2 f-70 Q^2 r^2 f (\phi^{(1)})^2+416 Q^2 f\}\phi^{(2)}
-r^6 (r^2-Q^2) r_{,U} f (\phi^{(1)})^7
\notag\\
&\hspace{1cm}
+24 Q^2 r^4 r_{,U} f(\phi^{(1)})^5
+32 r^2 (r^2-13 Q^2) r_{,U} f (\phi^{(1)})^3
\notag\\
&\hspace{1cm}
+128 r_{,U} (52 r^2 r_{,V} r_{,U}+r^2 f+25 Q^2 f)\phi^{(1)}
-6144r^2 r_{,U}^2 \phi_{,V}
\big]\ ,
\end{align}
}%
where we used evolution equations~(\ref{fUV}-\ref{phiUV}),  
constraint equations~(\ref{C1},\ref{C2}) and their $U$ derivatives 
to obtain the above expressions. 
Solving the above equations, we can evaluate the radial derivatives of the 
scalar field. Note that the right hand sides of above
equations are determined by $f$, $r$, $\phi$, $\phi^{(n)}$ and their
first derivatives. Thus, we do not have to evaluate the second
derivatives of these functions when we solve the time evolution for
$\phi^{(n)}$. 

\end{document}